\documentclass[%
preprint,
showpacs,preprintnumbers,
 amsmath,amssymb,
 aps,
pre,
]{revtex4-1}

\usepackage[colorlinks,linkcolor=blue,citecolor=blue]{hyperref}
\usepackage{graphicx}
\usepackage{subfigure}
\usepackage{color}
\usepackage{url}
\usepackage{hyperref}
\usepackage{multirow}
\usepackage{amsmath,amssymb}
\usepackage{caption}    
\usepackage{lineno,hyperref}
\usepackage{dcolumn}
\usepackage{bm}
\modulolinenumbers[5]

\begin{document}




\title{A simplified discrete unified gas kinetic scheme for incompressible flow}

\author{Mingliang ZHONG}
\email[Zhong, M. L.: ]{mzhongad@connect.ust.hk}
\author{Sen ZOU}
\email[Zou, S.: ]{2019100412@mail.nwpu.edu.cn}
\author{Dongxin Pan}
\email{chungou@mail.nwpu.edu.cn}
\author{Congshan Zhuo}%
\email{zhuocs@nwpu.edu.cn}
\author{Chengwen Zhong}%
\email{Corresponding author: zhongcw@nwpu.edu.cn}
\affiliation{School of Aeronautics, Northwestern Polytechnical University, Northwestern Polytechnical University, Xi'an, Shaanxi 710072, China.}

\date{\today}

\begin{abstract}
The discrete unified gas kinetic scheme (DUGKS) is a new finite volume (FV) scheme for continuum and rarefied flows which combines the benefits of both Lattice Boltzmann Method  (LBM) and unified gas kinetic scheme (UGKS). By reconstruction of gas distribution function using particle velocity characteristic line, flux contains more detailed information of fluid flow and more concrete physical nature. In this work, a simplified DUGKS is proposed with reconstruction stage on a whole time step instead of half time step in original DUGKS. Using temporal/spatial integral Boltzmann Bhatnagar-Gross-Krook (BGK) equation, the transformed distribution function with inclusion of collision effect is constructed. The macro and mesoscopic fluxes of the cell on next time step is predicted by reconstruction of transformed distribution function at interfaces along particle velocity characteristic lines. According to the conservation law, the macroscopic variables of the cell on next time step can be updated through its macroscopic flux. Equilibrium distribution function on next time step can also be updated. Gas distribution function is updated by FV scheme through its predicted mesoscopic flux in a time step. Compared with the original DUGKS, the computational process of the proposed method is more concise because of the omission of half time step flux calculation. Numerical time step is only limited by the Courant-Friedrichs-Lewy (CFL) condition and relatively good stability has been preserved. Several test cases, including the Couette flow, lid-driven cavity flow, laminar flows over a flat plate, a circular cylinder, and an airfoil, as well as micro cavity flow cases are conducted to validate present scheme. The numerical simulation results agree well with the references' results.
\end{abstract}


\maketitle
\section{\label{sec:level1}Introduction}
Lattice Boltzmann method (LBM) was developed from the lattice gas automation model. Its implementation can be divided into two steps: streaming and collision. Since the discrete velocity models are coupled with the computational mesh, the distribution function can be precisely streaming from one node of mesh to its neighbor node in one time-step \cite{Chen1998LATTICE}. The macroscopic flow variables can then be calculated using the distribution function. LBM requires a small amount of coding effort while the computational and memory cost for large scale problems are significant \cite{Guo2013Lattice}.

The standard LBM is constructed based on the Cartesian mesh, which leads to a shortcoming in adaption to extensive application. For sufficient resolution, mesh points have to be distributed in computational field with uniform spatial scale, which usually produces enormous mesh cells and huge computational cost. In practical applications, we have to dispose of geometry with complex boundaries. Arbitrary mesh nodes distribution cannot be well fitted by uniform lattices. These two issues reduce computational efficiency and limit its application in more complex flow cases.

In the past few decades, there have been many attempts to apply LBM to arbitrary geometrical computational domains. X. He, et al. proposed a mesh based mesoscopic interaction for LBM \cite{He1996Some}. The advantages of LBM are retained and simulation results agree well with experimental data. J. Yao, et al. introduced an adaptive mesh refinement (AMR) method which refines meshes by constructing a linked-list of geometries for mesh levels refinement. Linked-list mesh data and LBM calculation are combined \cite{Yao2017An}. X. Guo, et al. applied AMR to immersed boundary LBM and bubble function in interpolation \cite{Guo2015A}. These methods extend LBM to more universal cases. However, numerical time step is still restricted by particle relaxation time. O. Aursj\o, et al. proposed local lattice Boltzmann algorithm for including a general mass source term that results in Galilean invariant continuum equations \cite{AursjOn}. D. Wang, et al. used LBM to simulate viscoelastic drops \cite{Di2019A}.

Recently, finite volume LBM (FVLBM) is developed to apply LBM to hybrid meshes and overcome the time step restriction. In this method, the body-fitted mesh and hybrid mesh can be used. S. Succi, et al. were the first to propose a finite volume (FV) scheme for discrete Boltzmann equations \cite{Nannelli1992The}. G. Peng, et al. developed the FV scheme based on LBM which was first applied to the unstructured mesh \cite{Xi1999Finite}. M. Stiebler, et al. developed an upwind discrete scheme for the FVLBM \cite{Stiebler2006An}. W. Li, et al. proposed a grid-transparent FVLBM, which shows low dissipation and high accuracy in simulation of viscous flows \cite{Li2016Finite}. Y. Wang, et al. studied the performance of a FVLBM scheme which is discretized on a unstructured mesh and simulated the steady and unsteady flows at relatively high Reynolds number. Results show lower computational cost and good agreement with previous benchmark data \cite{Wang1,wang2}. L. Chen proposed a unified and preserved Dirichlet boundary treatment \cite{Chen2015A}.

Generally, in original FVLBM, numerical time step is still limited by relaxation time. On a large time step, the computational process is no longer stable and non-physical oscillation occurs. Due to temporal integration of collision term decoupled with advection term, the lack of real particle collision mechanism in flux evaluation restricts marching time step to a same order of magnitude with collision time \cite{Lianhua2016Discrete}. Discrete unified gas kinetic scheme (DUGKS) is a new kind of FV scheme for discrete velocity method (DVM), which combines the benefits of both lattice Boltzmann equation (LBE) and unified gas kinetic scheme (UGKS) methods \cite{Guo2013Discrete}. In DUGKS, the evolution of the flux on the cell interface is simplified by transformation of distribution function coupling with collision term. Half time step advection is used to replace the original one, which provides DUGKS semi-implicit property. Different from particle-based method \cite{zhang2019particle, fang2020dsmc}, multi-scale property in DUGKS is embodied in temporal/spatial reconstruction in flux evaluation. Using particle velocity characteristic line, particle transport and collision processes are accurately traced. Gas nature is also described with high fidelity. In the recent study, DUGKS has been widely extended. DUGKS has been applied to rarefied gas flow, X. Zhao, et al. proposed a reduced order modeling-based DUGKS, which apply the reduced velocity space to the conventional DUGKS \cite{Zhao2020}. Recently DUGKS has been extended to binary gas mixtures of Maxwell molecules, and Y. Zhang, et al. extended DUGKS to gas mixture flows based on the McCormack model \cite{Zhang2019Discrete}.

Many works have been done to develop efficient, accurate and robust methods based on DUGKS. P. Wang, et al developed a coupled DUGKS \cite{Wang2015A}. By implementation of kinetic boundary condition and solving velocity and temperature independently, convection flows from laminar to turbulent flows are accurately simulated. C. Wu, et al. retained particle acceleration term in original Boltzmann equation and considered force term in non-equilibrium gas distribution function \cite{Wu2016Discrete}. Good accuracy and efficiency are validated in force-driven flow cases. C. Shu, et al. constructed a third-order DUGKS \cite{Wu2018Third}. Using Runge-Kutta temporal marching and high-order spatial reconstruction, more accurate results can be obtained with lower-resolution mesh. C. Zhang, et al. introduced a finite volume discretization of the anisotropic gray radiation equation based on DUGKS \cite{song2020discrete}. Radiative transfer in anisotropic scattering media was accurately simulated. J. Chen, et al. developed a conserved DUGKS for multi-scale problems \cite{Chen2019A}. In the unstructured particle velocity space, macroscopic variables are updated by moments of mesoscopic gas distribution function flux. Conservation property is greatly improved. Based on DUGKS solver, Z. Yang, et al. applied phase-field method to simulation of two-phase flows \cite{Yang2019Phase}. Several cases in which interfaces undergo severe deformation are efficiently simulated and many delicate details are accurately captured. D. Pan, et al. proposed an implicit DUGKS based on Lower-Upper Symmetric Gauss Seidel (LU-SGS) iteration \cite{Pan2019An}. In all flow regimes, computational efficiency can be improved by one or two orders of magnitude in comparison with explicit method.

In present work, a kinetic flux reconstruction strategy is introduced in modeling flow evolution. Macroscopic flux is applied to flow evolution for improving macroscopic conservation \cite{zhang2020double}. Gas distribution function at next time step is applied to reconstruct particle motion in a mesh cell \cite{Yang2019An}. Different from original DUGKS, collision term is not implicitly included in evolutionary process. Instead, the Bhatnagar-Gross-Krook (BGK) operator is integrated in time and space using Maxwellian and gas distribution function at next time step. Gas distribution function is predicted by temporal and spatial reconstruction on particle velocity characteristic lines. Maxwellian is predicted by macroscopic variables which are updated by moments of mesoscopic gas distribution function flux. To extend present method to unstructured meshes, accurate and robust interpolation method is crucial. In previous gas kinetic model, linear least squares regression (LLSR) was proved to be an efficient and accurate way for interpolation \cite{Lenz2019An,Li2019A,Pan2016A}. In present work, LLSR is applied in spatial interpolation. The time step in this flux reconstruction strategy is only constrained by Courant-Friedrichs-Lewy (CFL) condition which is more than ten times larger than relaxation time, which is more efficient and robust than FVLBM. By using explicit discretization of collision term, present method is simpler than original DUGKS.

Our paper is organized as follows. Discretization of governing equation, reconstruction method and evaluation of both mesoscopic and macroscopic fluxes are presented in Sec.~\ref{algorithm}. Boundary conditions we used are presented in Sec.~\ref{BoundaryConditions}. In Sec.~\ref{cases}, several test cases, including Couette flow, lid-driven cavity flow, and the flows over a flat plate, circular cylinder, and the National Advisory Committee for Aeronautics (NACA) airfoil (NACA 0012 airfoil) are conducted to validate present method. Micro cavity flows are carried out to verify the all flow regimes simulation ability of present method. Some remarks and discussions are concluded in Sec.~\ref{Conclusion}.

\section{\label{algorithm}Simplified discrete unified gas kinetic scheme}
The starting point of the DUGKS is the Boltzmann equation with BGK collision model. The Boltzmann BGK equation reads
\begin{equation}\label{BoltzmannEquation}¡¡
\frac{{\partial f}}{{\partial t}} + {\bm{\xi }} \cdot \nabla f = \Omega {\text{ = }} - \frac{{f - {f^{eq}}}}{\tau },
\end{equation}
where $f = f\left( {{\bm{x}},{\bm{\xi }},t} \right)$ is the gas distribution function for particles moving with velocity ${\bm{\xi }}$ at position $\bm{x}$ on time $t$, $\tau$ is the relaxation time, and ${f^{eq}}$ is the Maxwellian distribution function,
\begin{equation}\label{Feq}
{f^{eq}} = \frac{\rho }{{{{\left( {2\pi RT} \right)}^{D/2}}}}\exp \left( { - \frac{{{{\left| {{\bm{\xi }} - {\bm{u}}} \right|}^2}}}{{RT}}} \right),
\end{equation}
where $\rho$ is the density, $R$ is the gas constant, $T$ is the temperature, $D$ is the spatial dimension, $\bm{u}$ is the flow velocity.
\begin{equation}\label{Macro}
\bm{W}\left( \bm{x},t \right)=\left( \begin{matrix}
   \rho   \\
   \rho \bm{u}  \\
   \rho E  \\
\end{matrix} \right)=\int{\bm{\psi }\left( \bm{\xi } \right)f\left( \bm{x},\bm{\xi },t \right)d\bm{\xi }},
\end{equation}
where $\rho E=\rho {u}^{2} / 2 + \rho e$ is the total energy, $e$ is the internal energy per unit mass, and ${\bm{\psi }} = {\left( {1,\bm{\xi} ,{\bm{\xi} ^2}/2} \right)^T}$ is the collision invariant. Eq.~\eqref{BoltzmannEquation} can be discretized with a semi-implicit form in which flux and collision terms are predicted with kinetic reconstruction \cite{yuan2020conservative}.

By integrating Eq.~\eqref{BoltzmannEquation} on the control volume ${V_j}$ from time $t_n$  to ${t_{n + 1}} = {t_n} + \Delta t$, discrete equation for solving gas distribution function $f\left( {{\bm{x}},{\bm{\xi }},t} \right)$ at every time step can be written as
\begin{equation}\label{Integration}
{f_j}^{n + 1} - {f_j}^n + {{\Delta t} \over {{V_j}}}F_{j,meso}^{n + 1} = \Delta t\left( { - {{{f_j}^{n + 1} - {f_j}^{eq,\left( {n + 1} \right)}} \over {{\tau _j}}}} \right),
\end{equation}
where $F_{j,meso}^{n + 1}$ is the flux of distribution function $f_{\rm{j}}^{n + 1}$ across every cell interfaces of the control volume $V_j$ at time $t_{n+1}$. Flux term $F_{j,meso}^{n + 1}$ is applied to simplify the evaluation of distribution function which is one of the differences between our method and original DUGKS.

By simple derivation, update scheme for gas distribution function on next time step ${f_j}^{n+1}$ is easily written as
\begin{equation}
\left( {1{\rm{ + }}{{\Delta t} \over {{\tau _j}}}} \right){f_j}^{n + 1} = {f_j}^n - {{\Delta t} \over {{V_j}}}F_{j,meso}^{n + 1} + {{\Delta t} \over {{\tau _j}}}{f_j}^{eq,\left( {n + 1} \right)},
\end{equation}
\begin{equation}\label{fn+1}
{f_j}^{n + 1} = {{\left[ {{f_j}^n - {{\Delta t} \over {{V_j}}}F_{j,meso}^{n + 1} + {{\Delta t} \over {{\tau _j}}}{f_j}^{eq,\left( {n + 1} \right)}} \right]} \mathord{\left/
 {\vphantom {{\left[ {{f_j}^n - {{\Delta t} \over {{V_j}}}F_{j,meso}^{n + 1} + {{\Delta t} \over {{\tau _j}}}{f_j}^{eq,\left( {n + 1} \right)}} \right]} {\left( {1{\rm{ + }}{{\Delta t} \over {{\tau _j}}}} \right)}}} \right.
 \kern-\nulldelimiterspace} {\left( {1{\rm{ + }}{{\Delta t} \over {{\tau _j}}}} \right)}}.
 \end{equation}
The flux $F_{j,meso}^{n + 1}$ and equilibrium distribution function ${f_j}^{eq,\left( {n + 1} \right)}$ of the cell at time ${t_{n + 1}}$ are unknown. For better description of particle motion, Boltzmann BGK model equation is discretized and integrated on particle velocity characteristic line within one time step at the cell interface $\bm{x}_b$. Within a time step, trapezoidal rule is applied to integrate Eq.~\eqref{BoltzmannEquation}
\begin{equation}\label{bar_eq_plus}
f\left( {{{\bm{x}}_b},{\bm{\xi}} ,{t_n} + \Delta t} \right) - f\left( {{{\bm{x}}_b} - {\bm{\xi }}\Delta t,{\bm{\xi}},{t_n}} \right) = {{\Delta t} \over 2}\left[ {\Omega \left( {{{\bm{x}}_b},{\bm{\xi}} ,{t_n} + \Delta t} \right) + \Omega \left( {{{\bm{x}}_b} - {\bm{\xi }}\Delta t,{\bm{\xi}} ,{t_n}} \right)} \right].
\end{equation}
Then we move the time ${t_{n + 1}}$ term to the left hand side as well as moving the term on time $t_n$ to the right hand side. Two new transformed distribution functions are obtained
\begin{equation}
\label{fbar}{\bar f_j}\left( {{{\bm{x}}_b},{\bm{\xi }},{t_{n + 1}}} \right) = \bar f_j^ + \left( {{{\bm{x}}_b} - {\bm{\xi }}\Delta t,{\bm{\xi }},{t_n}} \right),
\end{equation}
in which,
\begin{align}
\label{fbar1} {{\bar{f}}_{j}}\left( {{\bm{x}}_{b}},\bm{\xi },{{t}_{n+1}} \right)&={{f}_{j}}\left( {{\bm{x}}_{b}},\bm{\xi },{{t}_{n+1}} \right)-\frac{\Delta t}{2}\Omega \left( {{{\bm{x}}_b},{\bm{\xi}} ,{t_n} + \Delta t} \right) \notag\\
&= \frac{2\tau +\Delta t}{2\tau }{{f}_{j}}\left( {{\bm{x}}_{b}},\bm{\xi },{{t}_{n+1}} \right)-\frac{\Delta t}{2\tau }f{{_{j}^{eq,{{n+1}}}}}\left( {{\bm{x}}_{b}},\bm{\xi },{{t}_{n+1}} \right), \\
\label{fbarplus}\bar{f}_{j}^{+}\left( {{\bm{x}}_{b}}-\bm{\xi }\Delta t,\bm{\xi },{{t}_{n}} \right)&={{f}_{j}}\left( {{\bm{x}}_{b}}-\bm{\xi }\Delta t,\bm{\xi },{{t}_{n}} \right)+\frac{\Delta t}{2}\Omega \left( {{{\bm{x}}_b} - {\bm{\xi }}\Delta t,{\bm{\xi}} ,{t_n}} \right)\notag\\
 &=\frac{2\tau -\Delta t}{2\tau }{{f}_{j}}\left( {{\bm{x}}_{b}}-\bm{\xi }\Delta t,\bm{\xi },{{t}_{n}} \right)+\frac{\Delta t}{2\tau }f_{j}^{eq,{{n}}}\left( {{\bm{x}}_{b}}-\bm{\xi }\Delta t,\bm{\xi },{{t}_{n}} \right).
\end{align}
$\bar{f}_{j}$ at next time step can be spatially reconstructed by $\bar{f}_{j}^{+}$ at current time step. With the Taylor expansion around the cell center, for smooth flow regime, $\bar{f}_{j}^{+}\left( {{\bm{x}}_{b}}-\bm{\xi }\Delta t,\bm{\xi },{{t}_{n}} \right)$ can be approximated as
\begin{equation}\label{fbpsource}
\bar{f}_{j}^{+}\left( {{\bm{x}}_{b}}-\bm{\xi }\Delta t,\bm{\xi },{{t}_{n}} \right) = \bar{f}_{j}^{+}\left( {{\bm{x}}_{j}},\bm{\xi },{{t}_{n}} \right) + \left(\bm{x}_b-\bm{\xi }\Delta t - \bm{x}_j\right) \cdot {\bm{\sigma }_j}.
\end{equation}
To extend present method to more complex configuration, spatial derivative vector $\bm{\sigma }_j$ is evaluated by LLSR which is more efficient on unstructured mesh. The fitting formula takes following form
\begin{equation}
\bar{f}_{j}^{+}\left( {\bm{x}},\bm{\xi },{{t}_{n}} \right) = {\bar f^ + }\left( {{\bm{x}}_j,{\bm{\xi }},{t_n}} \right) + \left( {{\bm{x}} - {{\bm{x}}_j}} \right) \cdot {{\bm{\sigma }}_j}.
\end{equation}
For 2D cases, with mesh cells adjacent to cell $j$, the linear least squares regression equation is constructed as
\begin{equation}
\sum\limits_{k = 1}^{N_f} {\left[ {\begin{array}{*{20}{c}}
{{{\left( {x_k^{\left( 1 \right)} - x_j^{\left( 1 \right)}} \right)}^2}}&{\left( {x_k^{\left( 1 \right)} - x_j^{\left( 1 \right)}} \right)\left( {x_k^{\left( 2 \right)} - x_j^{\left( 2 \right)}} \right)}\\
{\left( {x_k^{\left( 1 \right)} - x_j^{\left( 1 \right)}} \right)\left( {x_k^{\left( 2 \right)} - x_j^{\left( 2 \right)}} \right)}&{{{\left( {x_k^{\left( 2 \right)} - x_j^{\left( 2 \right)}} \right)}^2}}
\end{array}} \right]} {{\bm{\sigma }}_j} = \sum\limits_{k = 1}^{N_f} {\left[ {\begin{array}{*{20}{c}}
{\left( {x_k^{\left( 1 \right)} - x_j^{\left( 1 \right)}} \right)\left( {\bar f_k^ +  - \bar f_j^ + } \right)}\\
{\left( {x_k^{\left( 2 \right)} - x_j^{\left( 2 \right)}} \right)\left( {\bar f_k^ +  - \bar f_j^ + } \right)}
\end{array}} \right]},
\end{equation}
where $k$ represent cells around cell $j$ and $N_f$ is the number of them. ${x^{\left( 1 \right)}}$ and ${x^{\left( 2 \right)}}$ are two components of space vector ${\bm{x}}$.\\
The distribution function ${{f}_{j}}\left( {{\bm{x}}_{b}},\bm{\xi },{{t}_{n}} \right)$ at current time step is adopted to calculate the flux of macro variables by following formula
\begin{equation}\label{MacroFlux}
{\bm{F}}_{j,m{\rm{acr}}o}^{n}{\rm{ = }}\oint\limits_{\partial {V_j}} {\int {{\bm{\psi }}\left( {{\bm{\xi }} \cdot {\bm{n}}} \right)} {{f}_j}\left( {{{\bm{x}}_b},{\bm{\xi }},{t_{n }}} \right)d{\bm{\xi }}} dl ,
\end{equation}
where $dl$ is circular integration variable around the interface of $\partial {{V}_{j}}$.\\
Macroscopic equations can be derived from moment of mesoscopic Boltzmann equation \cite{wu2020accuracy}. By integration of Eq.~(\ref{Integration}), update scheme of macro variables is written as
\begin{equation}\label{MacroUpdate}
{{\bm{W}}_{j}}^{n+1}-{{\bm{W}}_{j}}^{n}+\frac{\Delta t}{{{V}_{j}}}\bm{F}_{j,maro}^{n}=\bm{0}.
\end{equation}
After that, equilibrium distribution ${{f}_{j}}^{eq,\left( n+1 \right)}$ on ${{t}_{n+1}}$ is updated by macro variables ${{\bm{W}}_{j}}^{n+1}$.\\
According to Eq.~\eqref{fbar1}, distribution function $f\left( {{\bm{x}}_{b}},\bm{\xi },{{t}_{n+1}} \right)$ at the center of cell interface is written as.
\begin{equation}\label{f_t_n+1}
f\left( {{\bm{x}}_{b}},\bm{\xi },{{t}_{n+1}} \right)=\left( {{{\bar{f}}}_{j}}\left( {{\bm{x}}_{b}},\bm{\xi },{{t}_{n+1}} \right)+\frac{\Delta t}{2\tau }f{{_{j}^{eq,{{n+1}}}}}\left( {{\bm{x}}_{b}},\bm{\xi },{{t}_{n+1}} \right) \right)\cdot \frac{2\tau }{2\tau +\Delta t}.
\end{equation}
With distribution function $f\left( {{\bm{x}}_{b}},\bm{\xi },{{t}_{n+1}} \right)$, the flux $F_{j,meso}^{n+1}$ at the cell interface can be obtained.
\begin{equation}\label{MesoFlux}
F_{j,meso}^{n+1}\text{=}\oint\limits_{\partial {{V}_{j}}}{\left( \bm{\xi } \cdot \bm{n} \right)}f\left( {{\bm{x}}_{b}},\bm{\xi },{{t}_{n+1}} \right)dl.
\end{equation}
Finally, gas distribution function at the cell center on $t_{n+1}$ can be updated by Eq.~(\ref{fn+1}) since all the unknown terms are obtained. \\
In summary, the update rule of distribution function $f$ from $t_n$ to $t_n+1$ in our simplified discrete unified gas kinetic scheme (SDUGKS) is the following:

\begin{enumerate}
\item ${\bm{F}}_{j,macro}^{n}$ is calculated by Eq.~\eqref{MacroFlux}.
\item Using Eq.~\eqref{fbarplus} to obtain the $\bar f^ +$ at the cell center.
\item With Eq.~\eqref{fbpsource}, we can obtain $\bar{f}_{j}^{+}\left( {{\bm{x}}_{b}}-\bm{\xi }\Delta t,\bm{\xi },{{t}_{n}} \right)$ by LLSR.
\item According to Eq.~\eqref{fbar}, ${\bar f_j}\left( {{{\bm{x}}_b},{\bm{\xi }},{t_{n + 1}}} \right)$ is obtained.
\item Then the macro variables at cell interface on next time step can be updated using ${\bar f_j}\left( {{{\bm{x}}_b},{\bm{\xi }},{t_{n + 1}}} \right)$ by Eq.~\eqref{Macro} based on conservation constraint.
\item After the macro variables obtained, the equilibrium distribution function on the midpoint of interface at $t_{n+1}$ can be updated by Eq.~\eqref{Feq}
\item According to Eq.~\eqref{f_t_n+1}, using the ${\bar f_j}\left( {{{\bm{x}}_b},{\bm{\xi }},{t_{n + 1}}} \right)$ and ${f_j}^{eq, n+1}$ to get distribution function $f\left( {{\bm{x}}_{b}},\bm{\xi },{{t}_{n+1}} \right)$ at the center of cell interface.
\item Then the $F_{j,meso}^{n + 1}{}$ can be obtained by Eq.~\eqref{MesoFlux}.
\item Then update the macro variables at center of the cell on next time step by Eq.~\eqref{MacroUpdate}.
\item Update the equilibrium distribution function ${f^{eq}}\left( {{{\bm{x}}_j},{\bm{\xi }},{t_{n + 1}}} \right)$ at the cell center by Eq.~\eqref{Feq}.
\item All the unknown values in Eq.~\eqref{fn+1} are obtained, we can update the distribution function $f_{j}^{n+1}$ at the cell center on next time step.
\end{enumerate}

\section{\label{BoundaryConditions}Boundary conditions}
Boundary condition is decided by direction of particle velocity. Boundary outward normal vector is ${\bm{N}}_{bcf}$ and tangential vector is ${\bm{A}}_{bcf}$. Sign of dot product between $\bm{\xi }$ and ${\bm{N}}_{bcf}$ is applied to distinguish particle information. If $\bm{\xi }\cdot {{\bm{N}}_{bcf}}<0$, the implementation is as following.
\subsection{Nonslip wall boundary condition}
\begin{equation}
\begin{aligned}
\rho \left( {{\bm{x}}_{bcf}},{{t}_{n}} \right)&=\rho \left( {{\bm{x}}_{in}},{{t}_{n}} \right),\\
\bm{u}\left( {{\bm{x}}_{bcf}},{{t}_{n}} \right)&={{\bm{u}}_{wall}},\\
f\left( {{\bm{x}}_{bcf}},\bm{\xi },{{t}_{n}} \right)&={{f}^{eq}}\left( {{\bm{x}}_{bcf}},\bm{\xi },{{t}_{n}} \right)+\left( f\left( {{\bm{x}}_{in}},\bm{\xi },{{t}_{n}} \right)-{{f}^{eq}}\left( {{\bm{x}}_{in}},\bm{\xi },{{t}_{n}} \right) \right).\\
\end{aligned}
\end{equation}
Nonslip wall boundary condition is implemented based on non-equilibrium extrapolation method \cite{Guo2002Non}. In above formulas $bcf$ denotes interface on boundary and $in$ denotes its neighbor cell. Maxwellian on boundary is computed by macroscopic variables of solid wall.
\subsection{Inlet flow boundary condition}
\begin{equation}
\begin{aligned}
\rho \left( {{\bm{x}}_{bcf}},{{t}_{n}} \right)&={{\rho }_{\infty}},\\
\bm{u}\left( {{\bm{x}}_{bcf}},{{t}_{n}} \right)&={{\bm{u}}_{\infty}},\\
f\left( {{\bm{x}}_{bcf}},\bm{\xi },{{t}_{n}} \right)&={{f}^{eq}}\left( {{\bm{x}}_{bcf}},\bm{\xi },{{t}_{n}} \right)+\left( f\left( {{\bm{x}}_{in}},\bm{\xi },{{t}_{n}} \right)-{{f}^{eq}}\left( {{\bm{x}}_{in}},\bm{\xi },{{t}_{n}} \right) \right).\\
\end{aligned}
\end{equation}
where $\infty$ represents freestream.
\subsection{Outlet flow}
\begin{equation}
\begin{aligned}
\rho \left( {{\bm{x}}_{bcf}},{{t}_{n}} \right)&=\rho \left( {{\bm{x}}_{in}},{{t}_{n}} \right),\\
\bm{u}\left( {{\bm{x}}_{bcf}},{{t}_{n}} \right)&={{\bm{u}} \left( {{\bm{x}}_{in}},{{t}_{n}} \right)},\\
f\left( {{\bm{x}}_{bcf}},\bm{\xi },{{t}_{n}} \right)&={{f}^{eq}}\left( {{\bm{x}}_{bcf}},\bm{\xi },{{t}_{n}} \right)+\left( f\left( {{\bm{x}}_{in}},\bm{\xi },{{t}_{n}} \right)-{{f}^{eq}}\left( {{\bm{x}}_{in}},\bm{\xi },{{t}_{n}} \right) \right).\\
\end{aligned}
\end{equation}

\subsection{Symmetric boundary condition}
\begin{figure}
\centering
\includegraphics[width=0.5\textwidth]{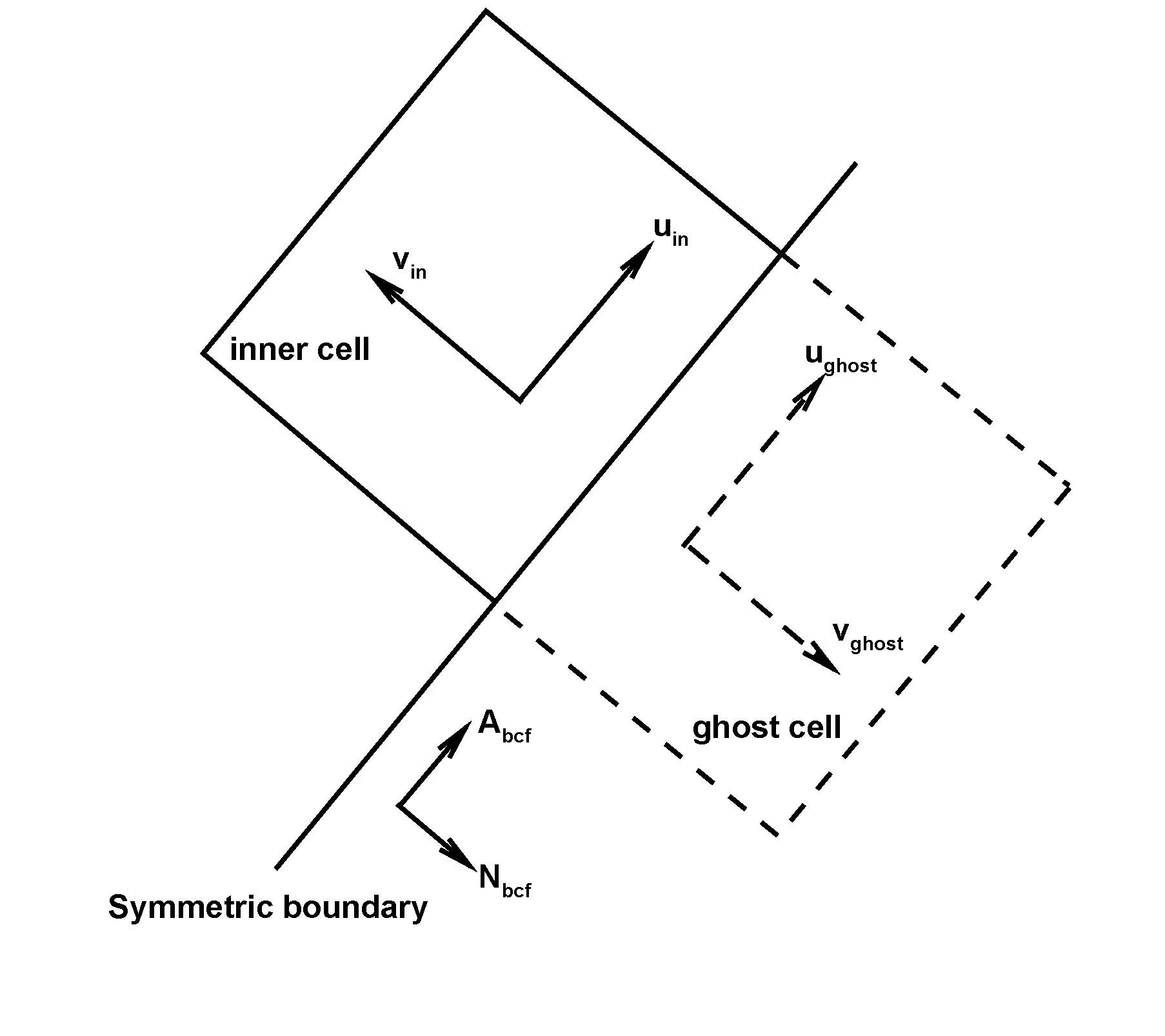}
\caption{\label{fig:Fig2001}Velocity vector relation of symmetric boundary condition.}
\end{figure}

For consistency and conservation on symmetric boundary, ghost cells symmetric to inner cells about boundaries are applied. Density in ghost cells should be equal to inner cells, which reads
\begin{equation}
\rho \left( {{\bm{x}}_{ghost}},{{t}_{n}} \right)=\rho \left( {{\bm{x}}_{in}},{{t}_{n}} \right).
\end{equation}
Velocity vector in ghost cells is presented as mirror-symmetry, which is shown in Fig.~\ref{fig:Fig2001}. Relation of macroscopic velocity between ghost cells and inner cells is written as
\begin{equation}
\begin{aligned}
{{\bm{u}}_{ghost}}\cdot {{\bm{N}}_{bcf}}&=-{{\bm{u}}_{in}}\cdot {{\bm{N}}_{bcf}},\\
{{\bm{u}}_{ghost}}\cdot {{\bm{A}}_{bcf}}&={{\bm{u}}_{in}}\cdot {{\bm{A}}_{bcf}}.\\
\end{aligned}
\end{equation}
Macroscopic variables on boundaries are interpolated using inner cells and ghost cells around boundaries. Gas distribution function on symmetric boundaries is reconstructed by LLSR.

\subsection{Periodic boundary condition}
\begin{figure}
\centering
\includegraphics[width=0.5\textwidth]{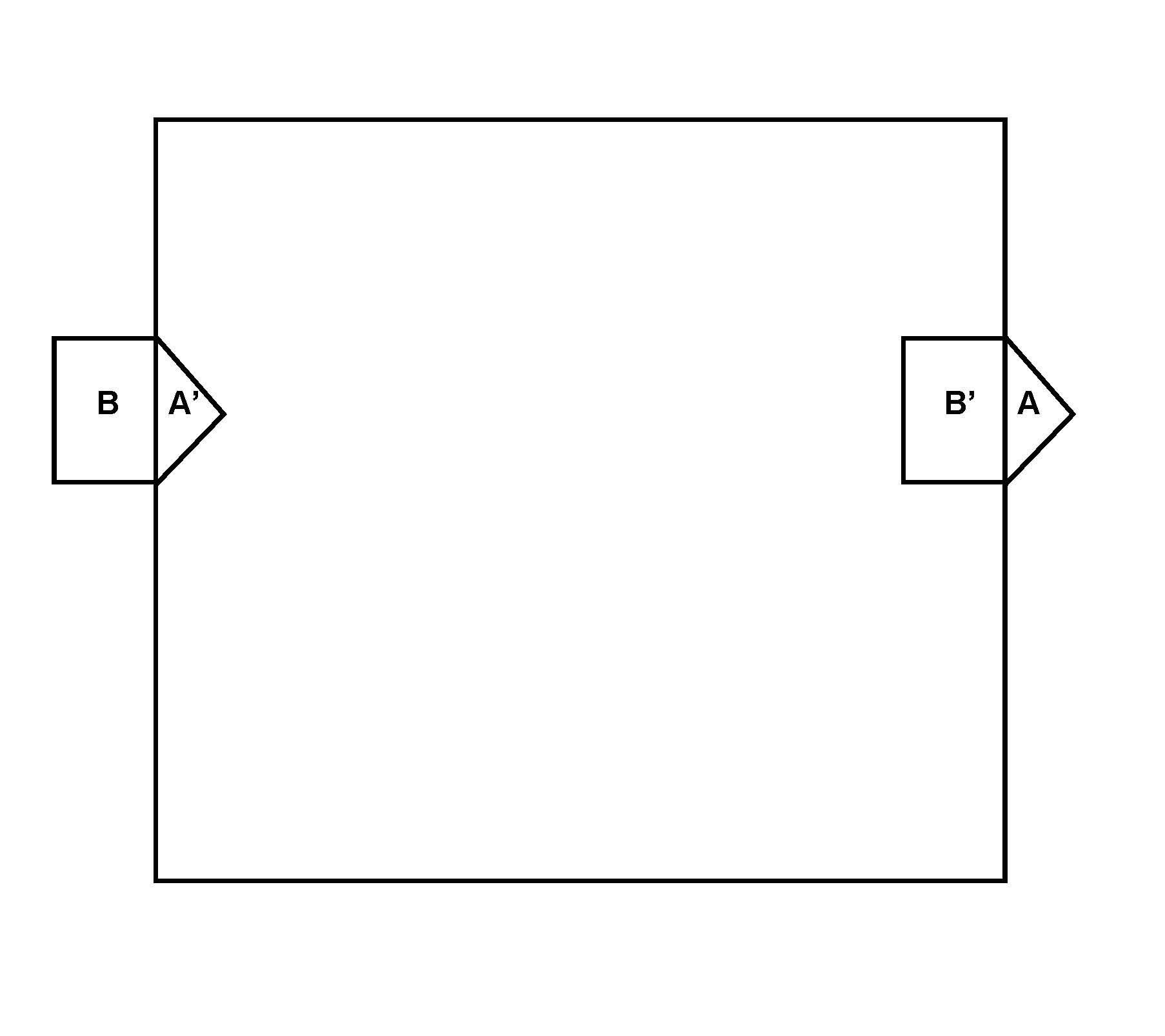}
\caption{\label{fig:Fig2000}Implementation of periodic boundary condition.}
\end{figure}

As gas flows away from computational field, it will flow into the same field from the opposite side simultaneously. As shown in Fig.~\ref{fig:Fig2000}, cells $A$ and $B$ are ghost cells corresponding to inner cells ${A}'$ and ${B}'$. The relation between them reads
\begin{equation}
\begin{aligned}
   \rho \left( {{\bm{x}}_{A}},{{t}_{n}} \right)&=\rho \left( {{\bm{x}}_{{{A}'}}},{{t}_{n}} \right),  \\
   \bm{u}\left( {{\bm{x}}_{A}},{{t}_{n}} \right)&=\bm{u}\left( {{\bm{x}}_{{{A}'}}},{{t}_{n}} \right),  \\
   f\left( {{\bm{x}}_{A}},\bm{\xi },{{t}_{n}} \right)&=f\left( {{\bm{x}}_{{{A}'}}},\bm{\xi },{{t}_{n}} \right).  \\
\end{aligned}
\end{equation}
\begin{equation}
\begin{aligned}
   \rho \left( {{\bm{x}}_{B}},{{t}_{n}} \right)&=\rho \left( {{\bm{x}}_{{{B}'}}},{{t}_{n}} \right),  \\
   \bm{u}\left( {{\bm{x}}_{B}},{{t}_{n}} \right)&=\bm{u}\left( {{\bm{x}}_{{{B}'}}},{{t}_{n}} \right),  \\
   f\left( {{\bm{x}}_{B}},\bm{\xi },{{t}_{n}} \right)&=f\left( {{\bm{x}}_{{{B}'}}},\bm{\xi },{{t}_{n}} \right).  \\
\end{aligned}
\end{equation}
Mesoscopic and macroscopic variables on periodic boundaries are equal to the values on the inner interface of the corresponding periodic boundary.

If $\bm{\xi }\cdot {{\bm{N}}_{bcf}}\ge 0$, variables on boundaries should be interpolated from inner flow field. In this way correct physical information is well preserved in reconstruction on the boundary. Consequently, evolution on boundaries is compatible with flow field, which increases robustness in computation.

\subsection{Diffuse boundary condition}
In rarefied flow simulation, diffuse-scattering rule is implemented on solid wall~\cite{Meng2014Diffuse}. Gas distribution function is computed by
\begin{equation}\label{Eq6001}
f\left( {{{\bm{x}}_{bcf}},{\bm{\xi }},t} \right) = {f^{eq}}\left( {{{\bm{x}}_{bcf}},{\bm{\xi }},t} \right)
\end{equation}
Density on wall is computed by~\cite{Guo2013Discrete}
\begin{equation}\label{Eq6002}
\rho \left( {{{\bm{x}}_{bcf}},t} \right) =  - {\left( {\sum\limits_{{\bm{\xi }} \cdot {{\bm{N}}_{bcf}} > 0} {\left( {{\bm{\xi }} \cdot {{\bm{N}}_{bcf}}} \right){f^{eq}}\left( {{{\bm{x}}_{bcf}},{\bm{\xi }},t} \right)} } \right)^{ - 1}}\left( {\sum\limits_{{\bm{\xi }} \cdot {{\bm{N}}_{bcf}} < 0} {\left( {{\bm{\xi }} \cdot {{\bm{N}}_{bcf}}} \right)f\left( {{{\bm{x}}_{bcf}},{\bm{\xi }},t} \right)} } \right)
\end{equation}

\section{\label{cases}Numerical results}
In this section, several cases are conducted to validate the proposed SDUGKS. The first Couette flow case verifies its spatial second-order accuracy. Lid-driven cavity flow is computed to validate present method in viscous flow simulation. On unstructured meshes, flows around NACA 0012 airfoil are accurately simulated. In unsteady cases, flows around circular cylinder are simulated and many flow details are captured. The same accuracy with previous work is proved with much larger time step and robustness is also pretty well. Micro cavity flow cases at different Knudsen numbers are simulated for testing multi-scale property and applications in all flow regimes of SDUGKS.
\subsubsection{Couette flow}
\begin{figure}
	\centering
	\includegraphics[width=0.16 \textwidth]{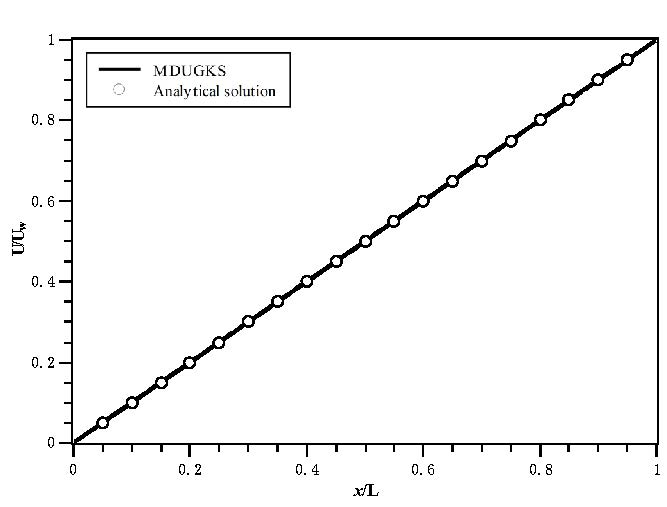}
    \caption{\label{CouetteGrid} Meshes of Couette flow for spatial accuracy test. (a) $4 \times 8$, (b) $4 \times 16$, (c) $4 \times 32$, (d) $4 \times 64$, (e) $4 \times 128$}
\end{figure}

Steady Couette flow case is computed for spatial accuracy validation. In this case, the geometry details are introduced as follow. Unstructured meshes with different resolutions of $4 \times 8$, $4 \times 16$, $4 \times 32$, $4 \times 64$, $4 \times 128$ are used in this case, which are shown in Fig.~\ref{CouetteGrid}. The height of computational domain is $1.0$, the width of mesh is depending on the number of cells it used.\\

Nonslip wall boundary condition is implemented on the top and bottom walls, while the top wall moves with constant velocity ${{u}_{w}} = 1.0$, and the bottom wall is a static wall with ${{u}_{w}} = 0.0$. The left and right boundaries are prescribed as periodic boundary condition.\\
The analytical solution of velocity distribution in Couette flow field is shown as follow:
\begin{equation}
\begin{aligned}
u&=u_w y,\\
v&=0.0 .
\end{aligned}
\end{equation}

\begin{figure}
	\centering
	\includegraphics[width=0.4 \textwidth]{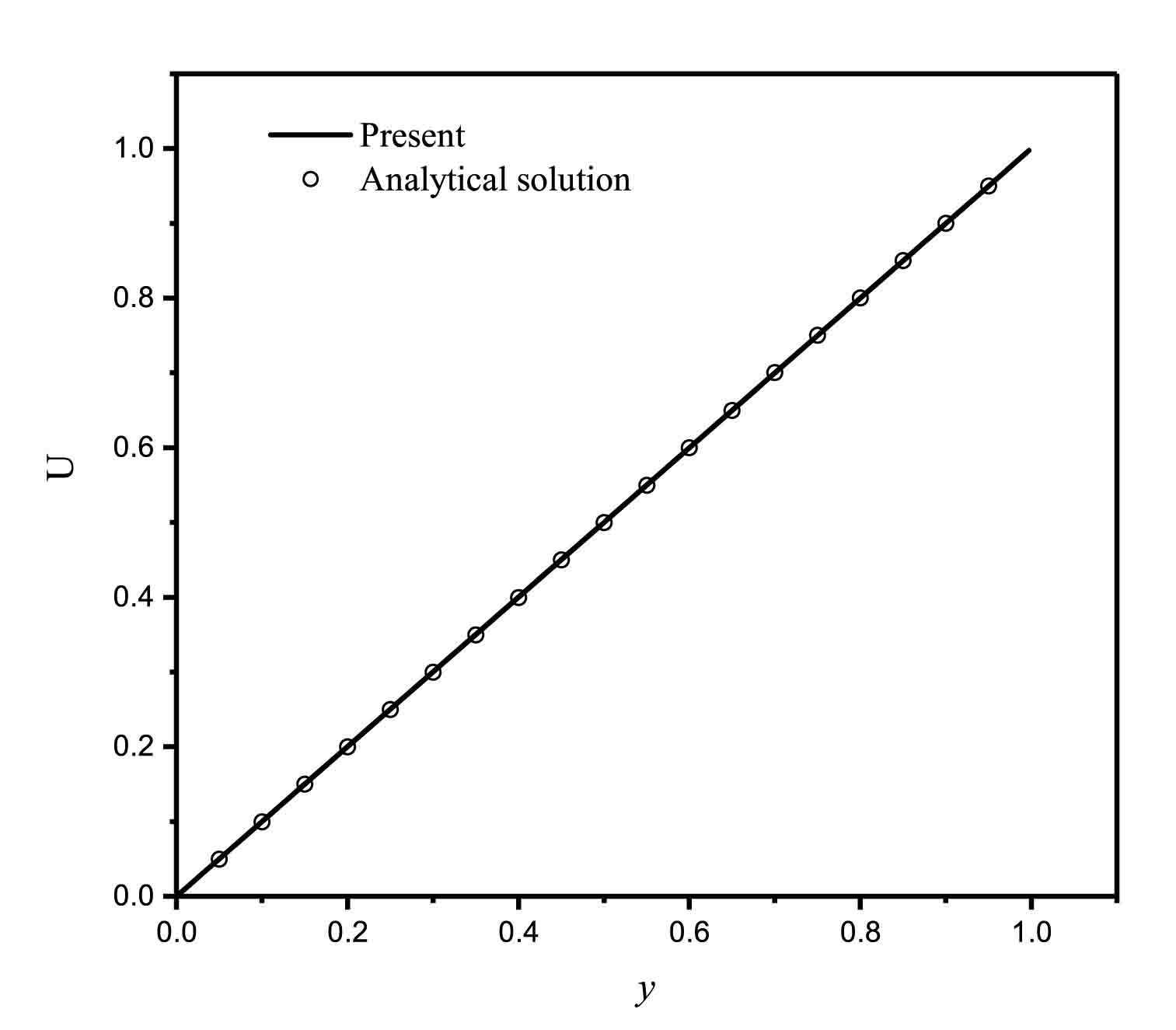}	
    \caption{\label{Couette_result} Comparison results for velocity profile of steady Couette flow case.}
\end{figure}

\begin{figure}
	\centering
	\includegraphics[width=0.4 \textwidth]{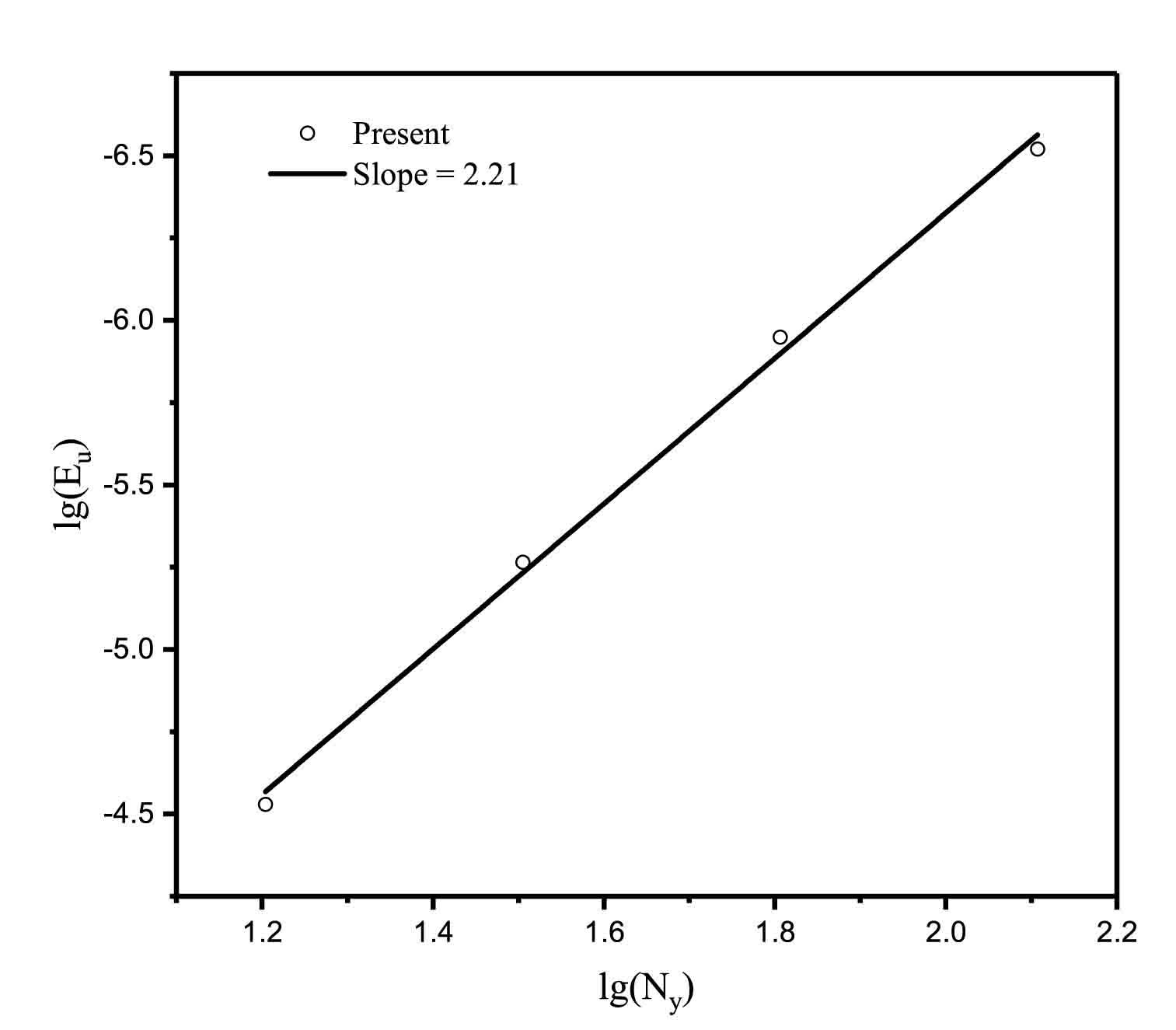}	
    \caption{\label{Couette_accuracy} Spatial accuracy validation of the present method.}
\end{figure}

Velocity profiles are plotted in Fig.~\ref{Couette_result} for comparison with analytical solution. In Fig.~\ref{Couette_accuracy}, second-order accuracy is proved by L2-norm of velocity errors varying with mesh resolution.

\subsubsection{Laminar flow over a flat plate}
\begin{figure}\label{laminarFlatPlateGrid}
	\centering
	\subfigure[]{
		\includegraphics[width=0.4 \textwidth]{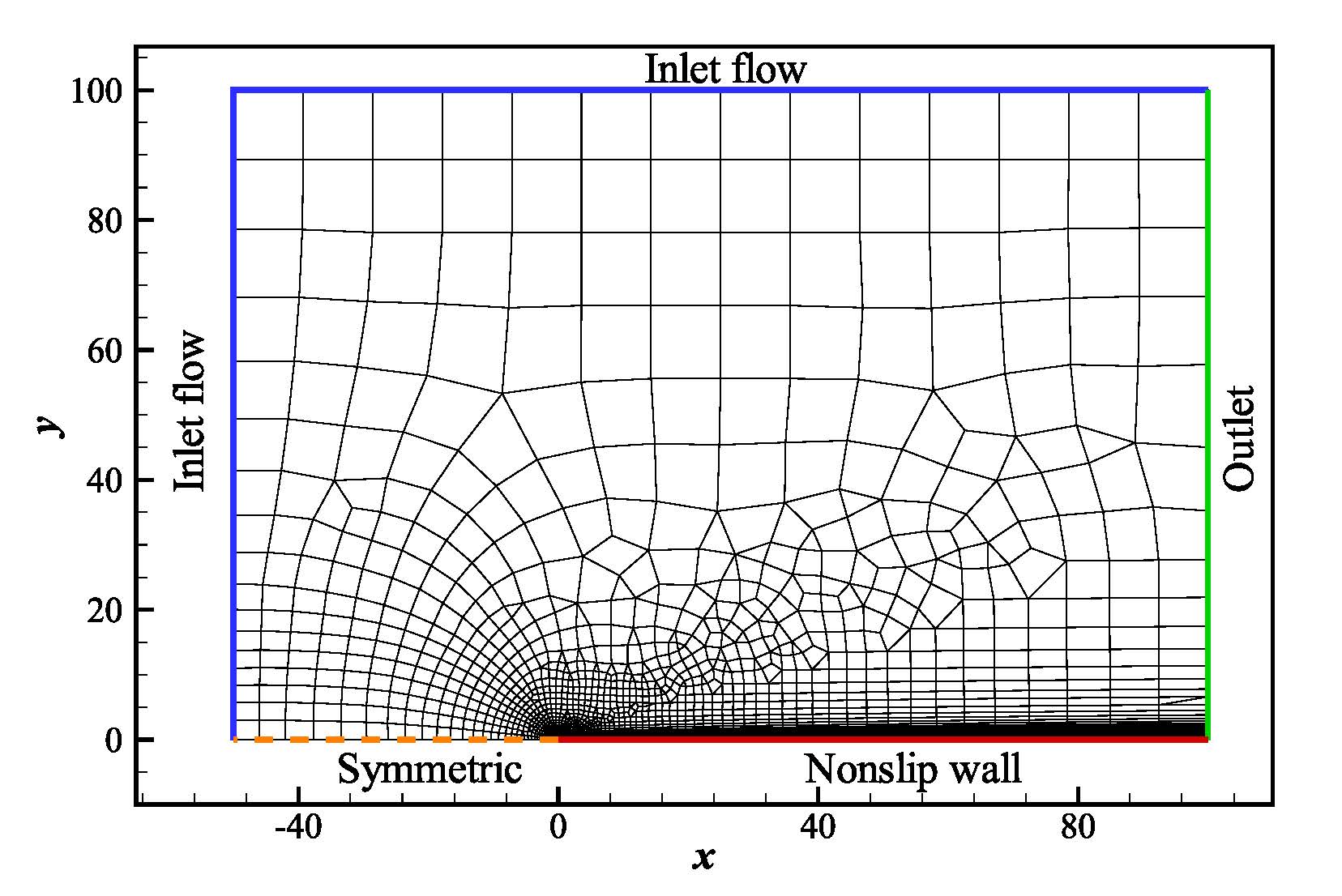}
		\label{laminarFlatPlateGridFull}
	}
	\subfigure[]{
		\includegraphics[width=0.4 \textwidth]{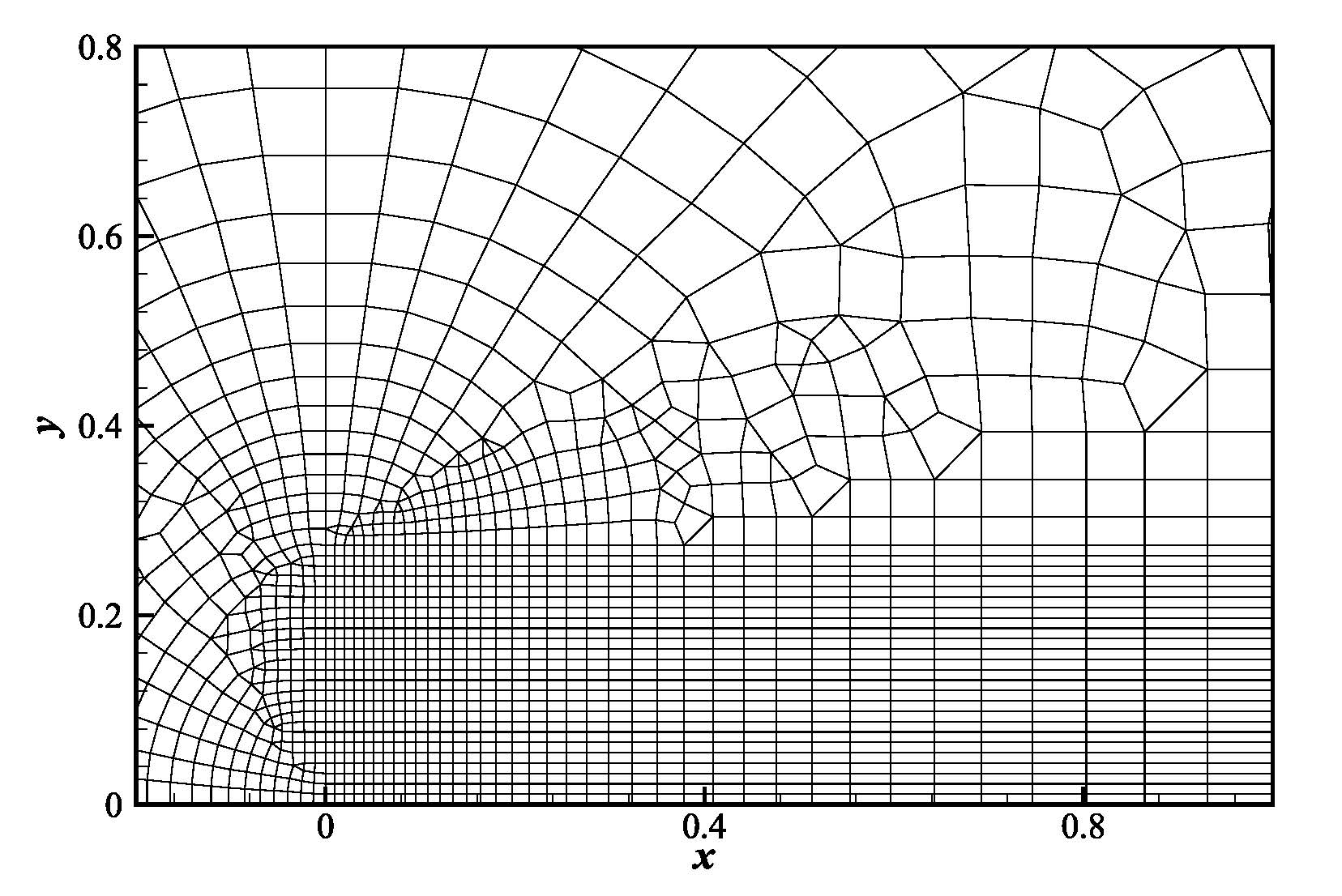}
		\label{LaminarFlatPlateGridPartial}
	}
\caption{ Mesh for flat plate. (a) Mesh and computational domain (b) Mesh near leading edge of flat plate.}
\end{figure}

\begin{figure}
	\centering
	\subfigure[]{
		\includegraphics[width=0.4 \textwidth]{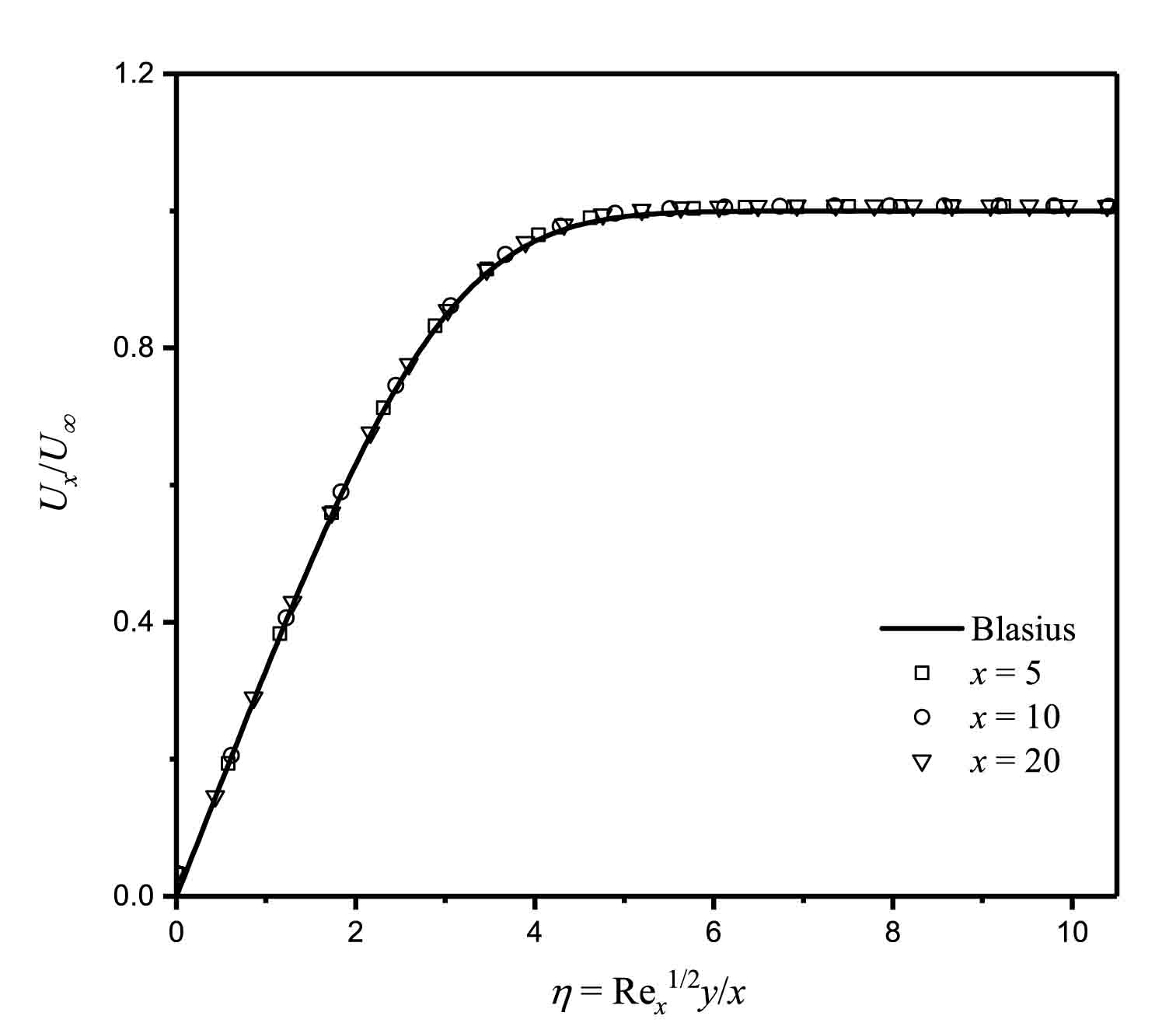}
		\label{laminarFlatPlateU}
	}
	\subfigure[]{
		\includegraphics[width=0.4 \textwidth]{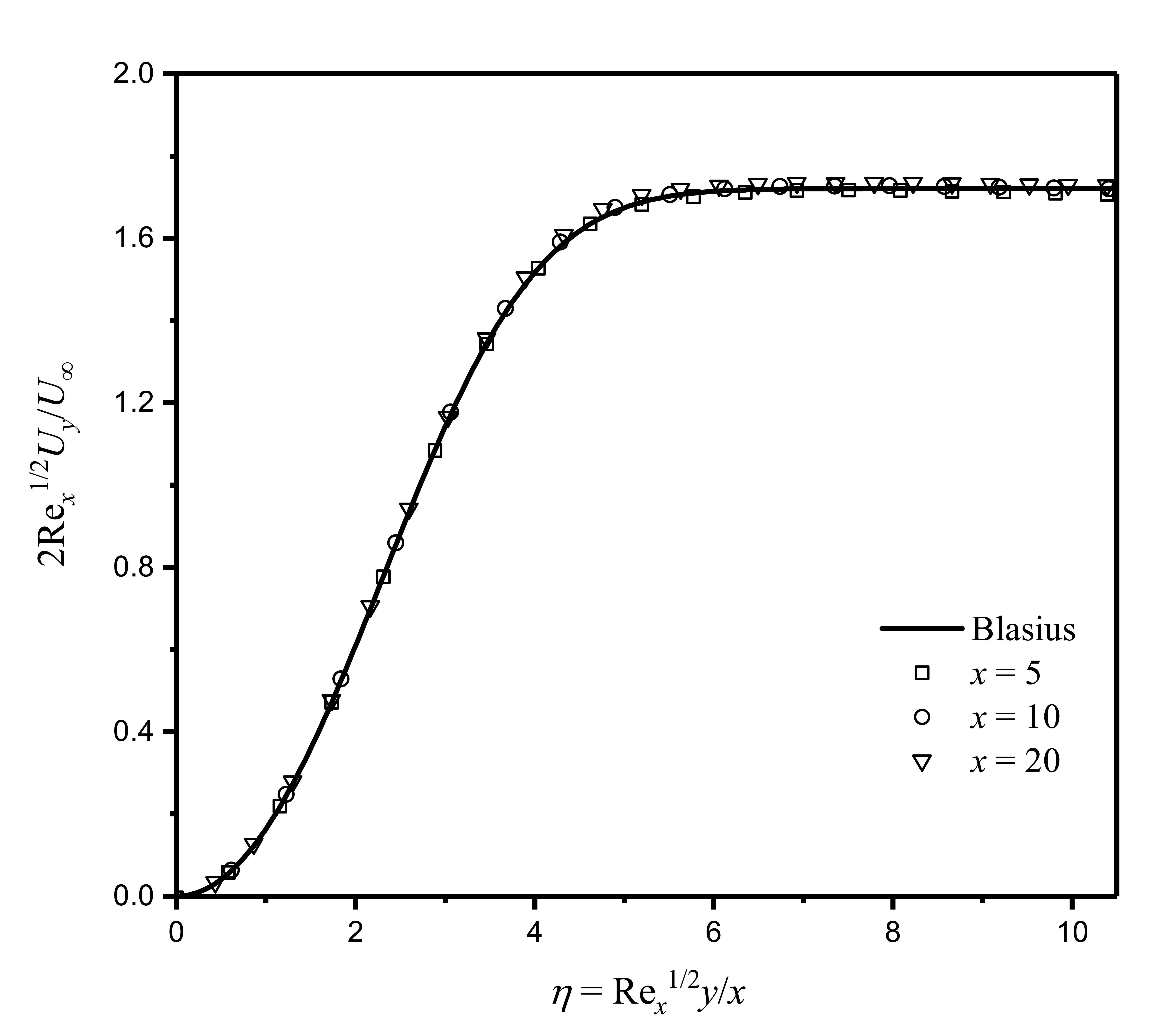}
		\label{laminarFlatPlateV}
	}
\emph{}	\caption{\label{VelocityPoriles} Comparison results for velocity profiles of flat plate flow. (a) $U_x$ and (b) $U_y$.}
\end{figure}

\begin{figure}
	\centering
	\includegraphics[width=0.4 \textwidth]{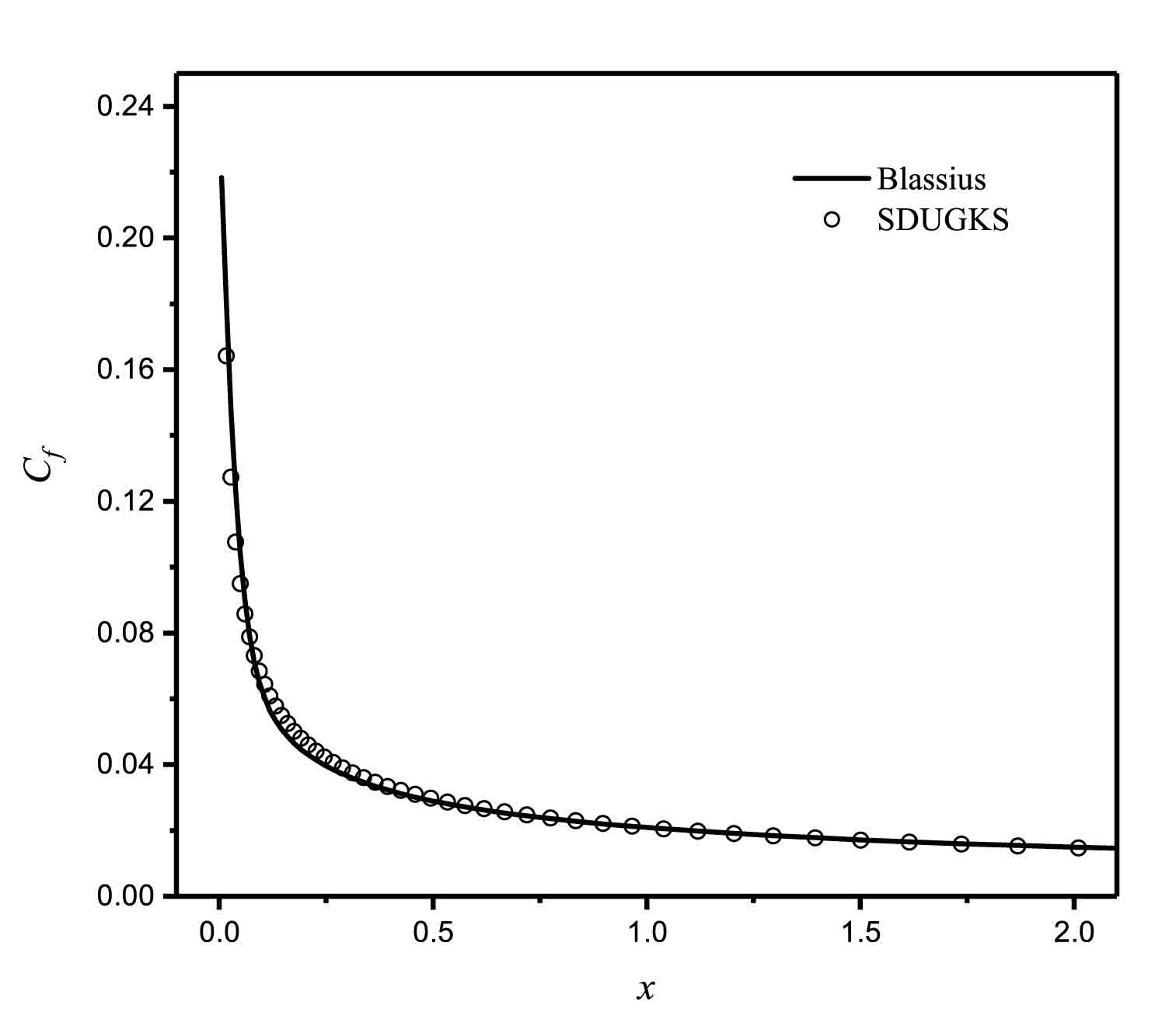}	
    \caption{\label{laminarFlatPlateCf} Comparison result for skin friction coefficients of flat plate.}
\end{figure}

As shown in Fig.~\ref{laminarFlatPlateGrid}, In this case, computational domain is set to be $x \in \left[ {-50,100} \right]$ and $y \in \left[ {0,100} \right]$. The number of mesh cells is 4876.
The top and left boundaries are prescribed as the inlet flow boundary condition. The right boundary is set as outflow condition. The bottom boundary in $x \in [-50, 0]$ in set as symmetric condition, and in $x \in [0, 100]$ is set as nonslip wall boundary condition.
The freestream velocity $u_w$ is set to $0.1$, $Re_{\infty} = 10^5$, $\rho = 1.0$, $\nu = U_{\infty}L / Re_{\infty}$, where $L = 100$ is the length of the flat plate.\\
The boundary layer velocity distributions at position $x = 5$, $x = 10$, $x = 20$ and comparisons with Blasius solution are shown in Fig.~\ref{VelocityPoriles}. The skin friction coefficients $C_f$ distribution along the flat plate is shown in Fig.~\ref{laminarFlatPlateCf}. Drag coefficient $C_d = 4.237 \times {10^{ - 3}}$ fits the Blasius solution well, which is $4.20 \times {10^{ - 3}}$.

\subsubsection{Lid-driven cavity flow}
\begin{figure}
	\centering
	\subfigure[]{
		\includegraphics[width=0.4 \textwidth]{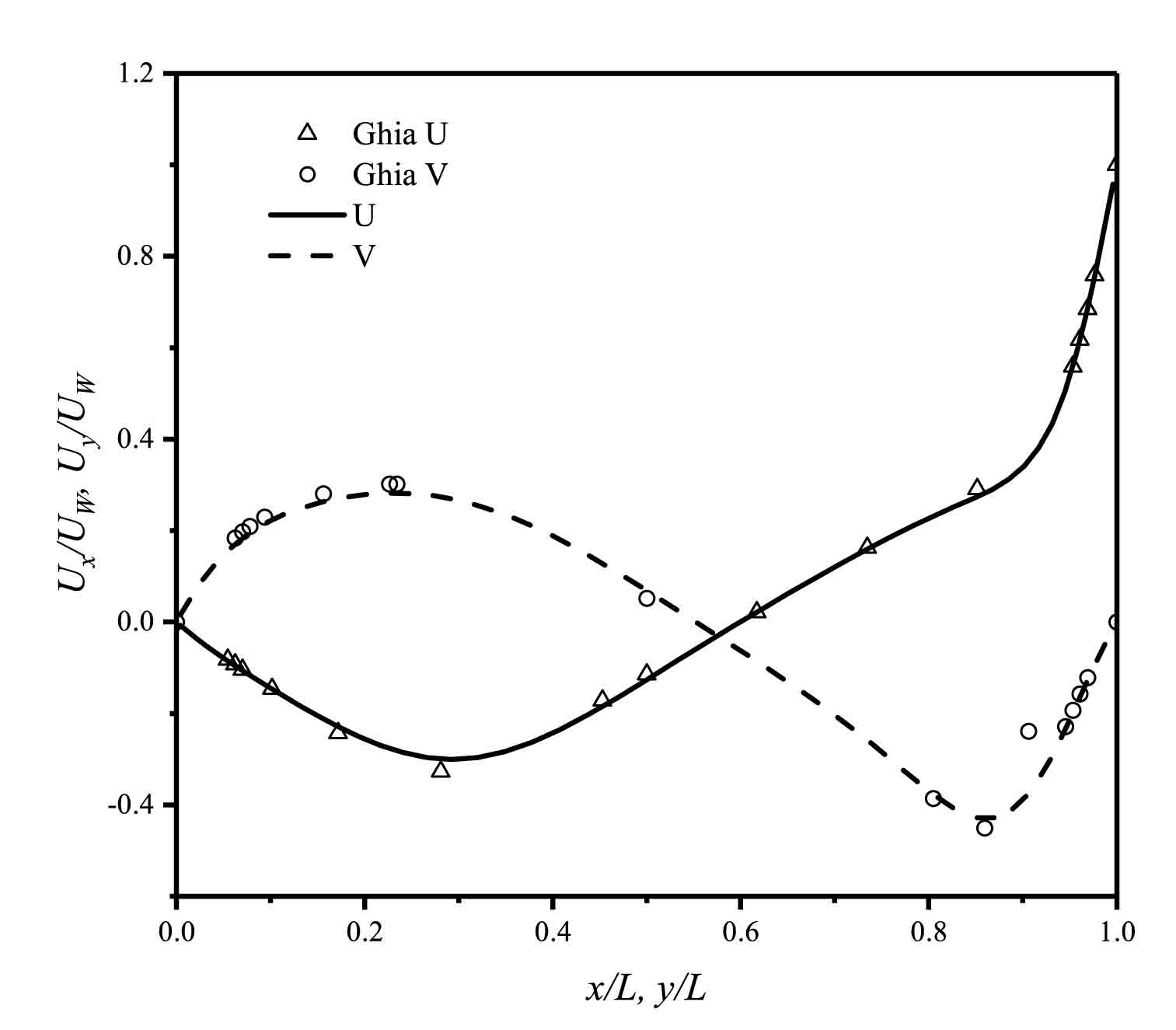}
		\label{Re400UV}
	}
	\subfigure[]{
		\includegraphics[width=0.4 \textwidth]{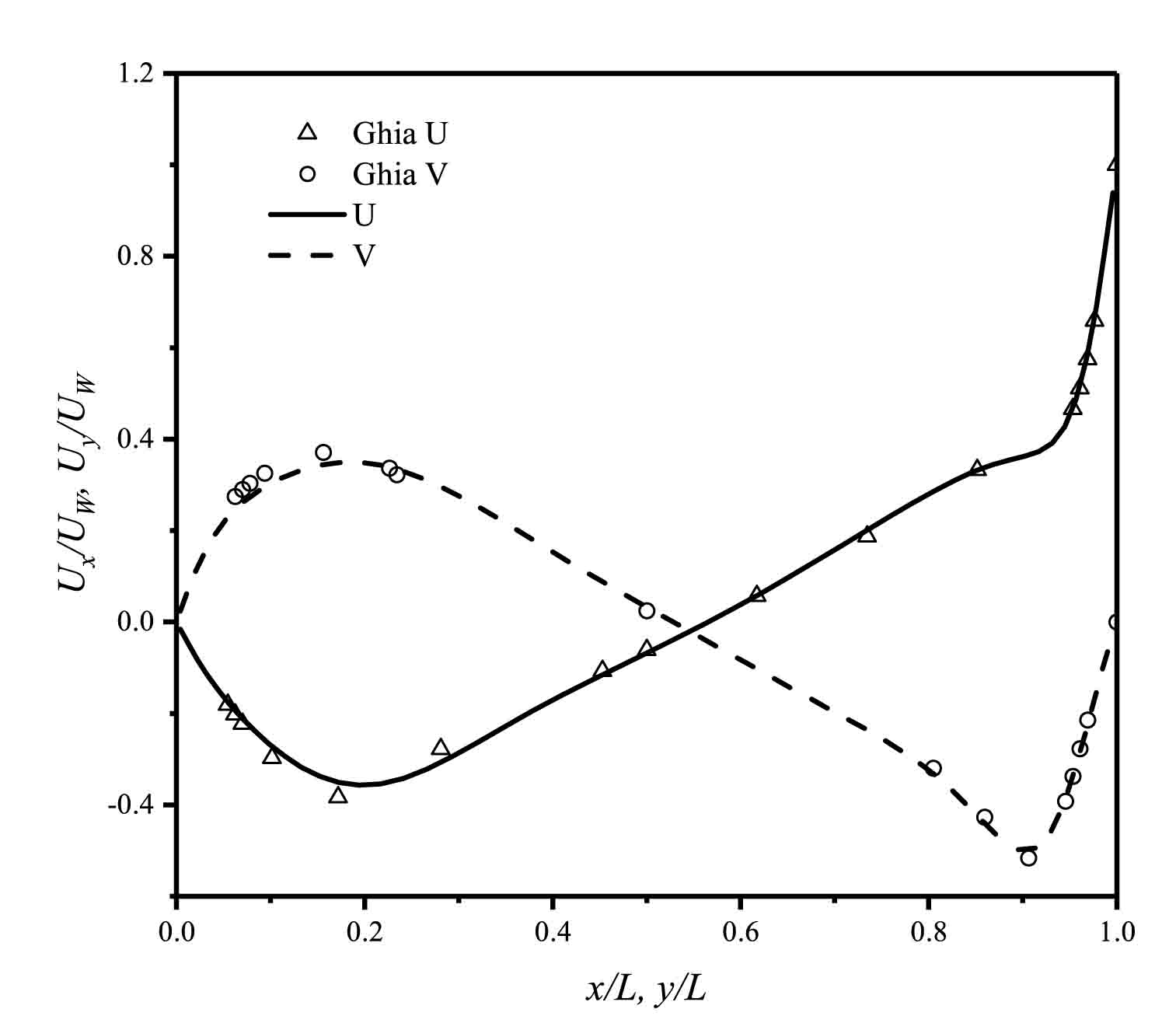}
		\label{Re1000UV}
	}
\emph{}	\caption{\label{cavityUV} Comparison results of lid-driven cavity flow for u-velocity along the central vertical line and v-velocity along the central horizontal line. (a) $Re = 400$ and (b) $Re = 1000$.}
\end{figure}

\begin{figure}
	\centering
	\subfigure[]{
		\includegraphics[width=0.4 \textwidth]{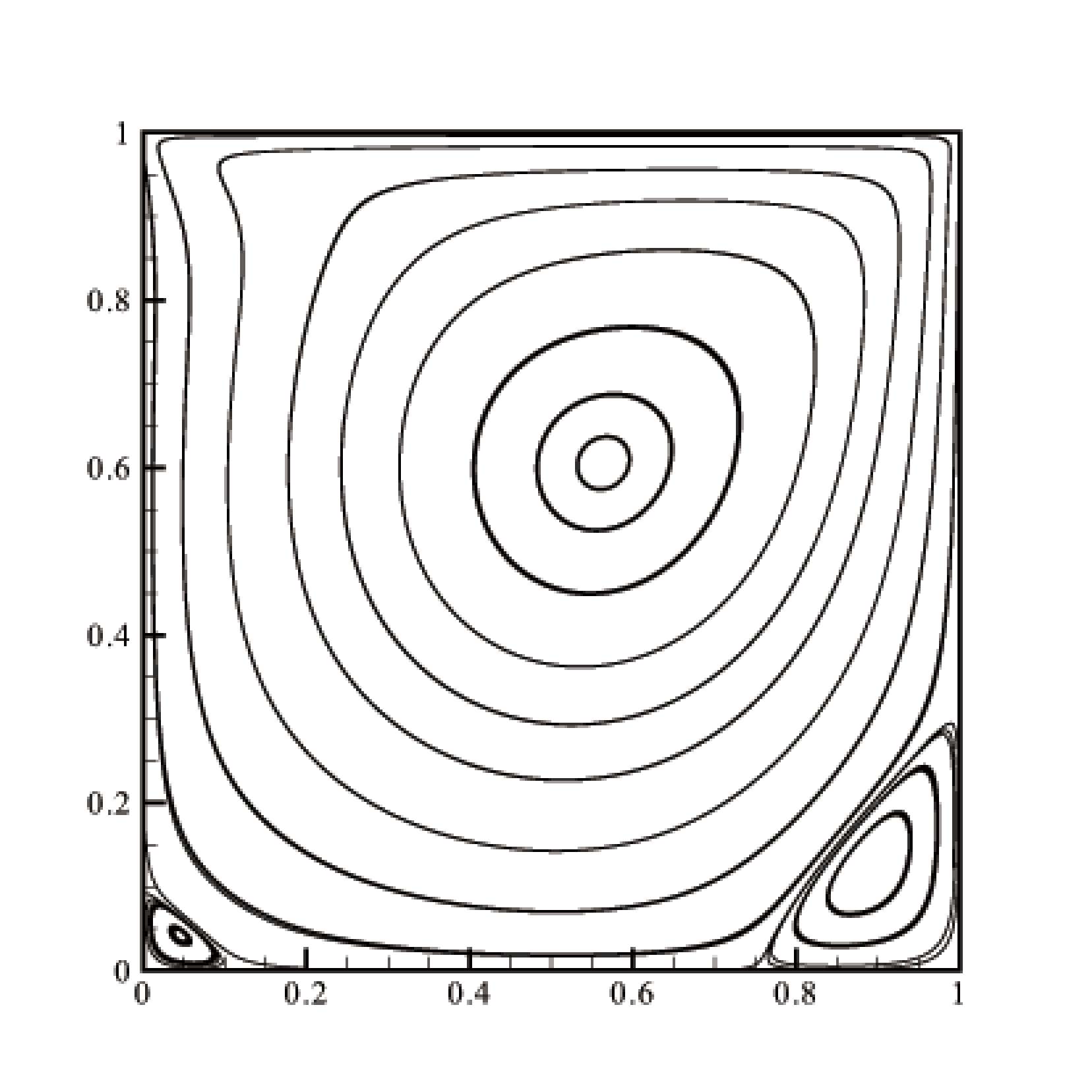}
		\label{Re400StreamLine}
	}
	\subfigure[]{
		\includegraphics[width=0.4 \textwidth]{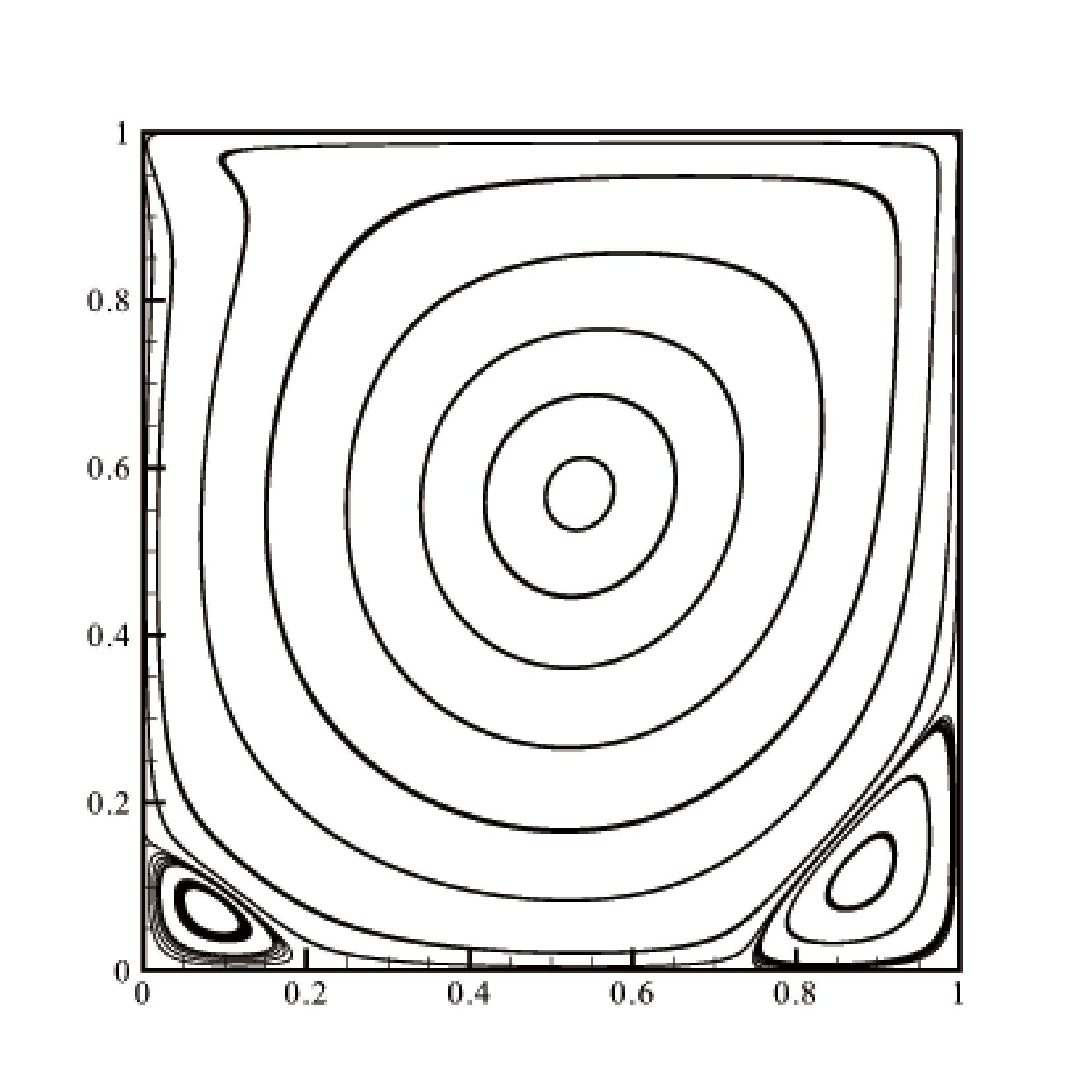}
		\label{Re1000StreamLine}
	}
	\caption{\label{cavity} Streamline of the lid-driven cavity flow at (a) $Re = 400$ and (b) $Re = 1000$.}
\end{figure}

Incompressible lid-driven cavity flow is a benchmark case for validation of numerical scheme in viscous flow simulation. Fluid in a square cavity is driven by a top moving wall with other three static walls keeping. The top wall is moving with a fix speed ${u_w}$. This incompressible case is only featured by Reynolds number.
In our simulation, in order to guarantee a nearly incompressible flow, the driven velocity is set to be $u_w = 0.1$, density $\rho$ is set to be $\rho = 1.0$, the length of the cavity is taken to be the number of cell in $x$ direction. The Reynolds number is defined as $Re = \frac{u_w L}{\nu}$. Time step in this case is set to be ten times of collision time $\tau $. Convergence is decided by residual of the horizontal velocity $U$ and vertical velocity $V$. The convergence criterion is ${10^{ - 10}}$ which is being decided in every $1000$ iteration steps.\\
\begin{equation}
\epsilon =\frac{\sqrt{\sum\nolimits_{i}{\left[ {{\left( U_{i}^{n+1000}-U_{i}^{n} \right)}^{2}}+{{\left( V_{i}^{n+1000}-V_{i}^{n} \right)}^{2}} \right]}}}{\sqrt{\sum\nolimits_{i}{\left( {{\left( U_{i}^{n} \right)}^{2}}+{{\left( V_{i}^{n} \right)}^{2}} \right)}}}
\end{equation}
The computational domain is discretized into $48 \times 48$ cells. For comparison, The Ghia's benchmark solutions are included~\cite{Ghia1982High}, in which $128 \times 128$ mesh is adopted. It shows in Fig.~\ref{cavityUV} that SDUGKS fits the benchmark data well. The streamline of the lid-driven cavity flow is shown in Fig.~\ref{cavity}.\\


\begin{figure}
\centering
\subfigure[]{
\includegraphics[width=0.45\textwidth]{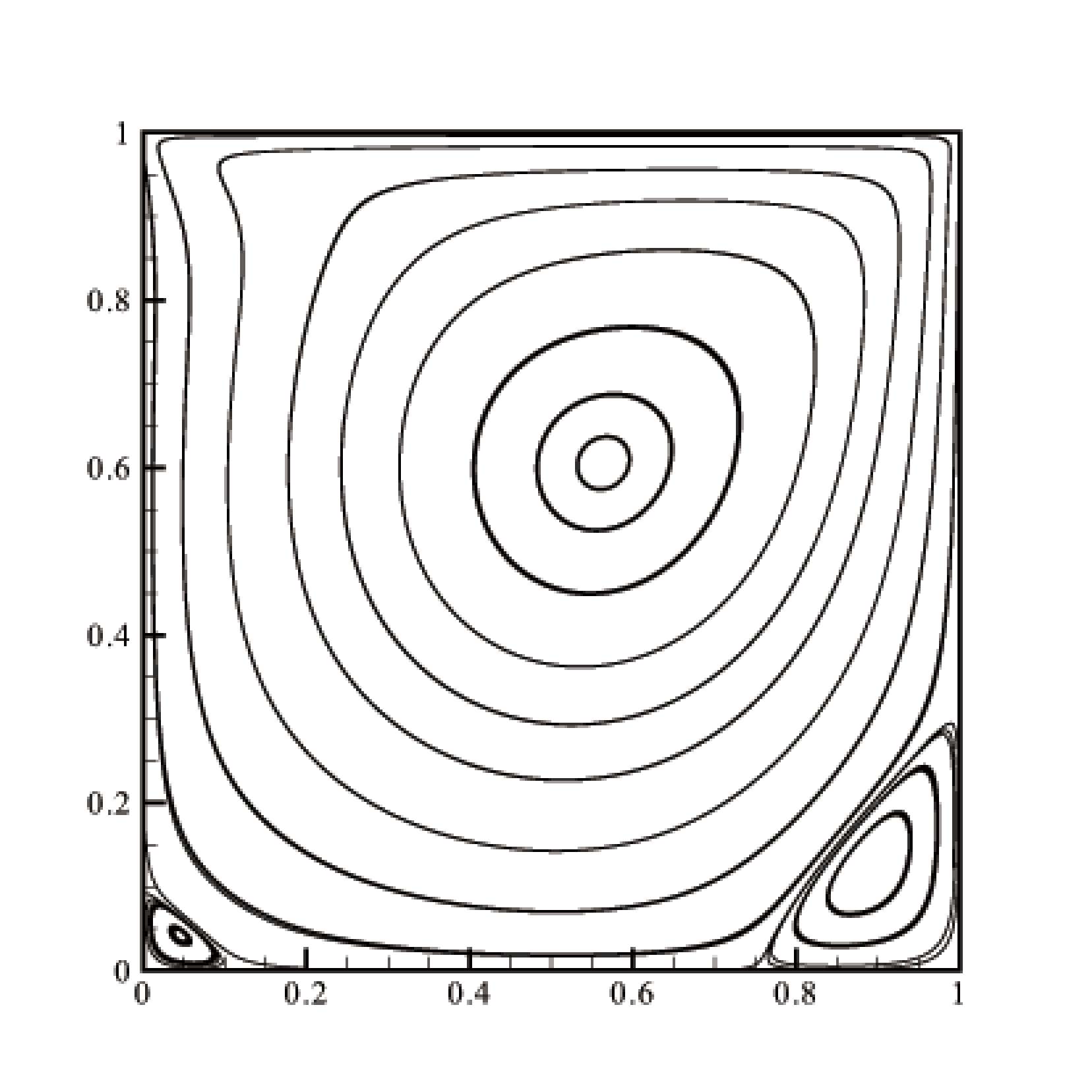}
}\hspace{0.05\textwidth}%
\subfigure[]{
\includegraphics[width=0.45\textwidth]{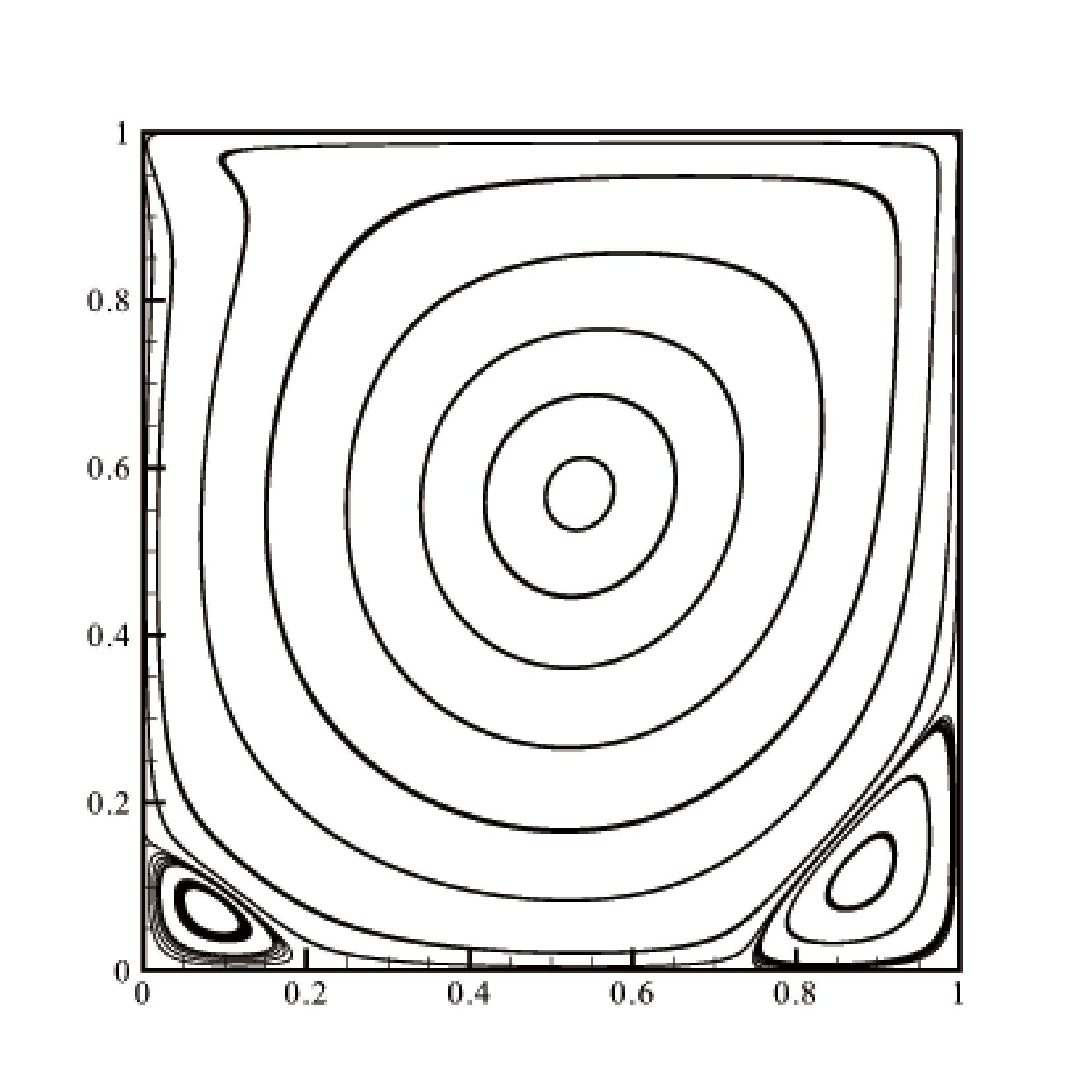}
}
\caption{\label{fig:test4_cpcf} (a) Pressure coefficient distributions and (b) skin friction coefficient distributions for flow around a NACA0012 airfoil at $\rm{Ma} = 0.8$, $\rm{Re} = 500$ and $\alpha  = 10^\circ $.}
\end{figure}

\subsubsection{Flows around circular cylinder}
\begin{figure}
	\centering
	\subfigure[]{
		\includegraphics[width=0.4 \textwidth]{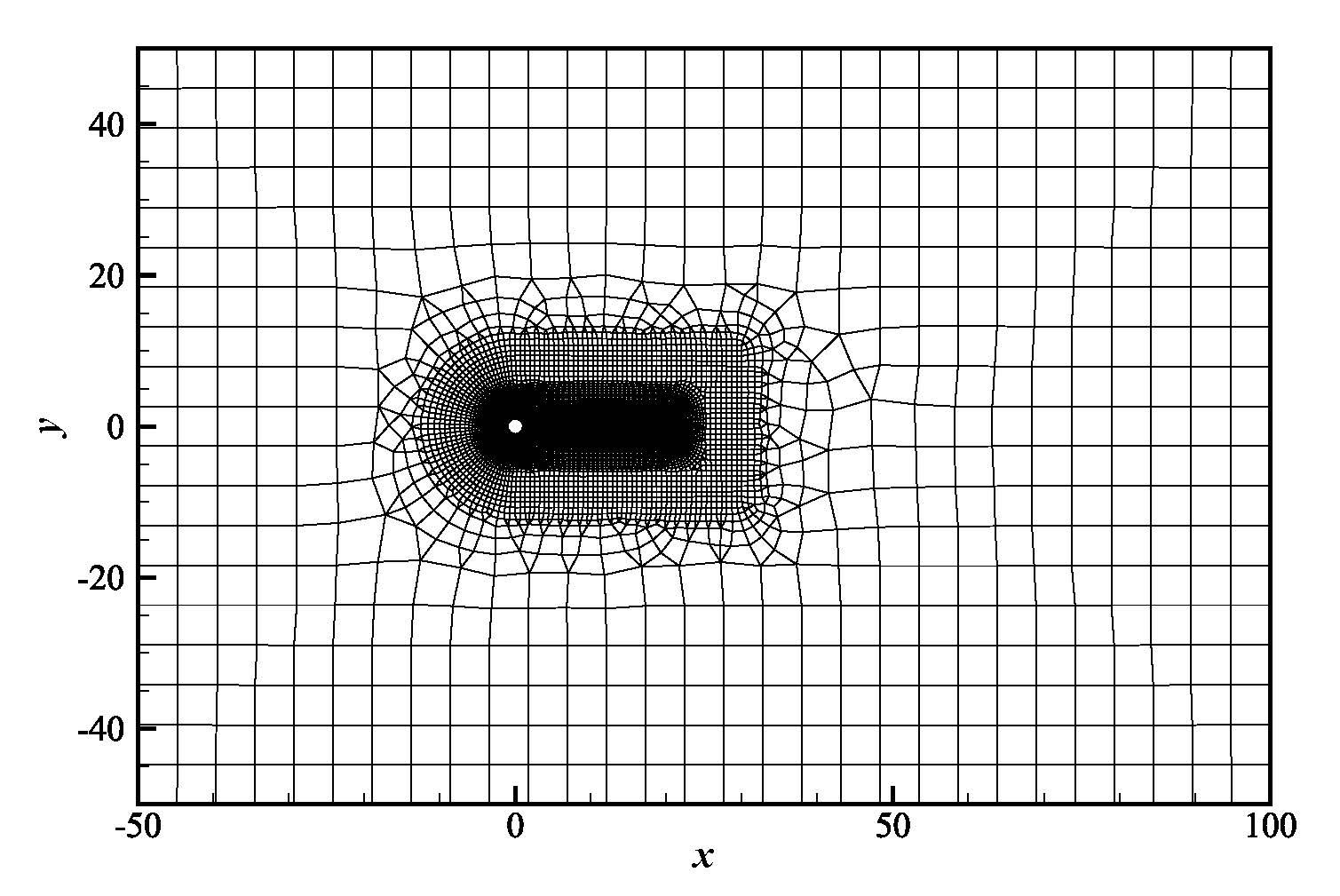}
		\label{cylinder_overallmesh_10_20_40}
	}
	\subfigure[]{
		\includegraphics[width=0.4 \textwidth]{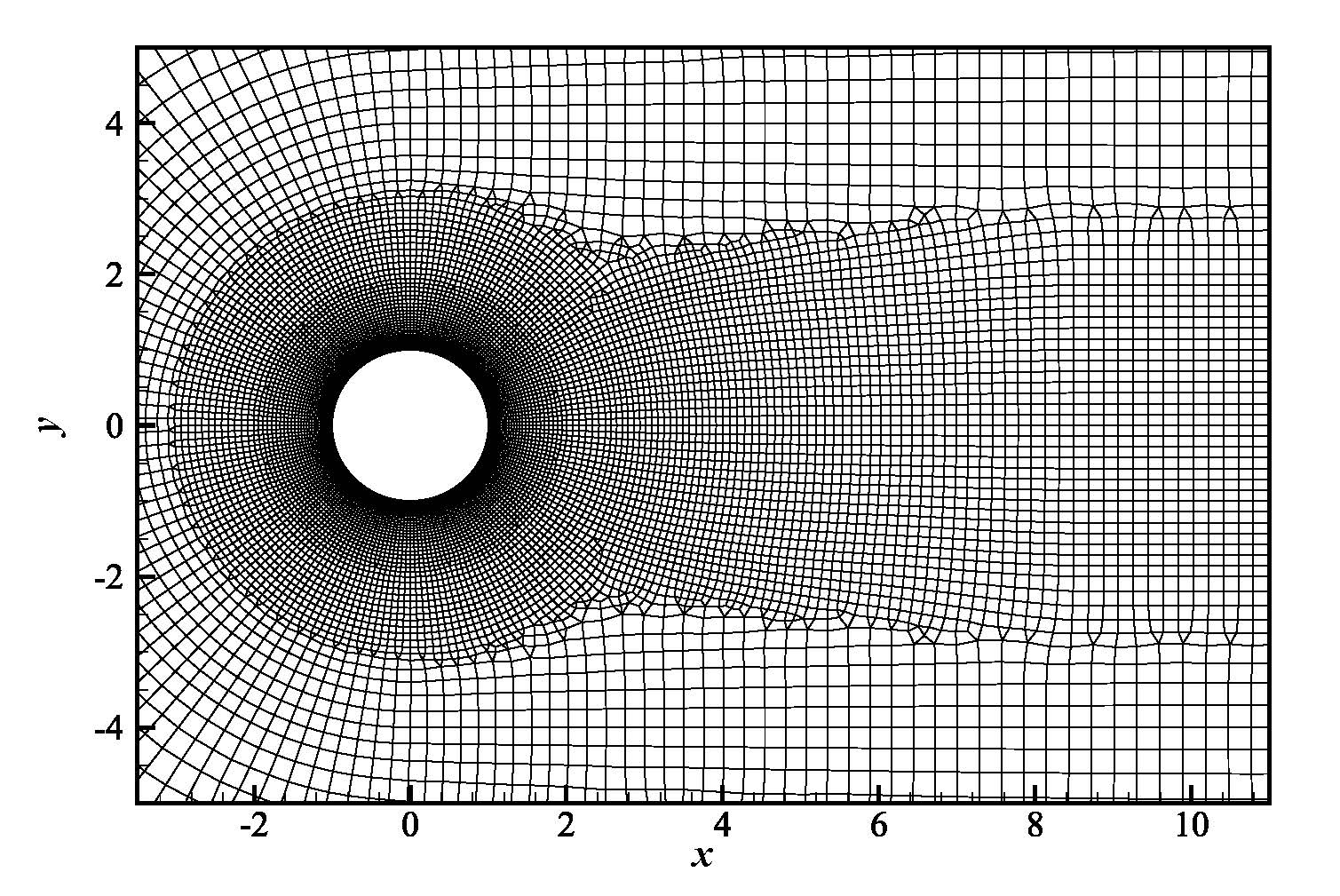}
		\label{cylinder_localmesh_10_20_40}
	}
	\subfigure[]{
		\includegraphics[width=0.4 \textwidth]{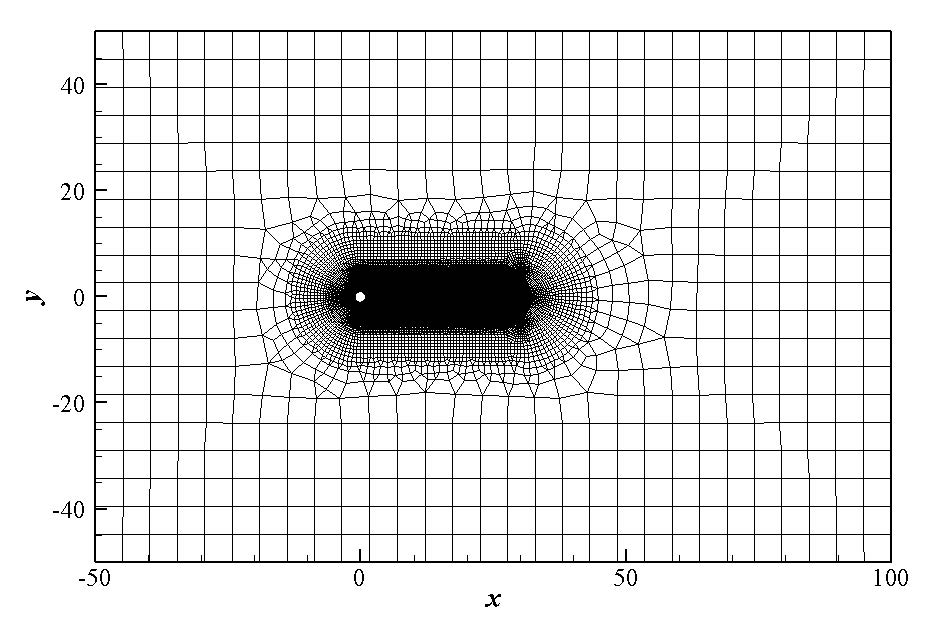}
		\label{cylinder_overallmesh_60_80_100}
	}
	\subfigure[]{
		\includegraphics[width=0.4 \textwidth]{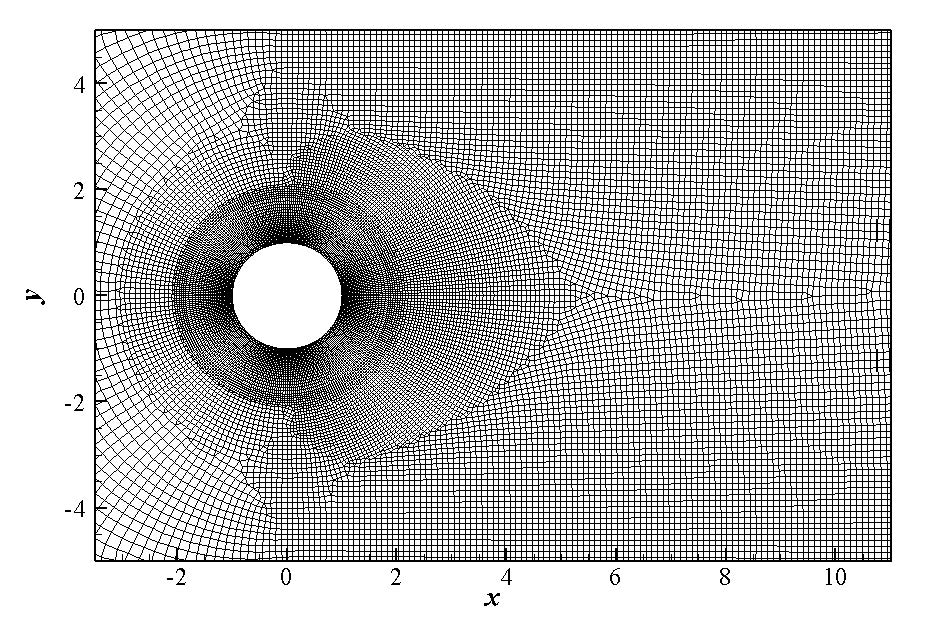}
		\label{cylinder_localmesh_60_80_100}
	}
	\caption{\label{CylinderMesh} Hybrid meshes for the simulation of laminar flow around circular cylinder: (a) Far field for $Re = 10, 20, 40$, and (b) mesh near wall of circular cylinder for $Re = 10, 20, 40$, and (c) far field for $Re = 60, 80, 100$, and (d) mesh near wall of circular cylinder for $Re = 60, 80, 100$.}
\end{figure}

The flow past a single stationary circular cylinder is a good test case to demonstrate present SDUGKS in capturing boundary layer, detached structures and wake flows. Computational domain is set as a $75d \times 50d$ square zone with $d$-diameter cylinder at the center. The geometry and mesh of computational domain are shown in Fig.~\ref{CylinderMesh}. In order to improve the resolution of the tail vortex area and obtain more details of the flow field, the grid is refined. The number of the mesh cells in ${\mathop{\rm Re}\nolimits}  \le 47$ cases is 17658, and the number of the cells in ${\mathop{\rm Re}\nolimits}  \ge 60$ cases is 40029. The minimum mesh spacing for both of the mesh is $0.008d$. On circular cylinder surface, nonslip wall boundary condition is implemented.\\
In order to approximate the incompressible flow, a constant velocity $u_{\infty}$ is specified at far-field boundaries. $u_{\infty}$ is set to be $0.1$, $v_{\infty}$ is set to be $0.0$, density $\rho = 1.0$. The Reynolds number is defined as $Re = \frac{\rho U_\infty d}{\mu}$, where $Re = 10, 20, 40, 45, 47, 60, 80, 100$ respectively.

The drag coefficient and lift coefficient are defined as
\begin{equation}
\begin{aligned}
C_d &= \frac{F_d}{0.5 \rho {u_{\infty}}^2 d},\\
\quad C_l &= \frac{F_l}{0.5 \rho {u_{\infty}}^2 d},
\end{aligned}
\end{equation}
where $F_d$ is drag force, $F_l$ is lift force. \\
The evolution of $C_d$ and $C_l$ at different Reynolds numbers is shown in Fig.~\ref{CylinderCLCD}. From the Fig.~\ref{CylinderCLCD}, we can get that flow is steady when $Re \leq 40$, and $C_d$ can converge to a constant, and $C_l$ can also converge to a constant near zero. And the flow is unsteady when $Re > 60$, $C_d$ and $C_l$ show sinusoidal fluctuation. In order to estimate the critical Reynolds number $Re_{cr}$ for the flow, we test the flow at $Re = 45$, and $47$. $C_l$ shows slowly convergence while $Re = 45$, and for $Re = 47$, $C_l$ will develop into sinusoidal fluctuation, so $45 < Re_{cr} < 47$.\\

\begin{figure}
	\centering
	\subfigure[]{
		\includegraphics[width=0.4 \textwidth]{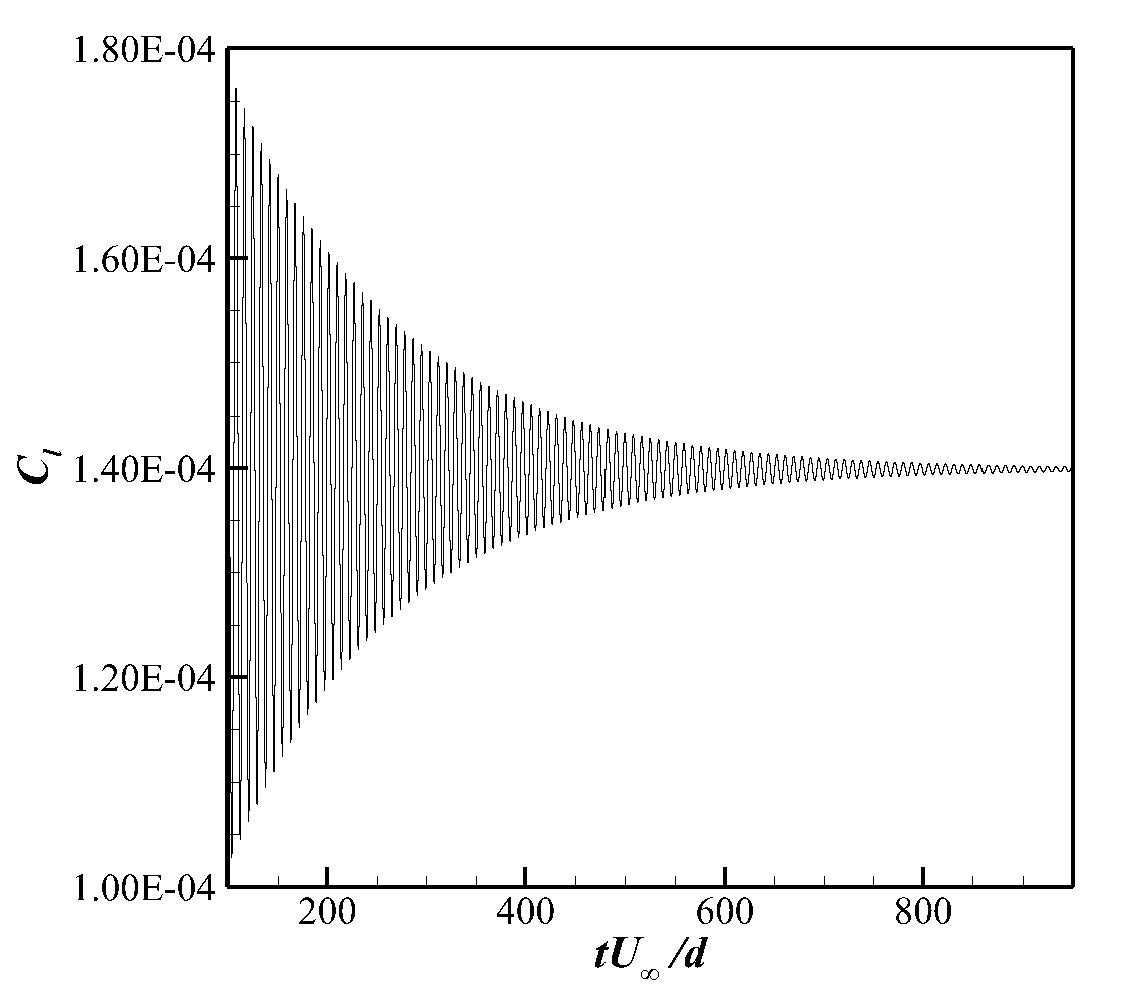}
		\label{CL45}
	}
	\subfigure[]{
		\includegraphics[width=0.4 \textwidth]{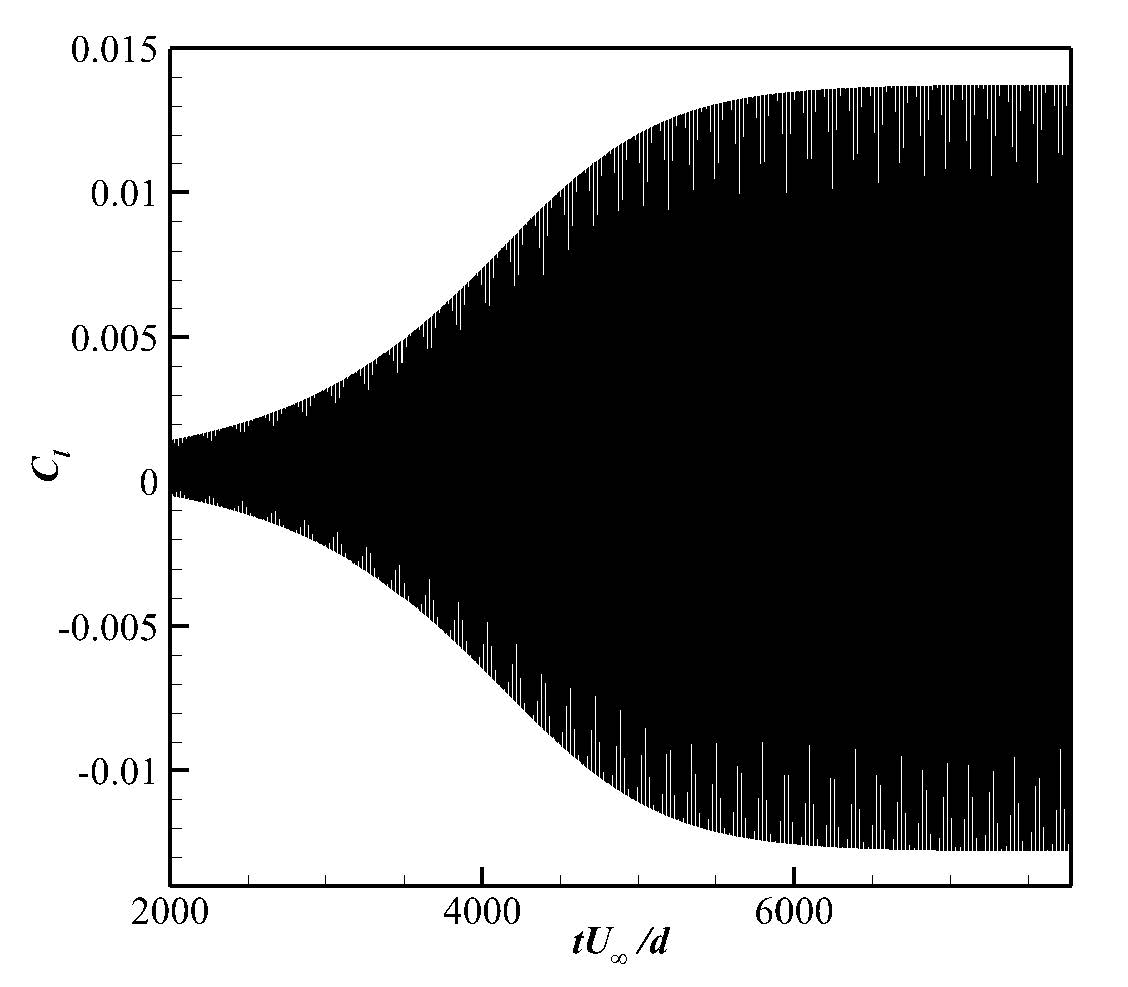}
		\label{CL47}
	}
	\caption{\label{CL4547} Evolution process of the lift coefficient for flows around circular cylinder at (a)Re=45, and (b) Re=47.}
\end{figure}

\begin{figure}[!htp]
    \centering
    \includegraphics[width=0.4 \textwidth]{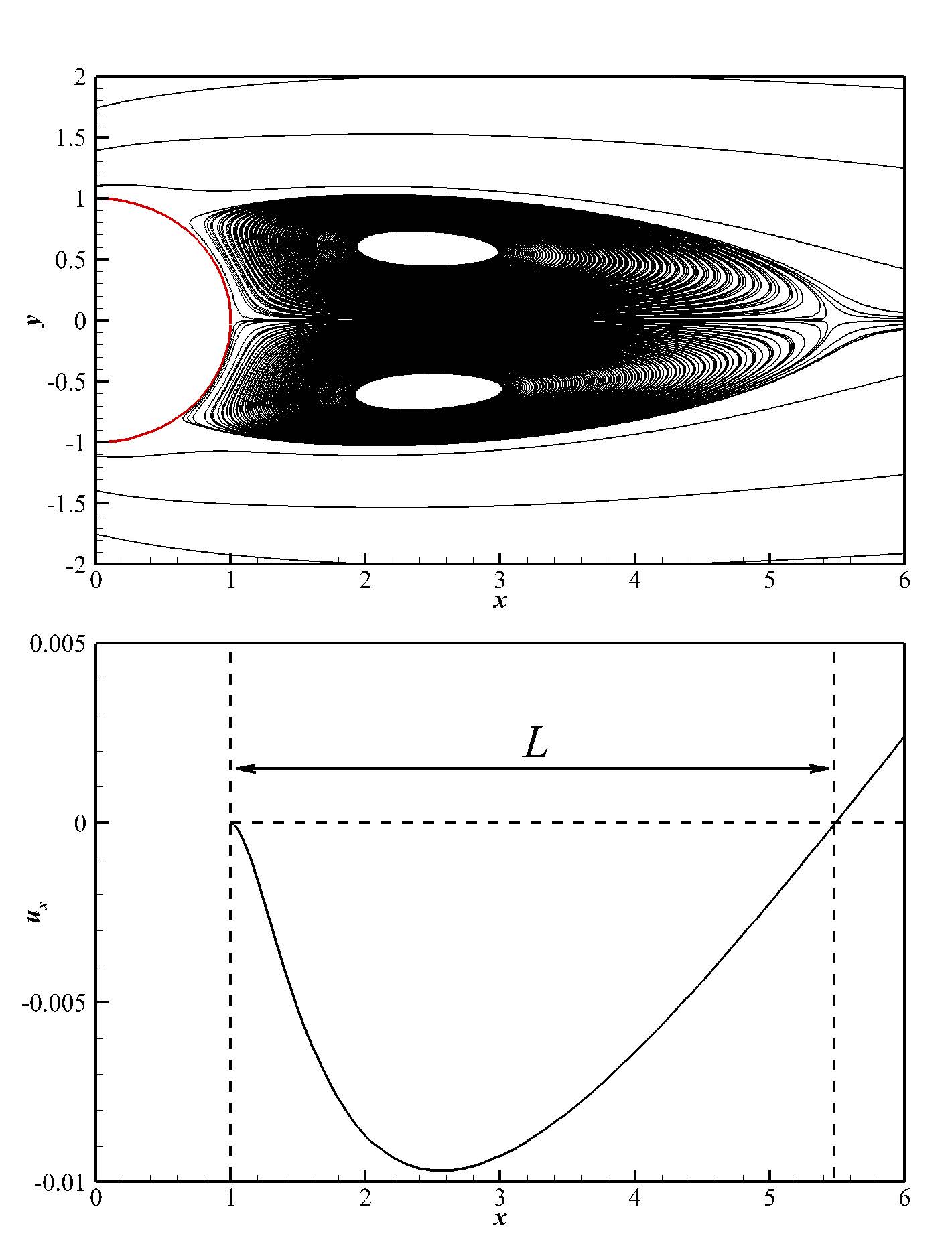}
    \caption{\label{BubbleLength} Definition of the separate vortex length $L$.}
\end{figure}

\begin{figure}
	\centering
	\subfigure[]{
		\includegraphics[width = 0.4 \textwidth]{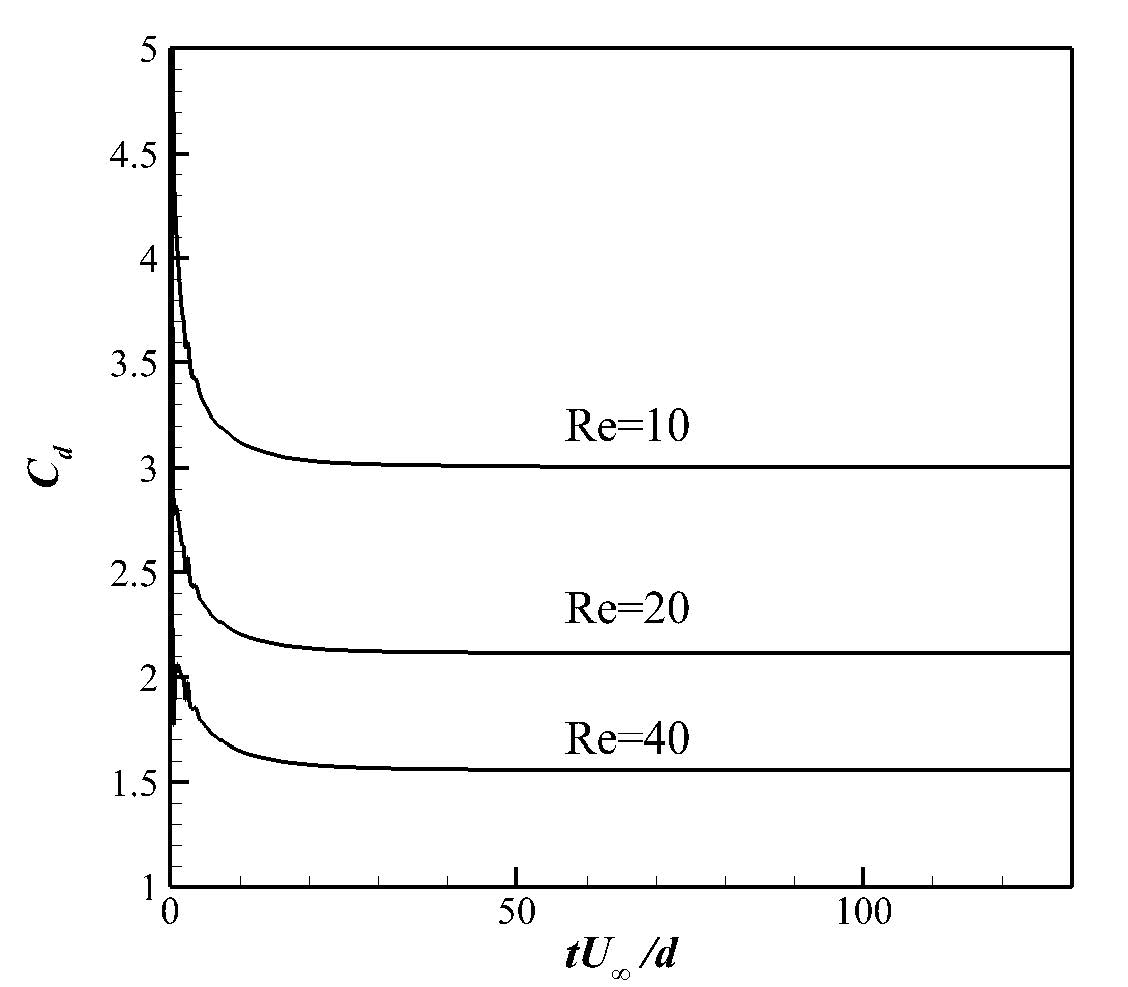}
		\label{CD6080100}
	}
	\subfigure[]{
		\includegraphics[width=0.4 \textwidth]{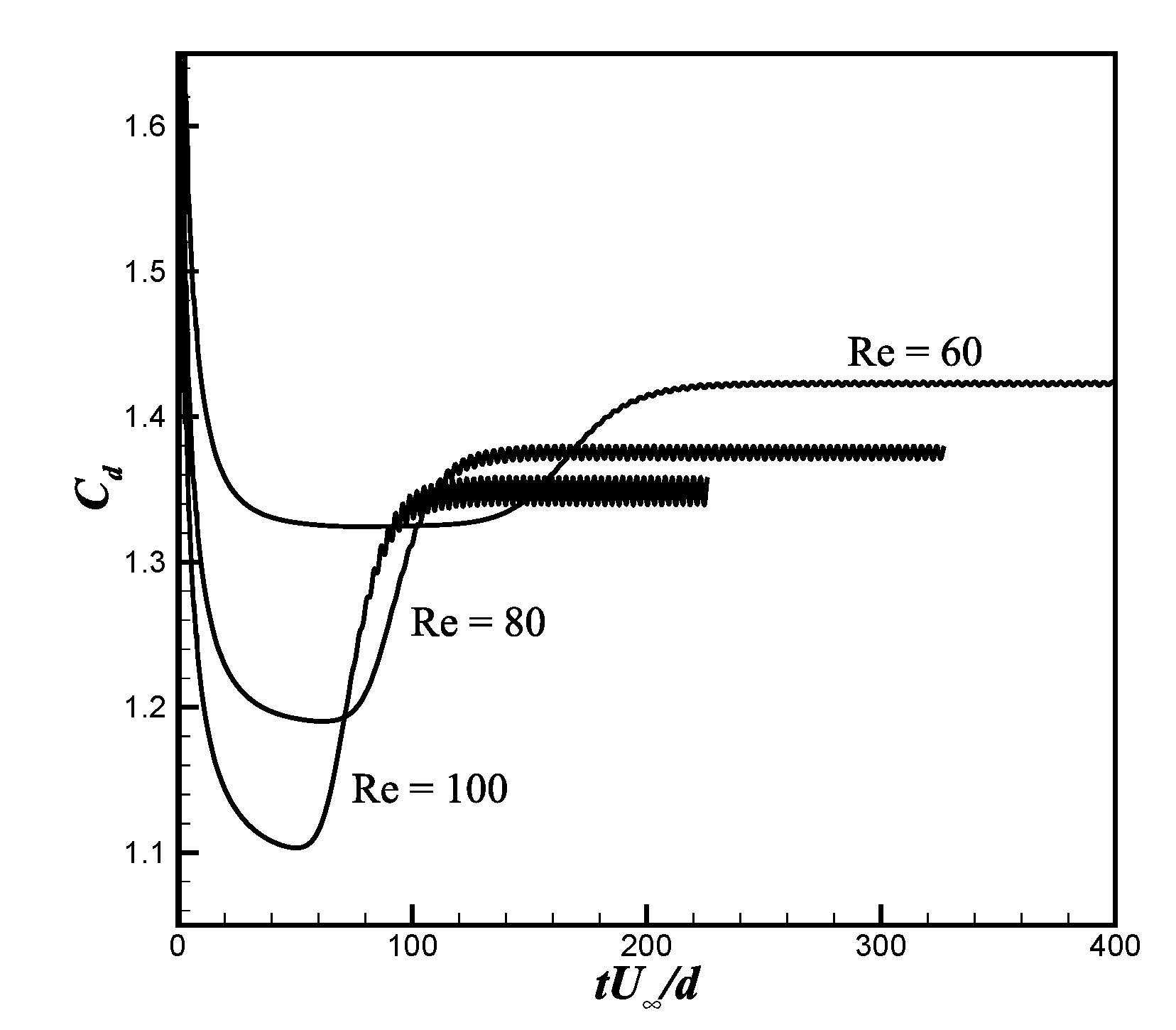}
		\label{CD6080100}
	}
	\subfigure[]{
		\includegraphics[width = 0.4 \textwidth]{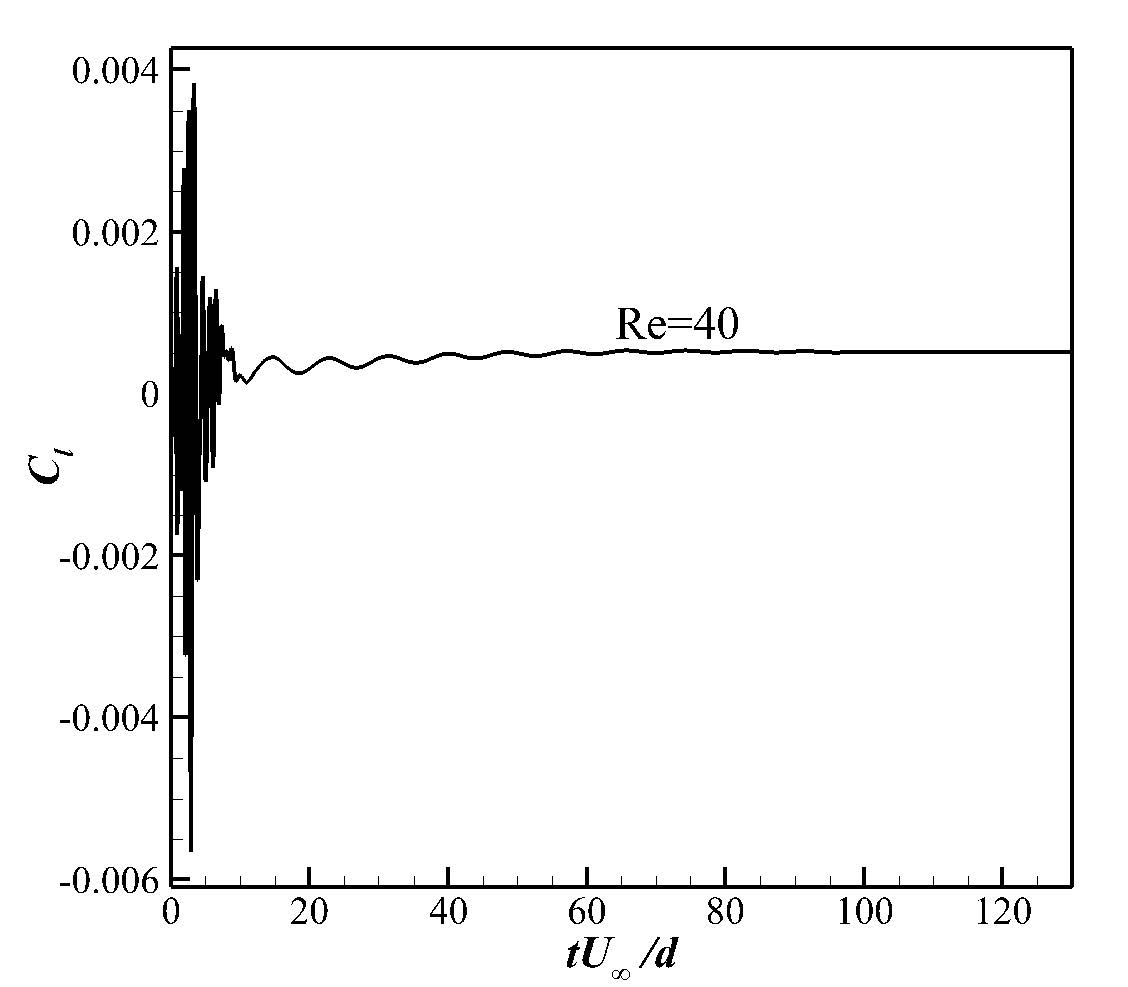}
		\label{CL6080100}
	}
	\subfigure[]{
		\includegraphics[width=0.4 \textwidth]{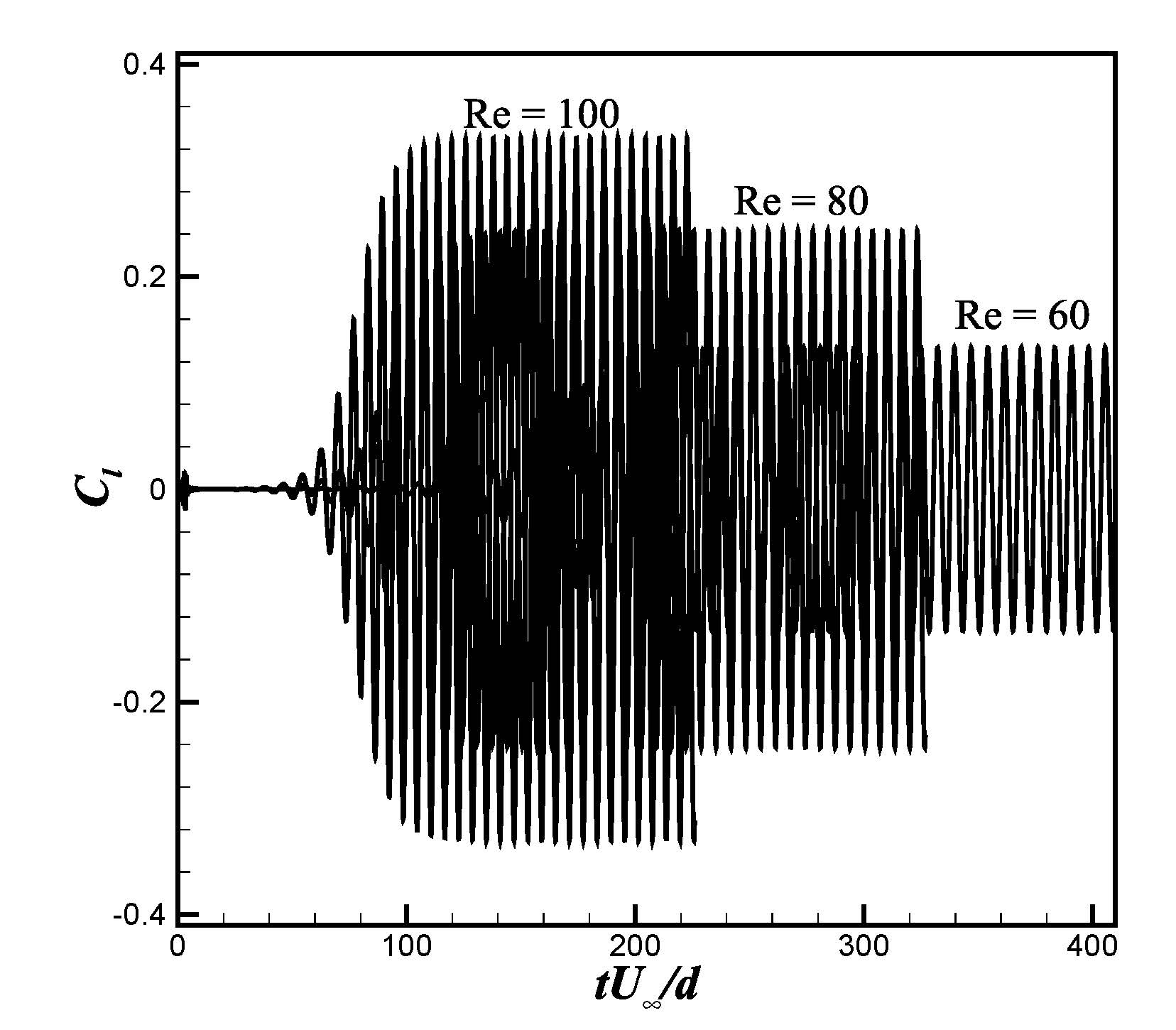}
		\label{CL6080100}
	}
	\caption{\label{CylinderCLCD} The evolutions at different Reynolds numbers of (a) (b) drag coefficient and (c) (d) lift coefficient.}
\end{figure}

The separate vortex length $L$ is defined as the distance between the base point of the cylinder and the stagnation point on the center axis downstream of the cylinder, which is shown in Fig.~\ref{BubbleLength}.\\

\begin{table}
    \centering
    \caption{\label{tab:cylinder_cd} Comparison of mean drag coefficient $\bar{C_d}$ and length of separate vortex $L$ for flow around circular cylinder at $Re = 10, 20, 40$.}
    \setlength{\tabcolsep}{2mm}
    \begin{ruledtabular}
    \begin{tabular}{*{12}{c}}
        \multirow{2}*{Re} &
        \multicolumn{1}{c}{Tritton~\cite{Tritton1959}} & \multicolumn{2}{c}{Park, et al.~\cite{Park1998Numerical}} & \multicolumn{2}{c}{Wang, et al.~\cite{Wang1}} & \multicolumn{2}{c}{Li, et al.~\cite{Li2016Finite}} & \multicolumn{2}{c}{He, et al.~\cite{He1997}} & \multicolumn{2}{c}{Present}\\
        \cline{2-2} \cline{3-4} \cline{5-6} \cline{7-8} \cline{9-10} \cline{11-12}
        & $\bar{C_d}$ & $\bar{C_d}$ & $L/R$ & $\bar{C_d}$ & $L/R$ & $\bar{C_d}$ & $L/R$ & $\bar{C_d}$ & $L/R$ & $\bar{C_d}$ & $L/R$  \\
        \hline
        10 & 2.926 & 2.78 & 0.476 & 2.88 & 0.505 & 3.003 & 0.6649 & 3.170 & 0.474 & 3.015 & 0.493\\
        20 & 2.103 & 2.01 & 1.814 & 2.072 & 1.866 & 2.118 & 2.0376 & 2.152 & 1.842 & 2.115 & 1.842 \\
        40 & 1.605 & 1.51 & 4.502 & 1.545 & 4.609 & 1.568 & 4.7027 & 1.499 & 4.490 & 1.557 & 4.490 \\
    \end{tabular}
    \end{ruledtabular}
\end{table}

\begin{figure}
	\centering
	\subfigure[]{
		\includegraphics[width = 0.4 \textwidth]{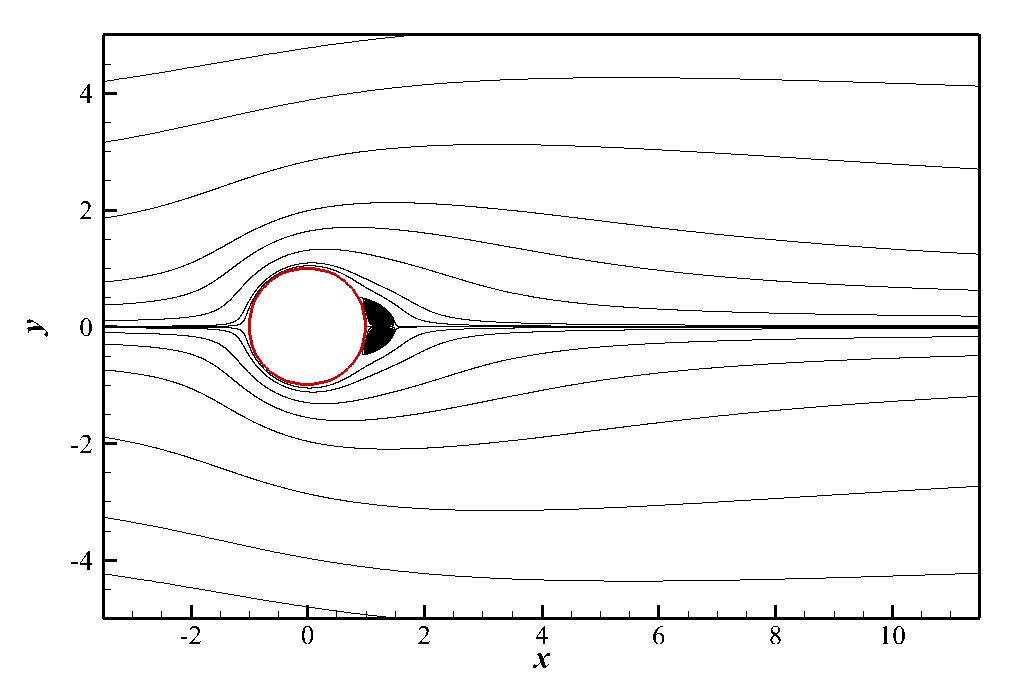}
		\label{Re10Streamline}
	}
	\subfigure[]{
		\includegraphics[width=0.4 \textwidth]{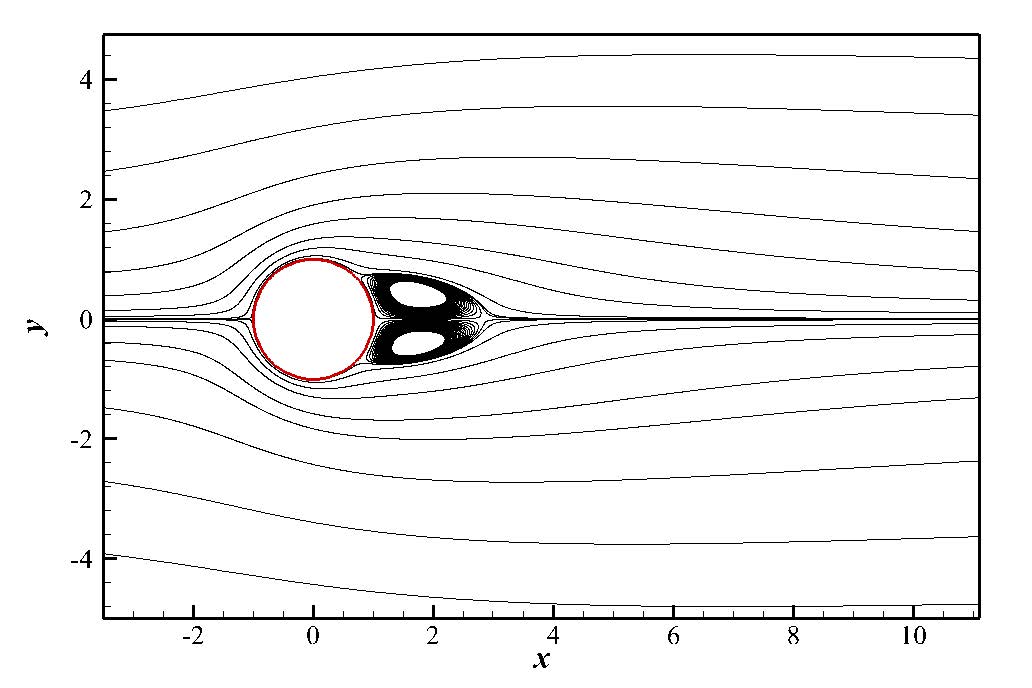}
		\label{Re20Streamline}
	}
	\subfigure[]{
		\includegraphics[width = 0.4 \textwidth]{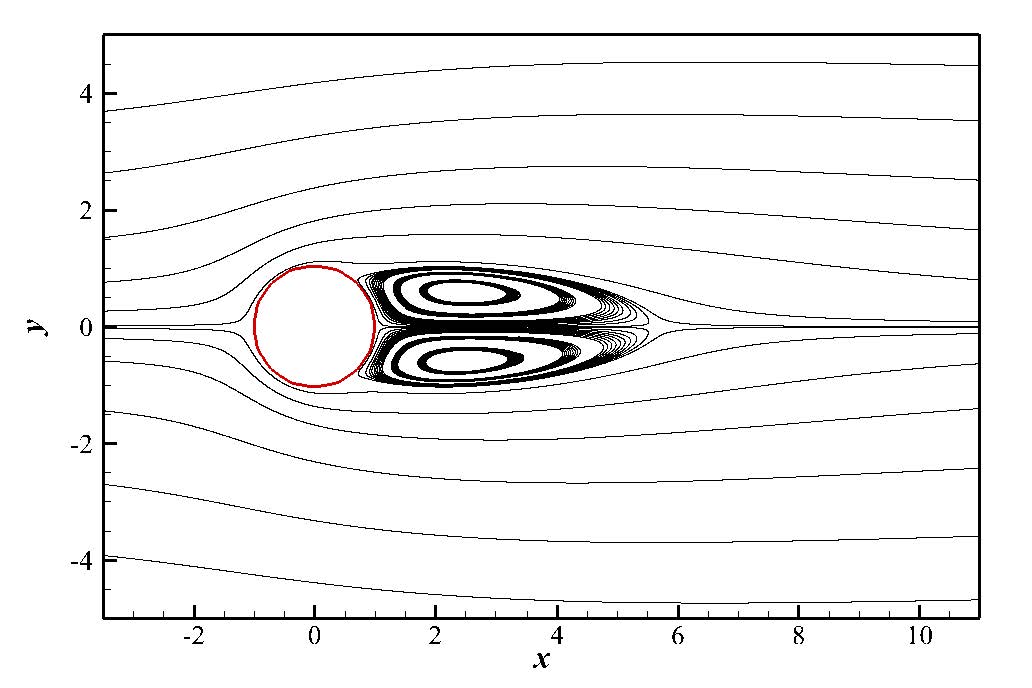}
		\label{Re40Streamline}
	}
	\subfigure[]{
		\includegraphics[width=0.4 \textwidth]{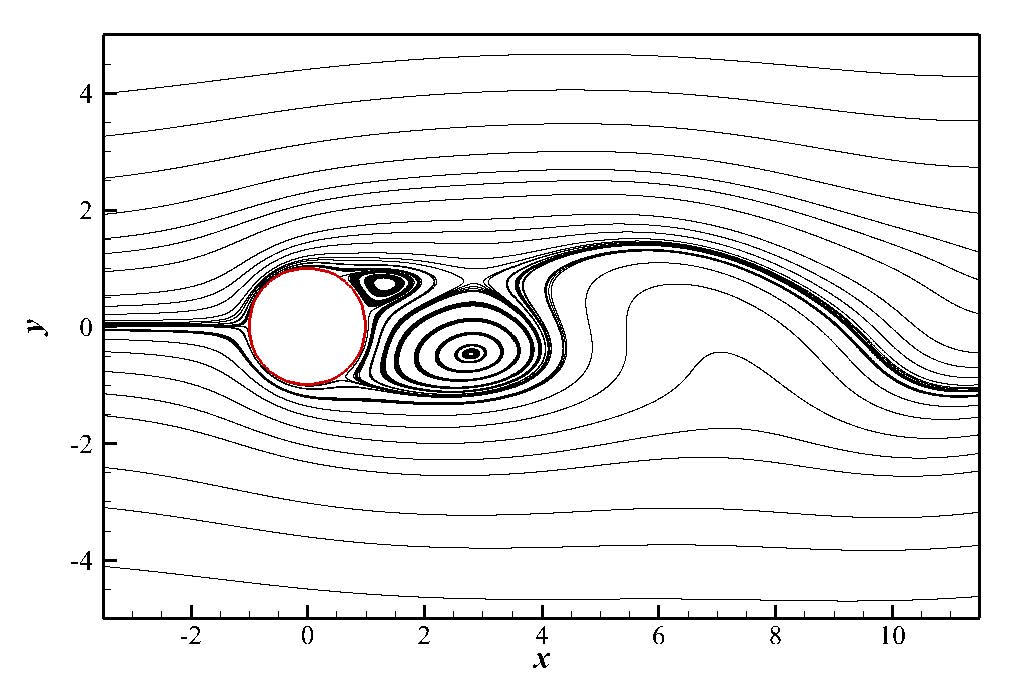}
		\label{Re60Streamline}
	}
	\caption{\label{Streamline} The streamlines at different Reynolds numbers: (a)$Re = 10$, (b)$Re = 20$, (c)$Re = 40$, (d)$Re = 60$.}
\end{figure}

To validate temporal property of present method, time-evolution parameters are also compared to benchmark results. The Strouhal number ($St$) is a dimensionless number for unsteady temporal variation, which is denoted by
\begin{equation}
St = \frac{fd}{U_{\infty}},
\end{equation}
where $f$ is the frequency of vortex shedding, $d$ is the diameter of cylinder.
The present $St$ fits well with the experimental fitting curve for the parallel vortex shedding model in reference paper~\cite{C1989Oblique}, and the experimental fitting curve of $St$ is defined as $St = -3.3265/Re + 0.1816 + 1.6 \times 10^{-4} Re$. The comparison of $C_d$, $St$, and $L/R$ are shown in Tab.~\ref{tab:cylinder_cd} (for steady flows) and Tab.~\ref{tab:cylinder_cd_st} (for unsteady flows).

\begin{figure}
	\centering
	\subfigure[]{
		\includegraphics[width = 0.4 \textwidth]{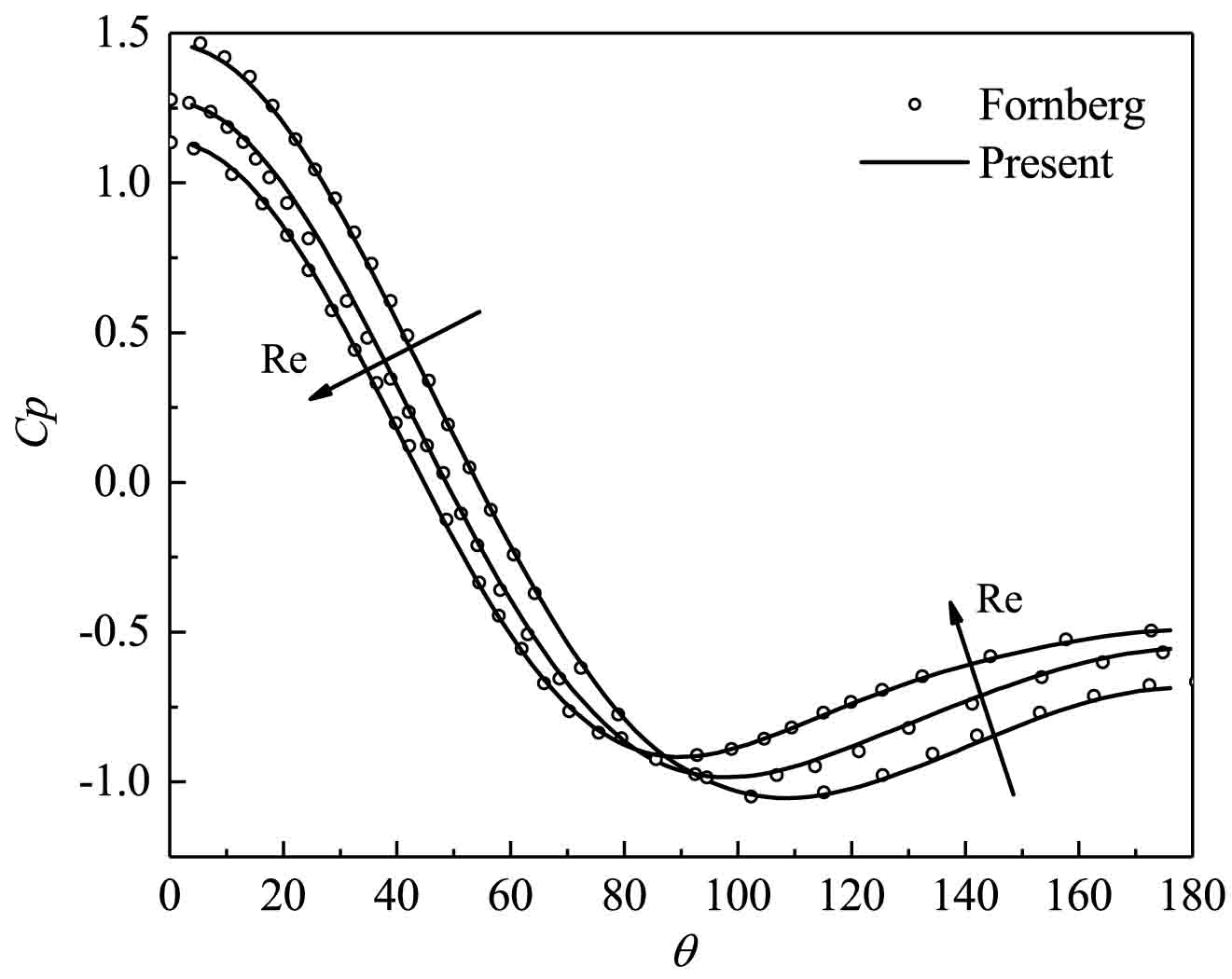}
		\label{cp102040}
	}
	\subfigure[]{
		\includegraphics[width=0.4 \textwidth]{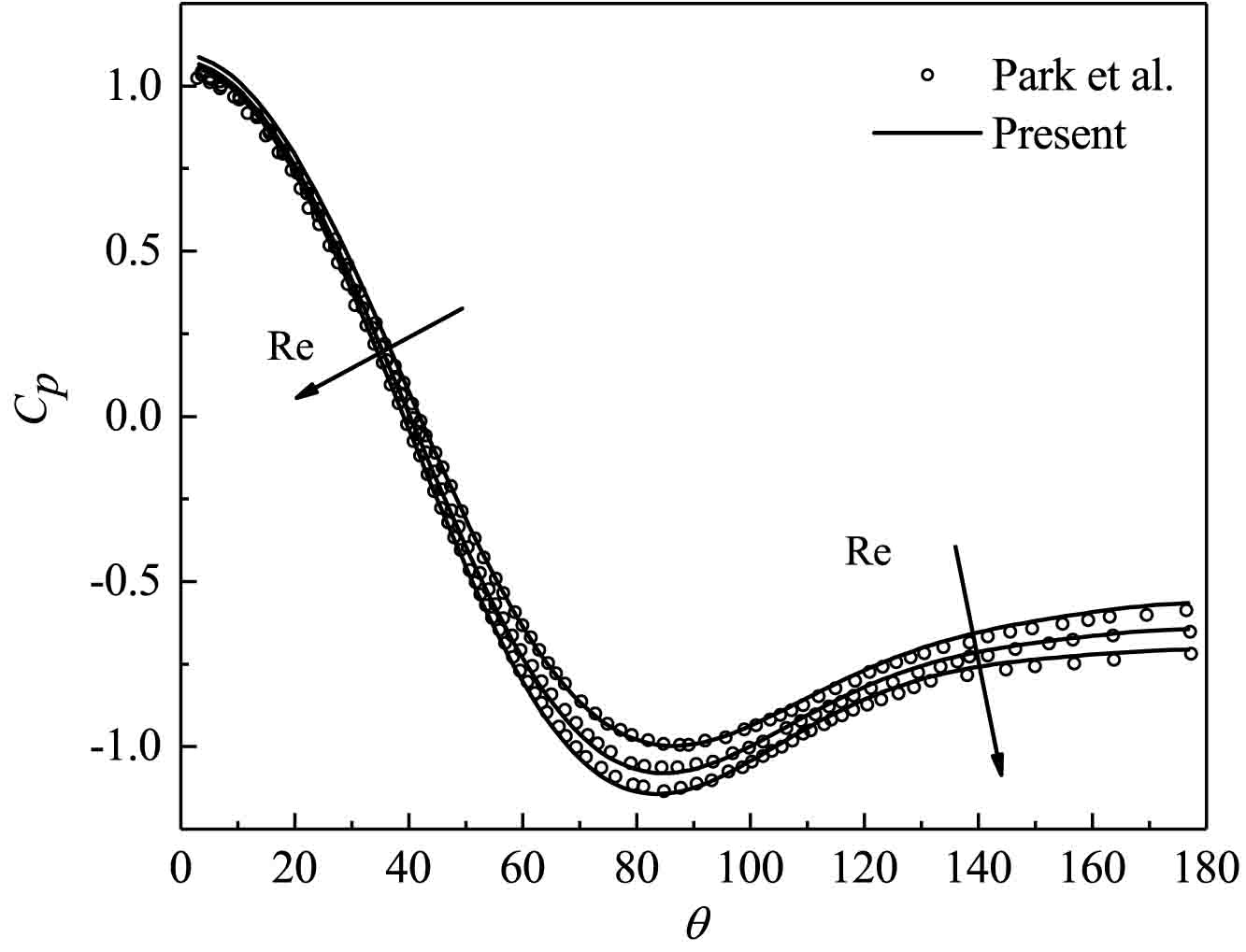}
		\label{cp6080100}
	}
	\caption{\label{CylinderCp} Comparison results of mean pressure coefficients for flow around circular cylinder at (a) $Re = 10, 20, 40$ and (b) $Re = 60, 80, 100$.}
\end{figure}

\begin{table}
    \centering
    \caption{\label{tab:cylinder_cd_st} Comparison of mean drag coefficient $\bar{C_d}$, length of separate vortex $L$, and Strouhal number $St$ for flows around circular cylinder at $Re = 60, 80, 100$.}
    \setlength{\tabcolsep}{2mm}
    \begin{ruledtabular}
    \begin{tabular}{*{11}{c}}
        \multirow{2}*{Re} &
        \multicolumn{1}{c}{Tritton~\cite{Tritton1959}} & \multicolumn{3}{c}{Park, et al.~\cite{Park1998Numerical}} & \multicolumn{3}{c}{Wang, et al.~\cite{Wang1}} & \multicolumn{3}{c}{Present}\\
        \cline{2-2} \cline{3-5} \cline{6-8} \cline{9-11}
        & $\bar{C_d}$ & $\bar{C_d}$ & $L/R$ & $St$ & $\bar{C_d}$ & $L/R$ & $St$ & $\bar{C_d}$ & $L/R$ & $St$ \\
        \hline
        60 & 1.398 & 1.39 & 4.132 & 0.1353 & 1.422 & 4.155 & 0.1375 & 1.4228 & 4.2008 & 0.1377\\
        80 & 1.316 & 1.35 & 3.312 & 0.1528 & 1.379 & 3.306 & 0.1550 & 1.375 & 3.314 & 0.1523\\
        100 & 1.271 & 1.33 & 2.782 & 0.1646 & 1.358 & 2.796 & 0.1670 & 1.349 & 2.832 & 0.1653\\
    \end{tabular}
    \end{ruledtabular}
\end{table}
Finally, mean flow pressure distribution along circular cylinder surface is extracted and compared with previous work. Pressure coefficient is defined as
\begin{equation}
C_{p} = \frac{p - p_{\infty}}{0.5 \rho_{\infty} U_{\infty}^2},
\end{equation}
$C_p$ distribution around the cylinder surface and its comparison with the reference data~\cite{Fornberg1980,Park1998Numerical} is shown in Fig.~\ref{CylinderCp}. And it shows that all the results are in good agreement with the references' data.

\begin{figure}
    \centering
    \includegraphics[width = 0.4 \textwidth]{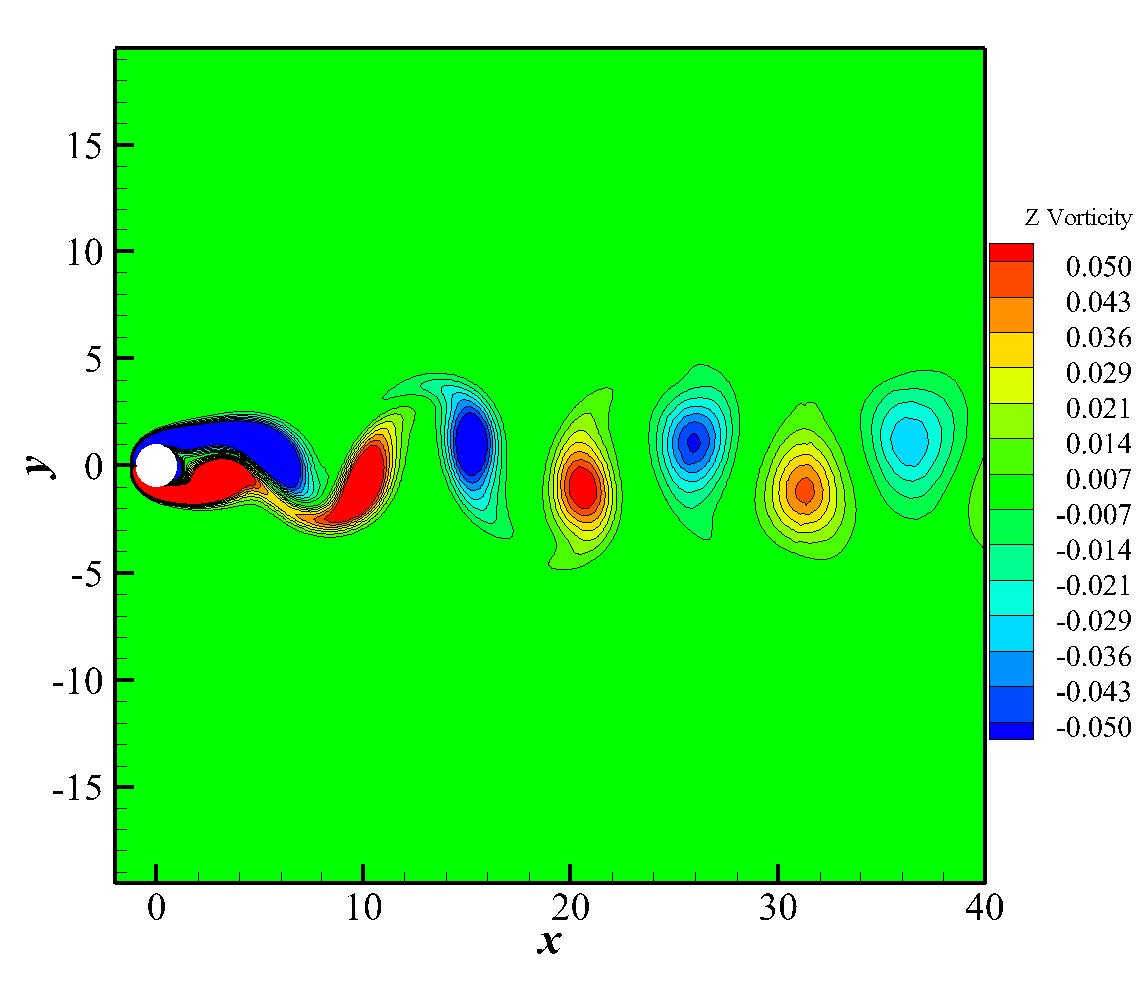}
    \caption{\label{VelocityContoursRe100} Instantaneous vorticity contours for flows around cylinder at $Re = 100$.}
\end{figure}

\subsubsection{Flows around NACA 0012 airfoil}
\begin{figure}\label{NACA0012Mesh}
	\centering
	\subfigure[]{
		\includegraphics[width=0.4 \textwidth]{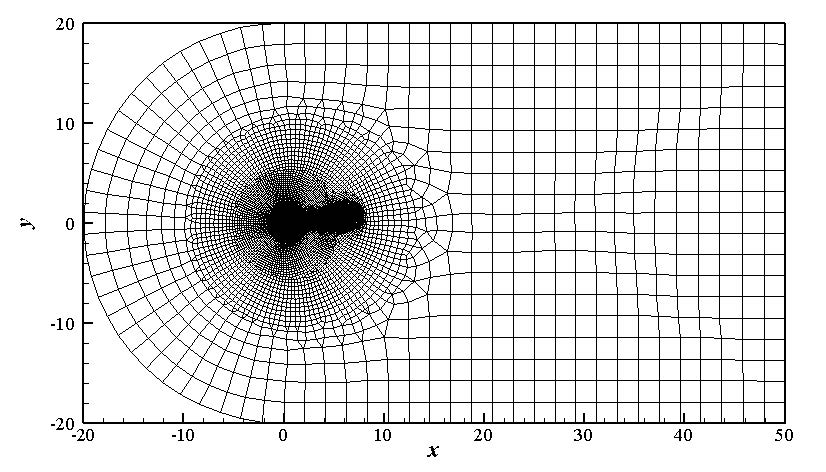}
		\label{NACA0012_overallmesh_AOA0}
	}
	\subfigure[]{
		\includegraphics[width=0.4 \textwidth]{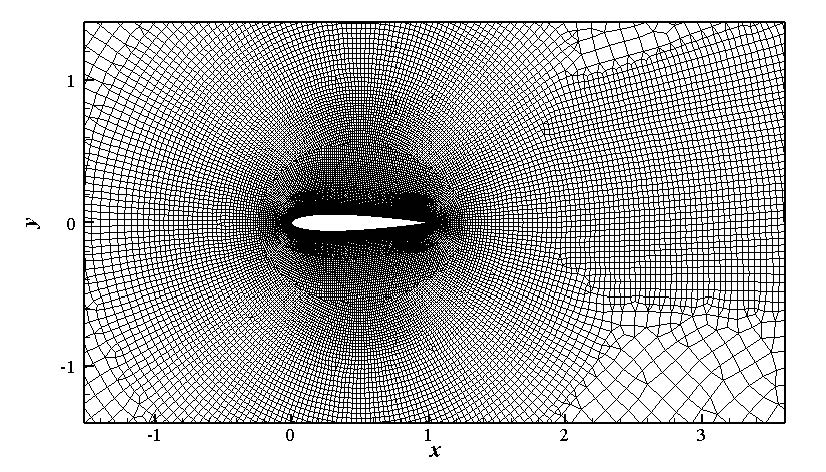}
		\label{NACA0012_localmesh_AOA0}
	}
	\caption{ Hybrid mesh for the simulation of laminar flow around a NACA 0012 airfoil: (a) Far field, and (b) mesh near wall of airfoil.}
\end{figure}

To extend present SDUGKS to extensive applications, flows around a NACA 0012 airfoil are simulated on hybrid meshes. The geometry and meshes of computational domain are shown in 
Fig.~\ref{NACA0012Mesh}. Simulations were conducted with the angles of attack $AOA = 0^{\circ}$ and $AOA = 8^{\circ}$. The total cells of the mesh is 38910, and minimum mesh spacing near the wall of airfoil is $4.5 \times 10^{-4}$. The geometry and mesh of computational domain are shown in Fig.~\ref{NACA0012Mesh}.
In airfoil case, meshes around wall and wake are refined to capture shear flow structures. In farfield, rather coarse mesh is distributed for less computational cost. Different resolutions of mesh are smoothly jointed by quadrilateral/triangle hybrid mesh. Rather good accuracy is obtained with rather few mesh cells. Boundary layer and detached trailing vortex are captured. On hybrid meshes, present method is accurate, robust and efficient in computational fluid dynamics (CFD) research and fluids engineering.

For the initial conditions, The $u_{\infty}$ is set to be $0.1$. The chord length of the airfoil is set to be $L = 1.0$, Reynolds number $Re$ is set to be 500. The density of fluid is set as $\rho = 1.0$, and the velocity components are prescribed as $u = U_\infty cos(AOA \cdot \pi /180),  v = U_\infty sin(AOA \cdot \pi /180)$.

\begin{figure}
	\centering
	\subfigure[Cross sections]{
		\includegraphics[width=0.4 \textwidth]{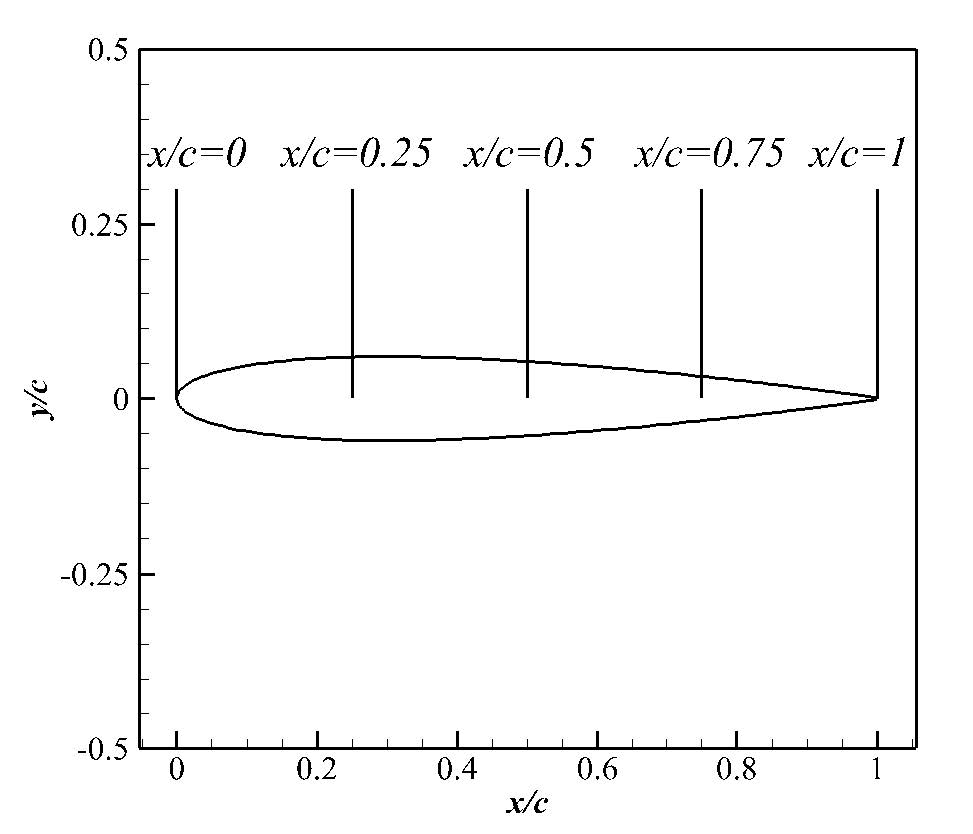}
	}
	\subfigure[$x = 0$]{
		\includegraphics[width=0.4 \textwidth]{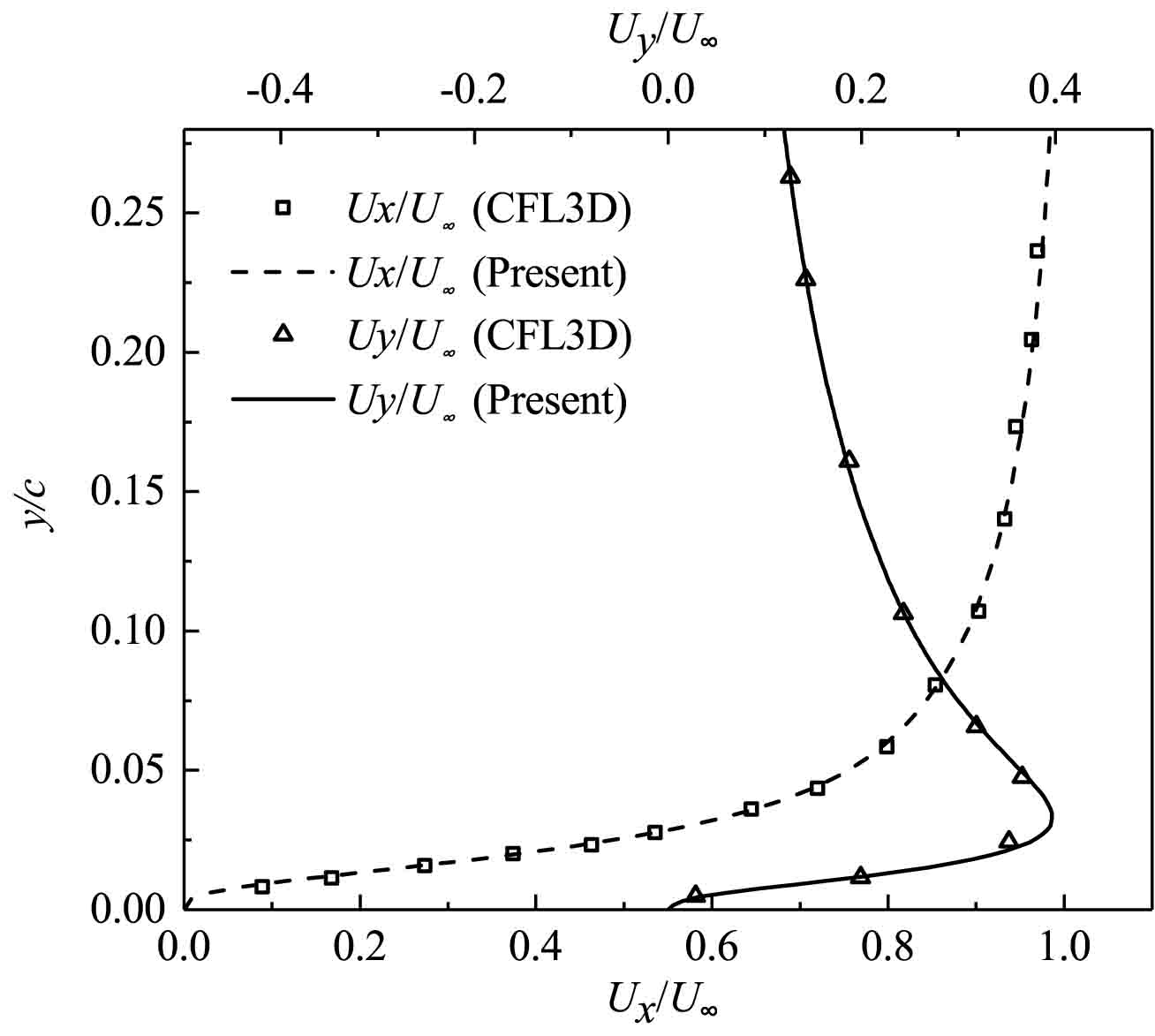}
    }
	\subfigure[$x = 0.25$]{
		\includegraphics[width=0.4 \textwidth]{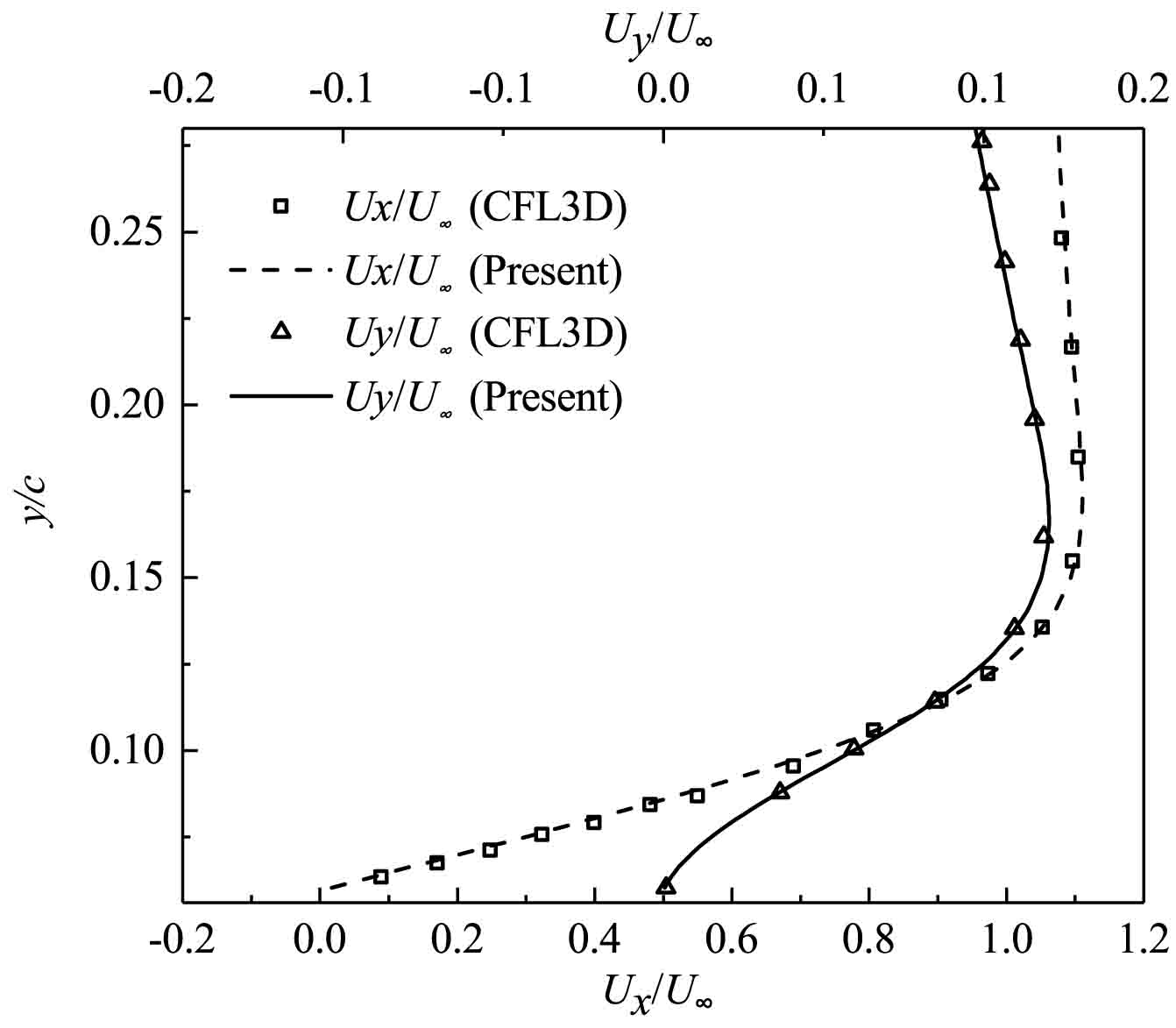}
	}
	\subfigure[$x = 0.5$]{
		\includegraphics[width=0.4 \textwidth]{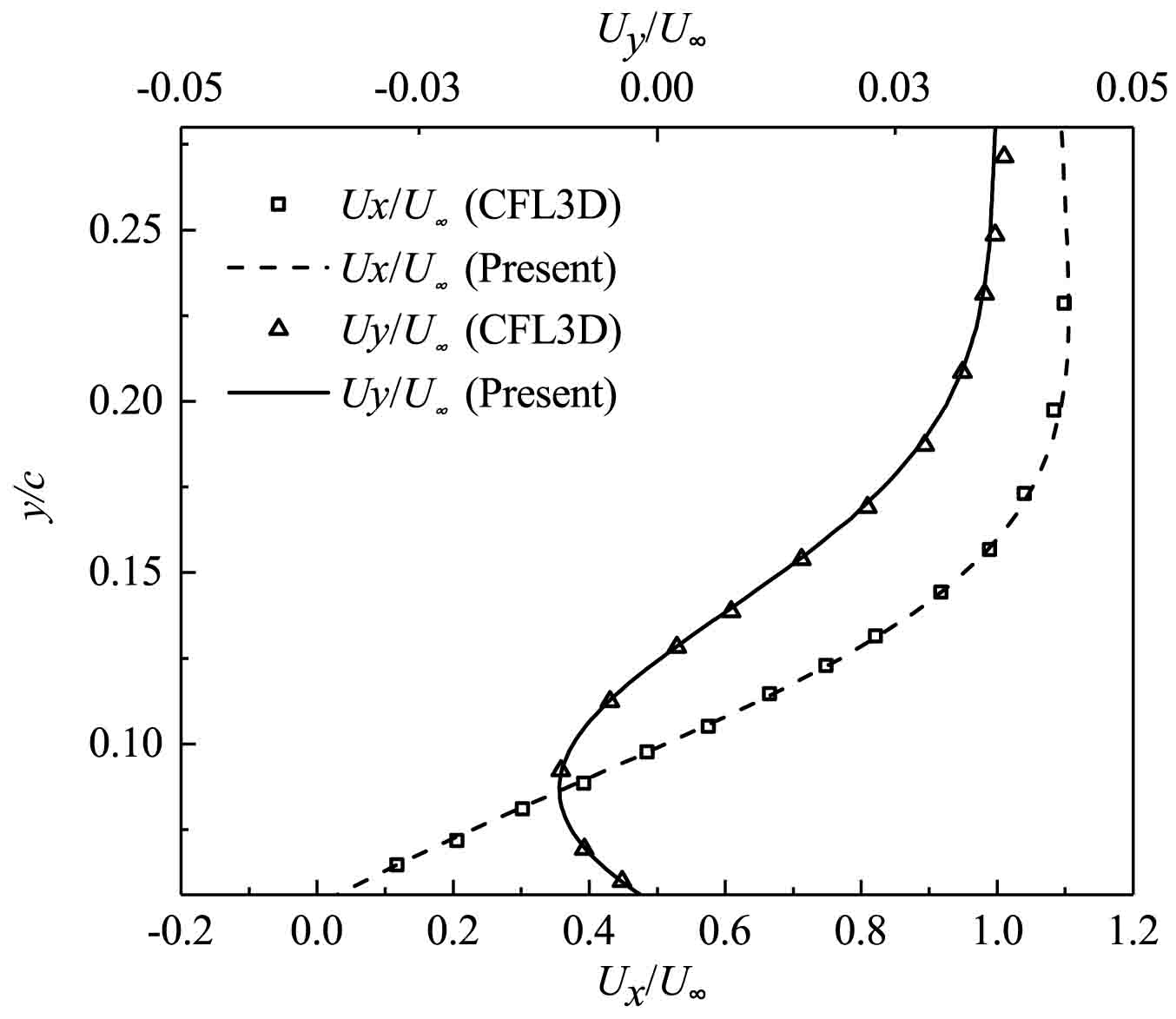}
    }
	\subfigure[$x = 0.75$]{
		\includegraphics[width=0.4 \textwidth]{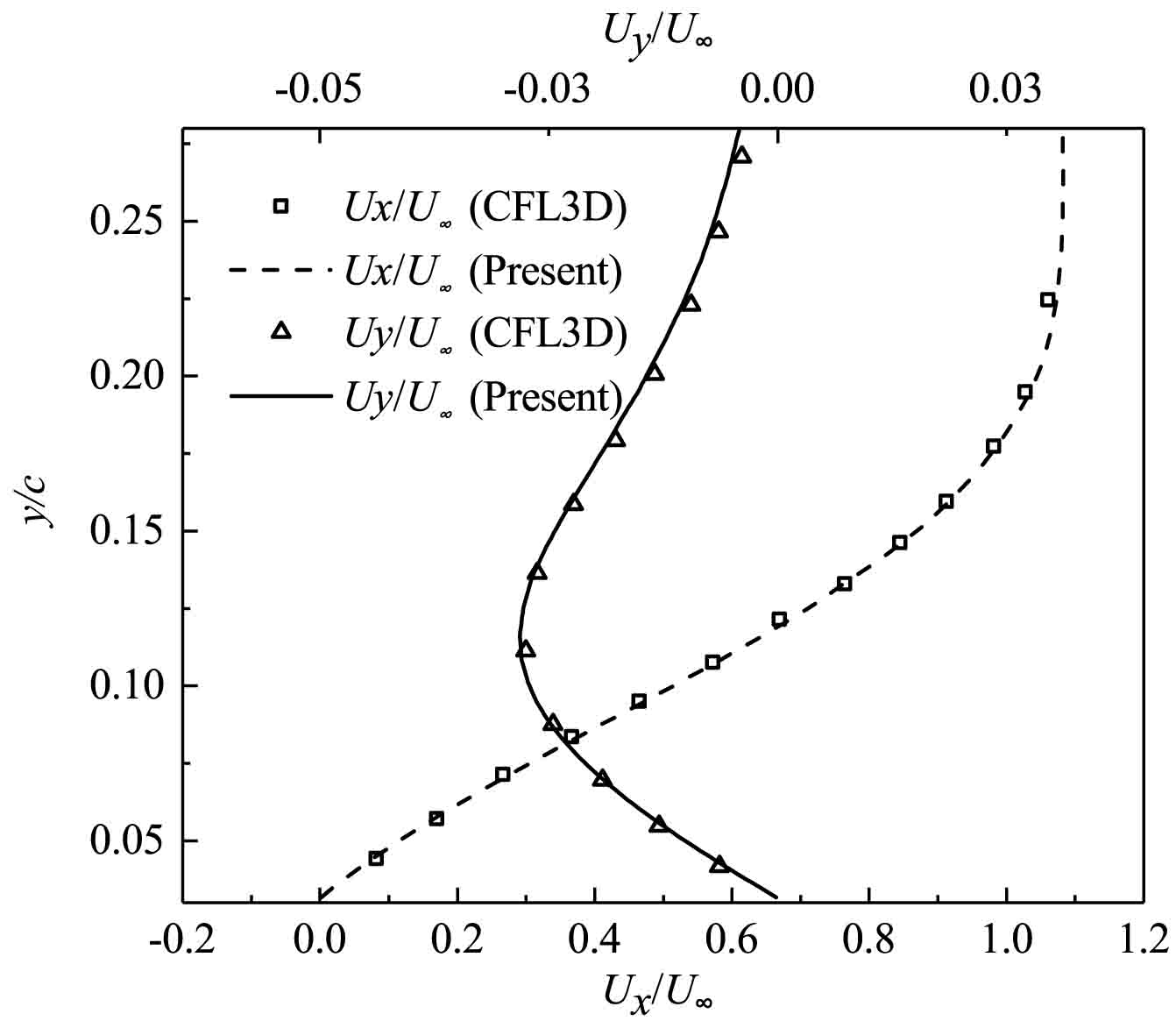}
    }
    \subfigure[$x = 1.0$]{
		\includegraphics[width=0.4 \textwidth]{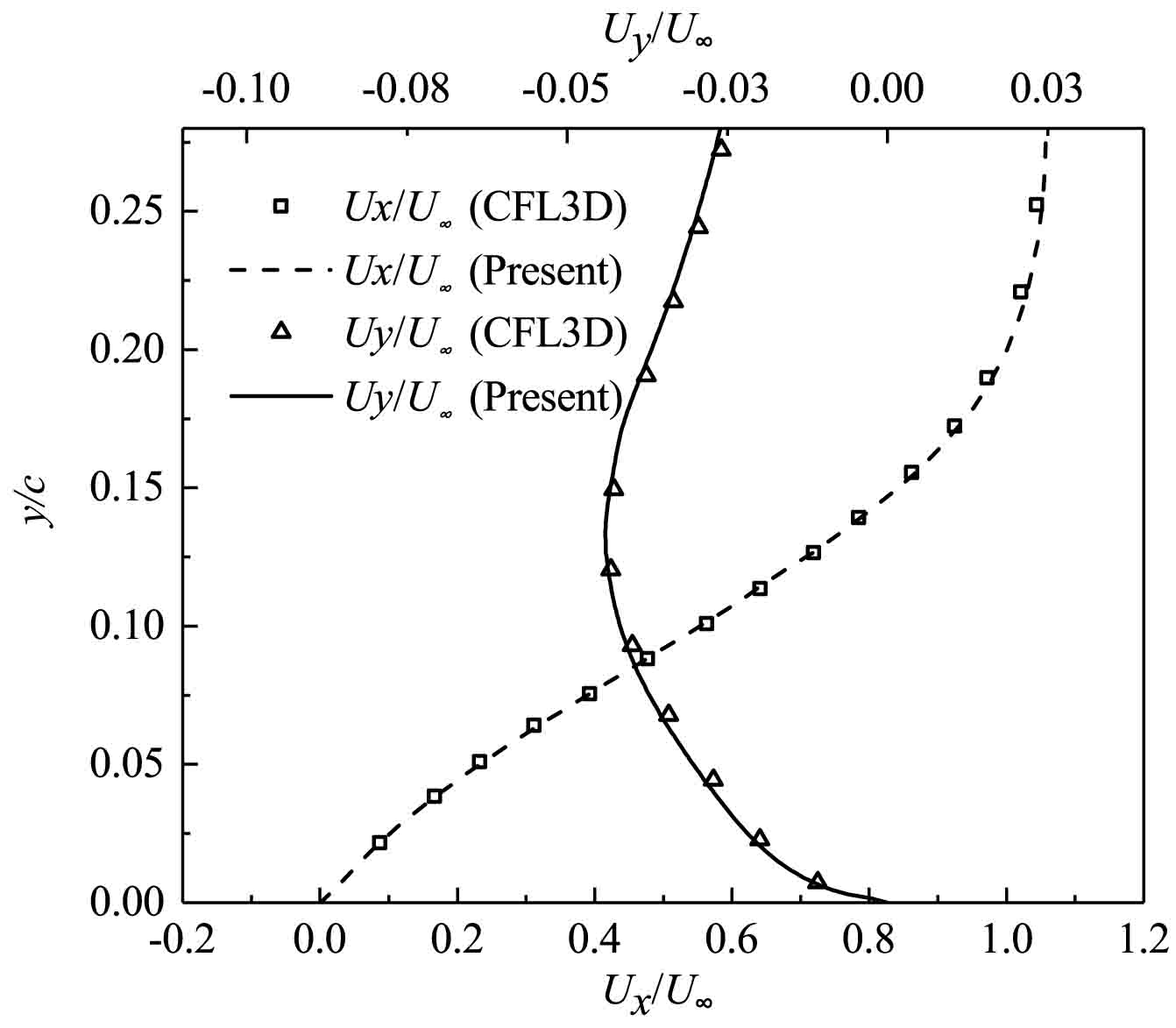}
    }
	\caption{\label{Velocity_AOA0} Comparison results for velocity profiles of laminar flow around a NACA 0012 airfoil at $AOA = 0^{\circ}$ at various cross sections $x$.}
\end{figure}

\begin{figure}
	\centering
	\subfigure[]{
	   \includegraphics[width=0.4 \textwidth]{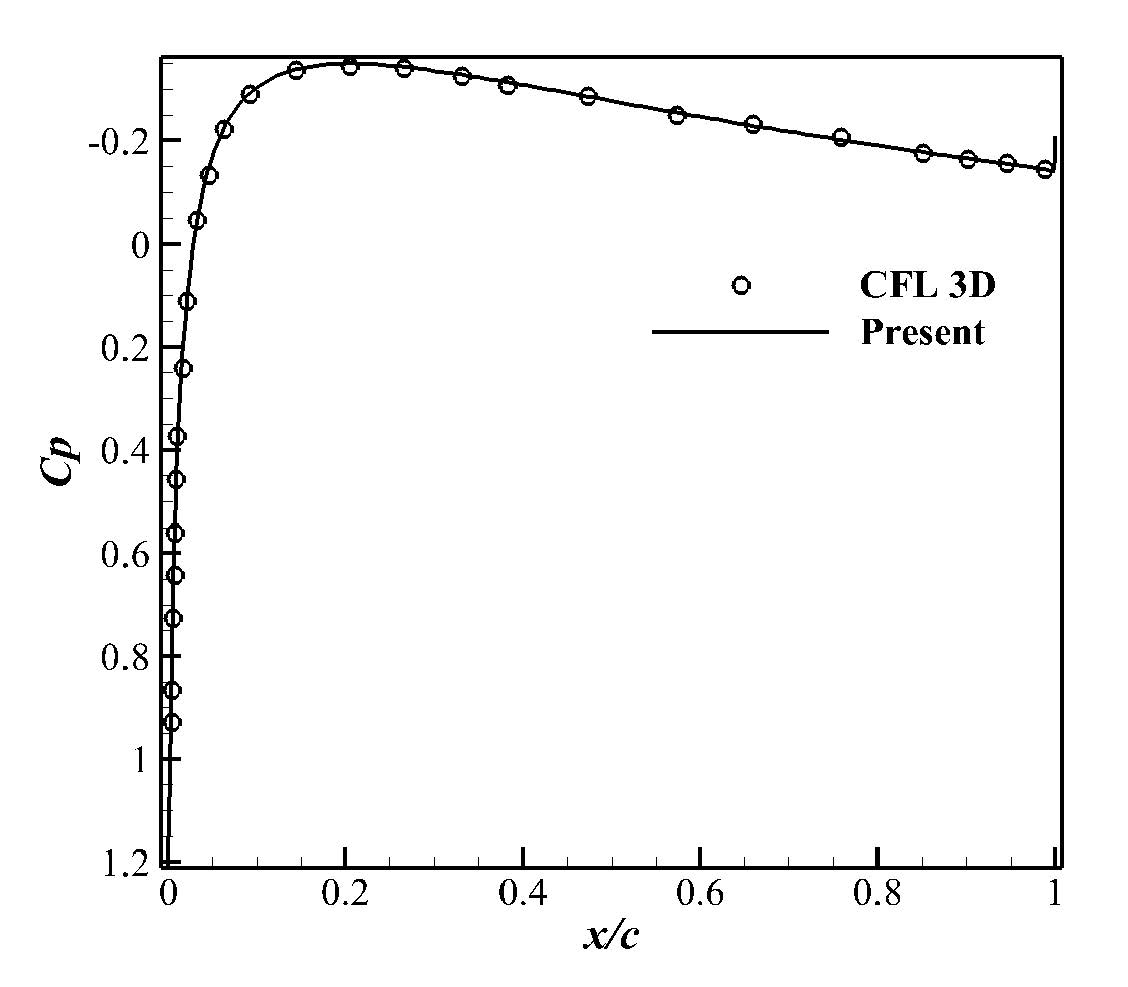}
	   \label{AOA0_cp}
    }
	\subfigure[]{
	   \includegraphics[width=0.4 \textwidth]{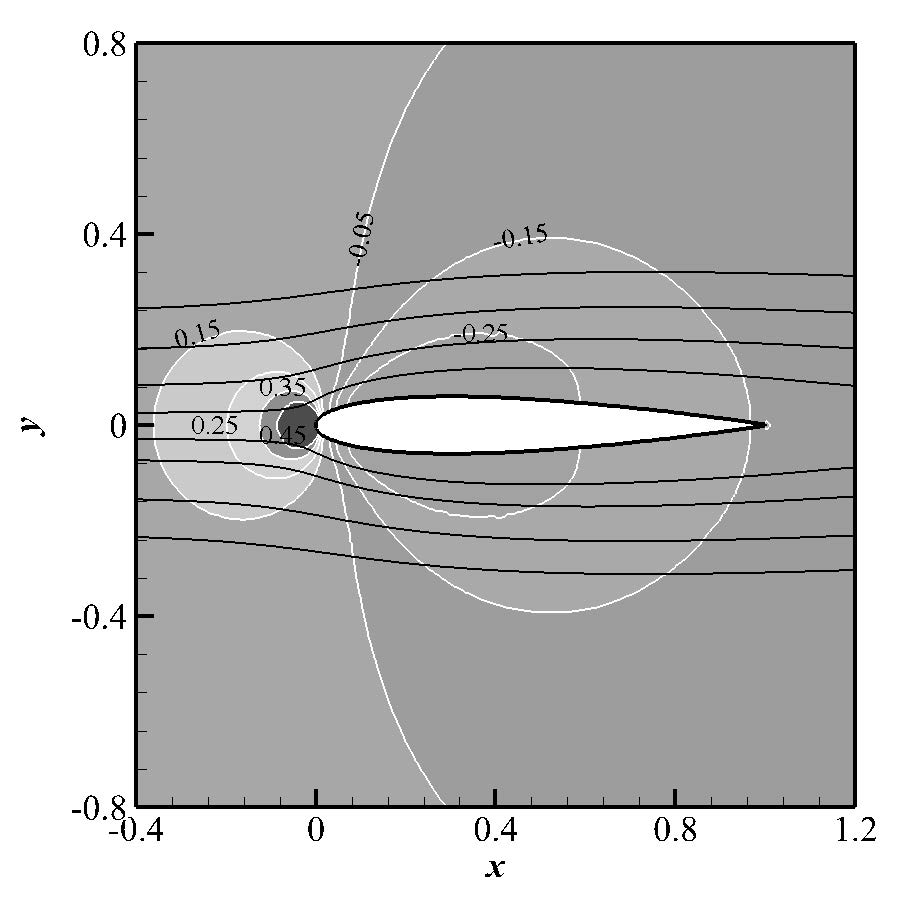}
	   \label{AOA0_cpcontour_and_streamline}
    }
    \caption{\label{AOA0_cp_Cpcontour_and_streamline} (a) The pressure coefficient distribution over NACA 0012 airfoil at $AOA = 0^{\circ}$. (b) Pressure coefficient contour and streamline around NACA 0012 airfoil at $AOA = 0^{\circ}$.}
\end{figure}

First we test NACA 0012 at $AOA = 0^{\circ}$, the drag and lift coefficient have been calculated, these calculation results are consistent with GILBM~\cite{Imamura2005Flow} and CFL3D, which are shown in Tab.~\ref{tab:NACA0012_cd_cl_AOA0}. The pressure coefficient, pressure coefficient contour, and streamline around NACA 0012 airfoil are shown in Fig.~\ref{AOA0_cp_Cpcontour_and_streamline}. The velocity profiles at different cross sections are shown in Fig.~\ref{Velocity_AOA0}, which fits the CFL3D calculation results well in every cross section.

\begin{figure}
	\centering
	\subfigure[]{
	   \includegraphics[width=0.4 \textwidth]{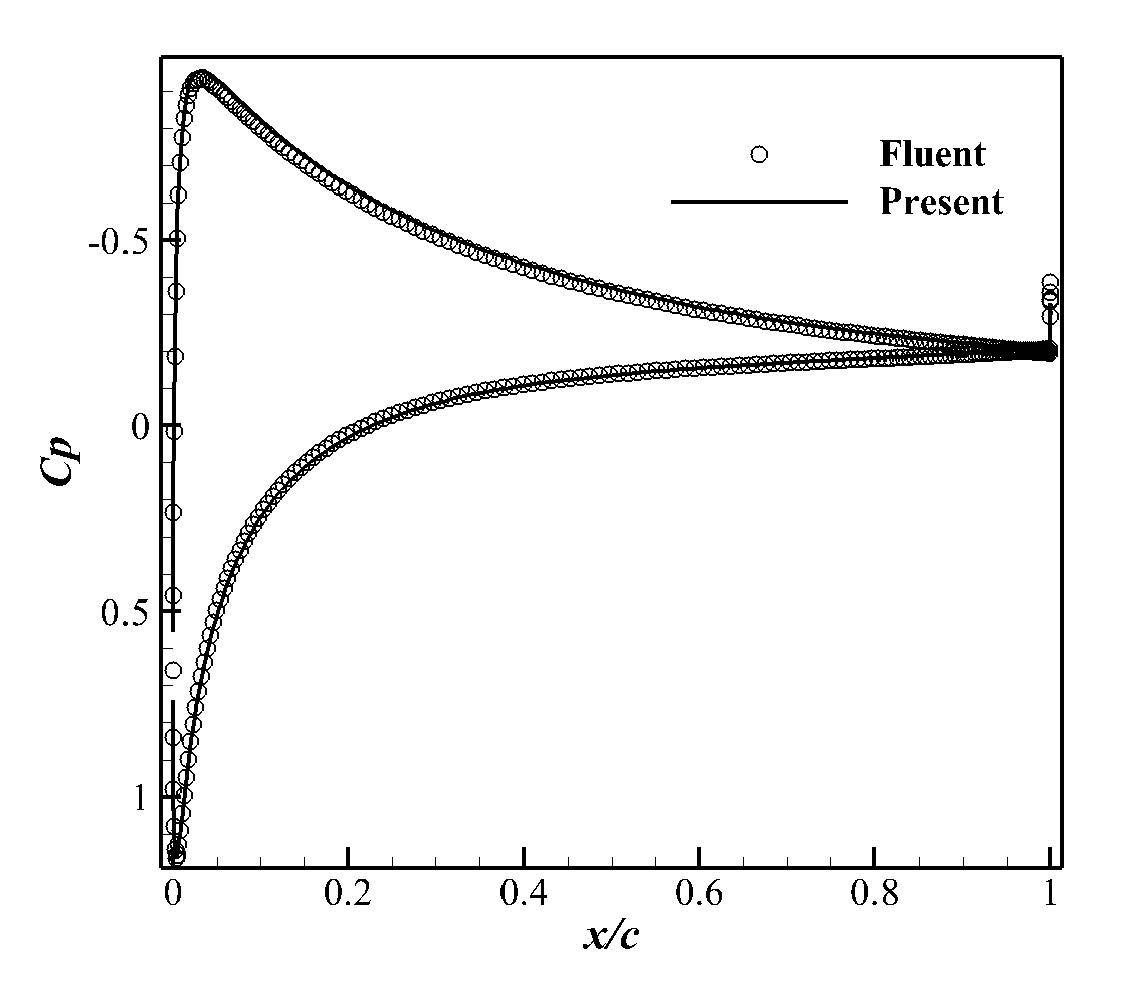}
	   \label{AOA8_cp}
    }
	\subfigure[]{
	   \includegraphics[width=0.4 \textwidth]{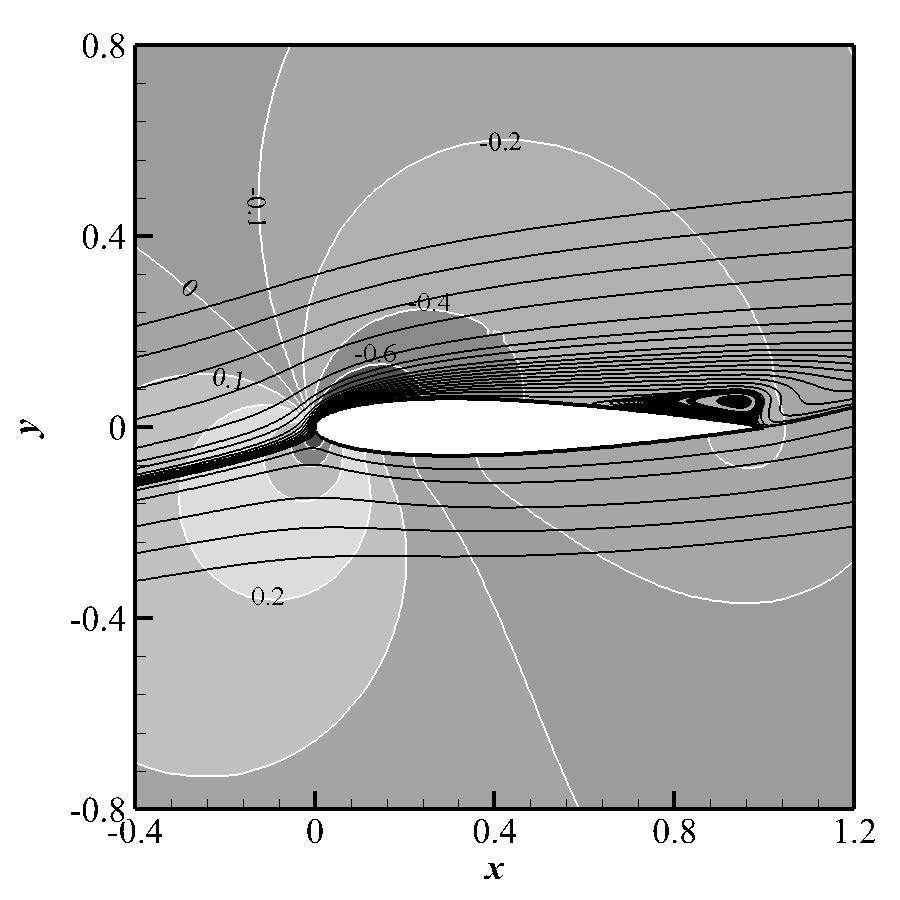}
	   \label{AOA8_cpcontour_and_streamline}
    }
    \caption{\label{AOA8_cp_Cpcontour_and_streamline} (a) The pressure coefficient distribution over NACA 0012 airfoil at $AOA = 8^{\circ}$. (b) Pressure coefficient contour and streamline around NACA 0012 airfoil at $AOA = 8^{\circ}$.}
\end{figure}

\begin{figure}
	\centering
	\subfigure[Cross sections]{
		\includegraphics[width=0.4 \textwidth]{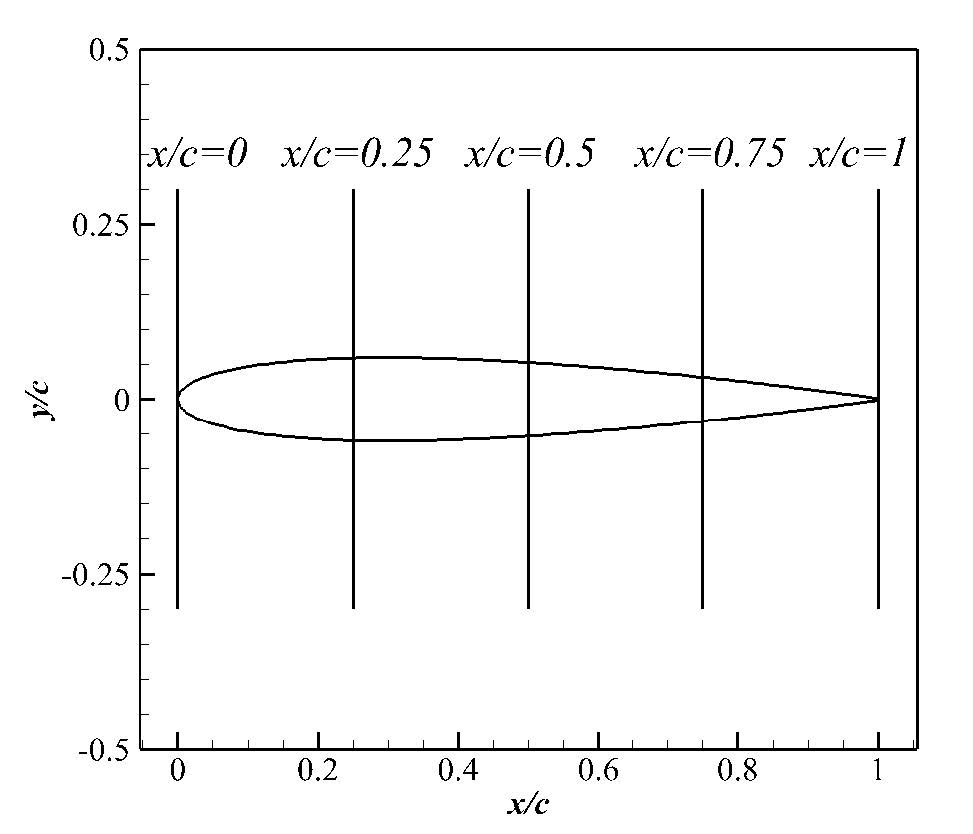}
	}
	\subfigure[$x = 0$]{
		\includegraphics[width=0.4 \textwidth]{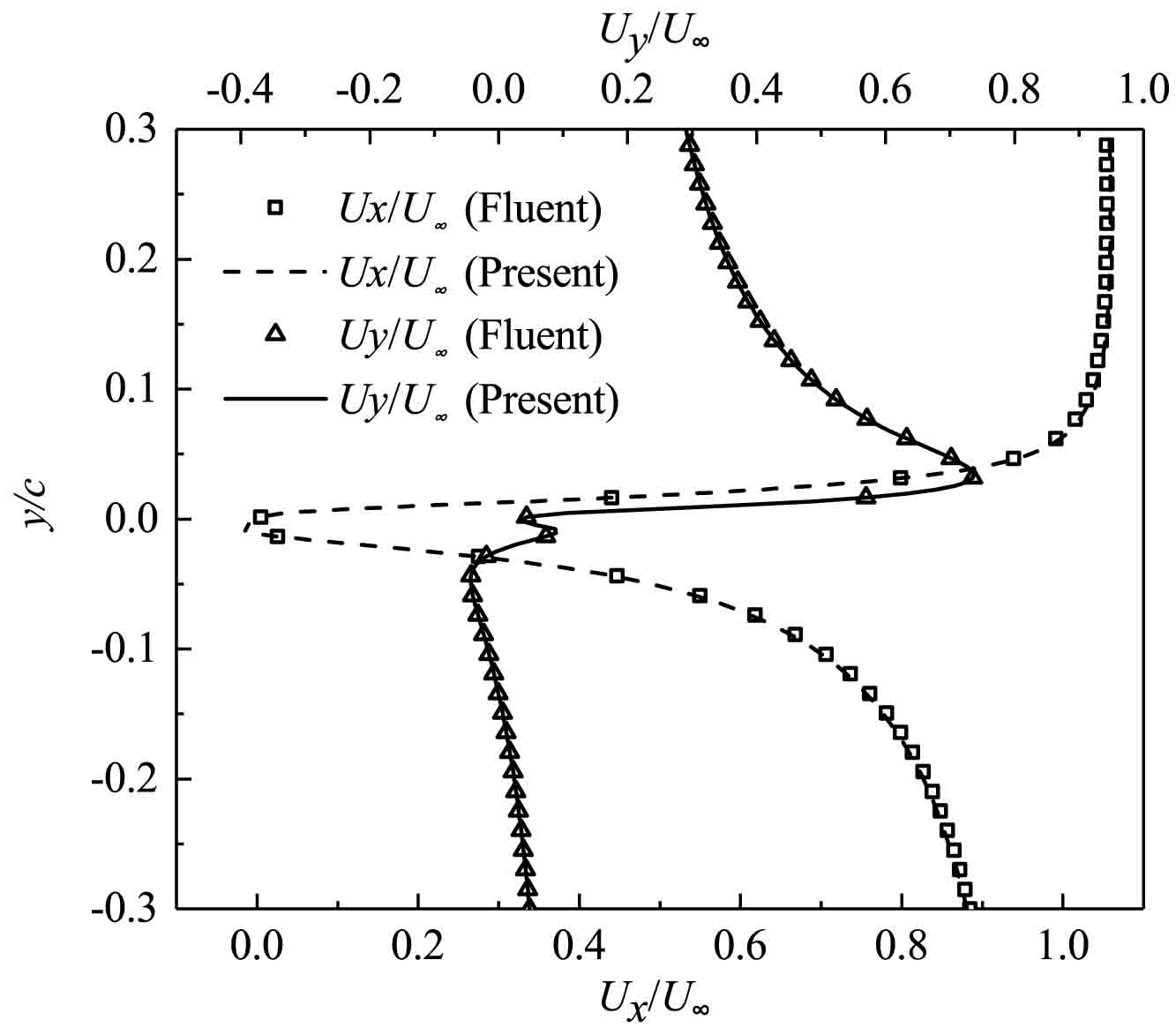}
    }
	\subfigure[$x = 0.25$]{
		\includegraphics[width=0.4 \textwidth]{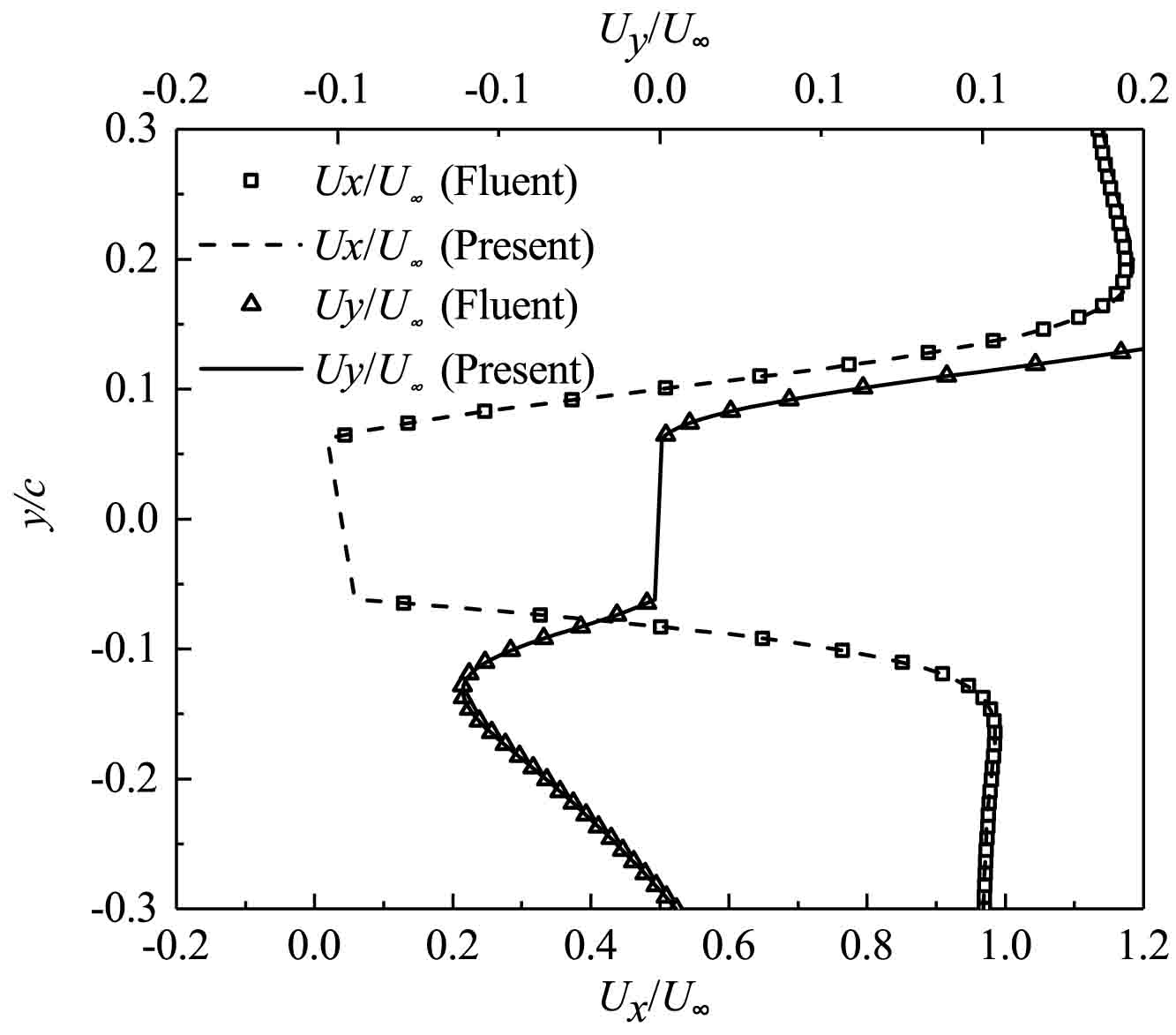}
	}
	\subfigure[$x = 0.5$]{
		\includegraphics[width=0.4 \textwidth]{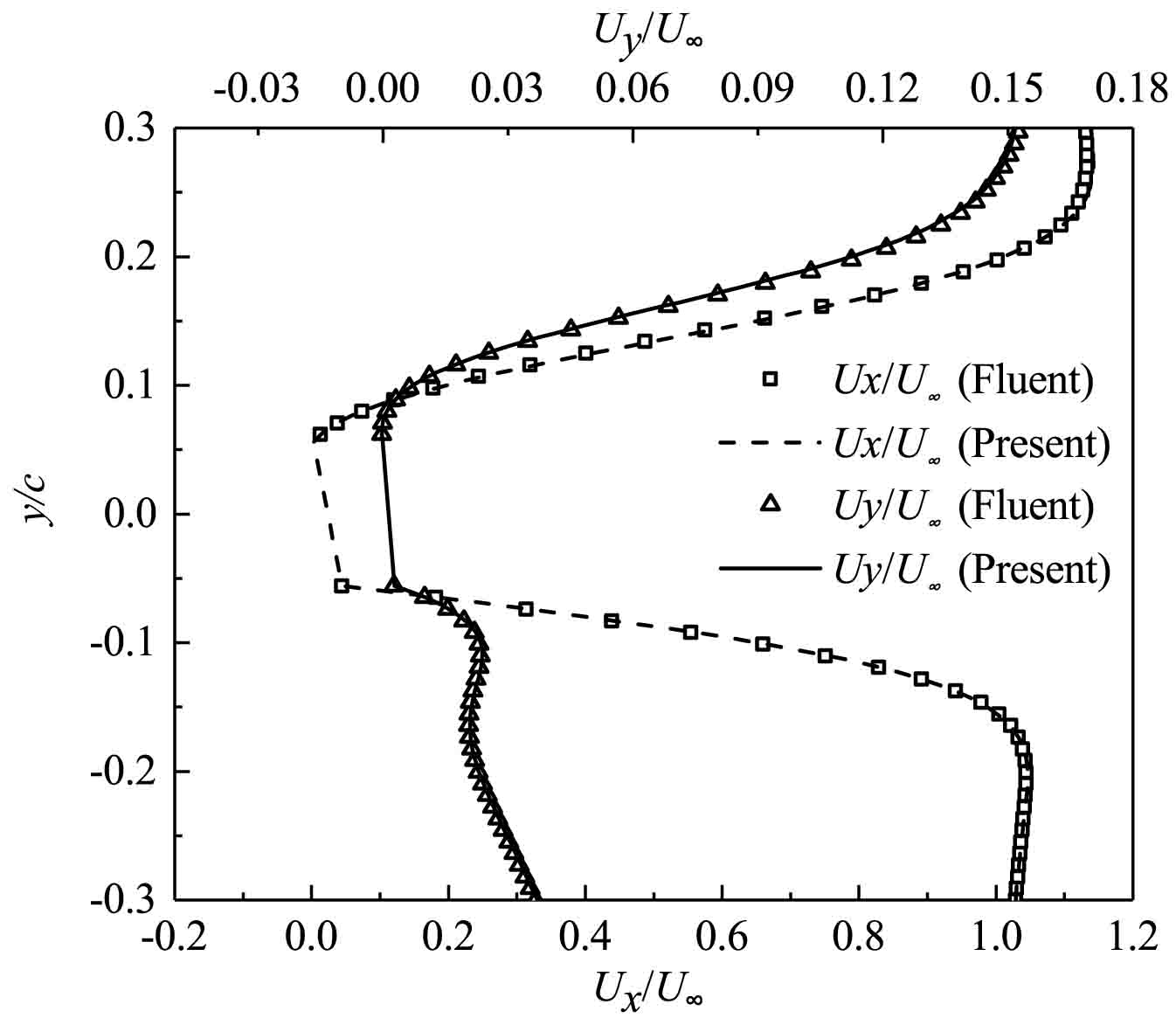}
    }
	\subfigure[$x = 0.75$]{
		\includegraphics[width=0.4 \textwidth]{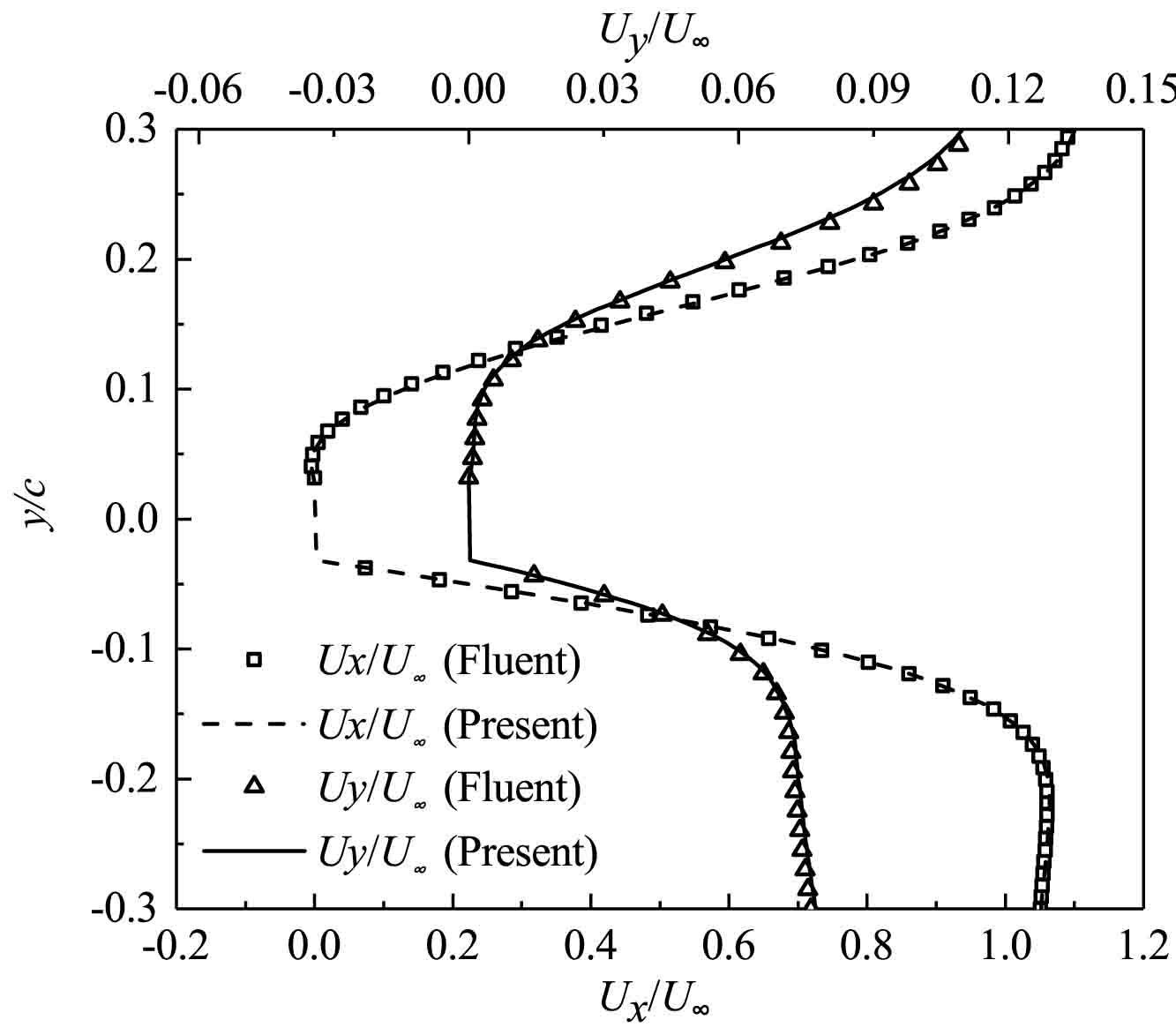}
    }
    \subfigure[$x = 1.0$]{
		\includegraphics[width=0.4 \textwidth]{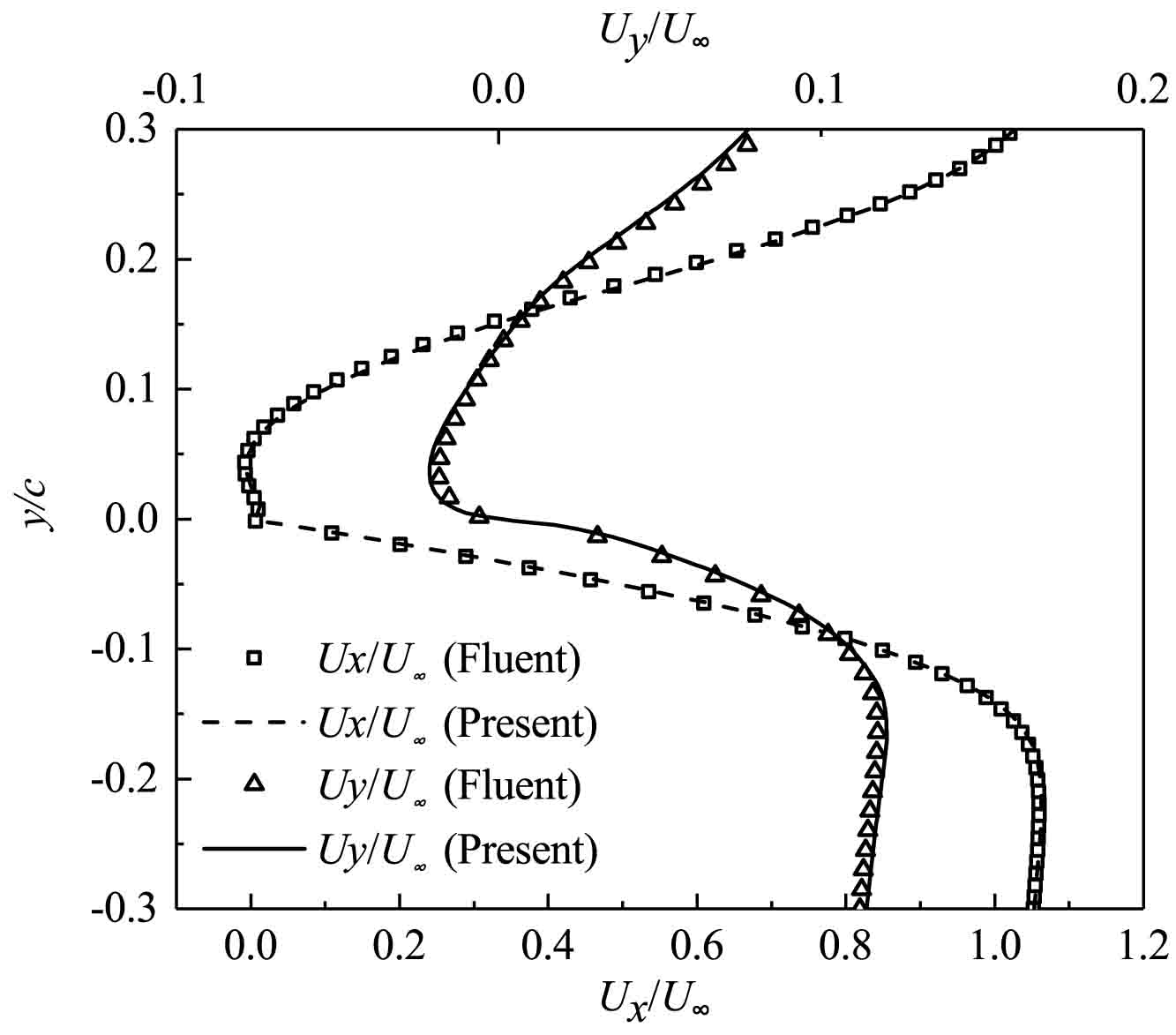}
    }
	\caption{\label{Velocity_AOA8} Comparison results for velocity profiles of laminar flow around a NACA 0012 airfoil at $AOA = 8^{\circ}$ at various cross sections $x$.}
\end{figure}

Similar to $AOA = 0^{\circ}$ condition, we simulate $AOA = 8^{\circ}$ case. The drag and lift coefficient and its comparison with Fluent are shown in Tab.~\ref{tab:NACA0012_cd_cl_AOA8}. The pressure coefficient $C_p$, pressure coefficient contour, and streamline around NACA 0012 airfoil are shown in Fig.~\ref{AOA8_cp_Cpcontour_and_streamline}. The velocity profiles at different cross section are shown in Fig.~\ref{Velocity_AOA8}.
\begin{table}
    \centering
    \caption{\label{tab:NACA0012_cd_cl_AOA0} Drag and lift coefficient at $AOA = 0^{\circ}$ compared with GILBM and CFL3D}
    \setlength{\tabcolsep}{2mm}
    \begin{ruledtabular}
    \begin{tabular}{*{4}{c}}
        Resolution (on airfoil) & $\Delta x_{min}$ & $C_d$ & $C_l$\\
        \hline
        \multicolumn{4}{c}{GILBM}\\
        16705 (173) & $4.5E-03$ & 0.1682 & $1.0E-13$\\
        16705 (173) & $4.5E-04$ & 0.1736 & $1.0E-13$\\
        52539 (251) & $4.5E-03$ & 0.1672 & $1.0E-13$\\
        52539 (251) & $4.5E-04$ & 0.1725 & $1.0E-13$\\
        \multicolumn{4}{c}{CFL3D}\\
        52539 & $1.2E-04$ & 0.1741 & $5.38E-06$\\
        16705 &   & 0.1762 & $1.15E-07$\\
        \multicolumn{4}{c}{Present}\\
        38910 (325) & $4.5E-4$ & 0.1757 & $9.52E-05$
    \end{tabular}
    \end{ruledtabular}
\end{table}

\begin{table}
    \centering
    \caption{\label{tab:NACA0012_cd_cl_AOA8} Drag and lift coefficient at $AOA = 8^{\circ}$ compared with Fluent}
    \setlength{\tabcolsep}{2mm}
    \begin{ruledtabular}
    \begin{tabular}{*{4}{c}}
        Resolution (on airfoil) & $\Delta x_{min}$ & $C_d$ & $C_l$\\
        \hline
        \multicolumn{4}{c}{Fluent}\\
        38910(325) & $4.5E-4$ & 0.19857 & 0.39619\\
        \multicolumn{4}{c}{Present}\\
        38910(325) & $4.5E-4$ & 0.19476 & 0.38127
    \end{tabular}
    \end{ruledtabular}
\end{table}

\subsubsection{Micro cavity flow}
\begin{figure}
	\centering
	\subfigure[]{
		\includegraphics[width=0.4 \textwidth]{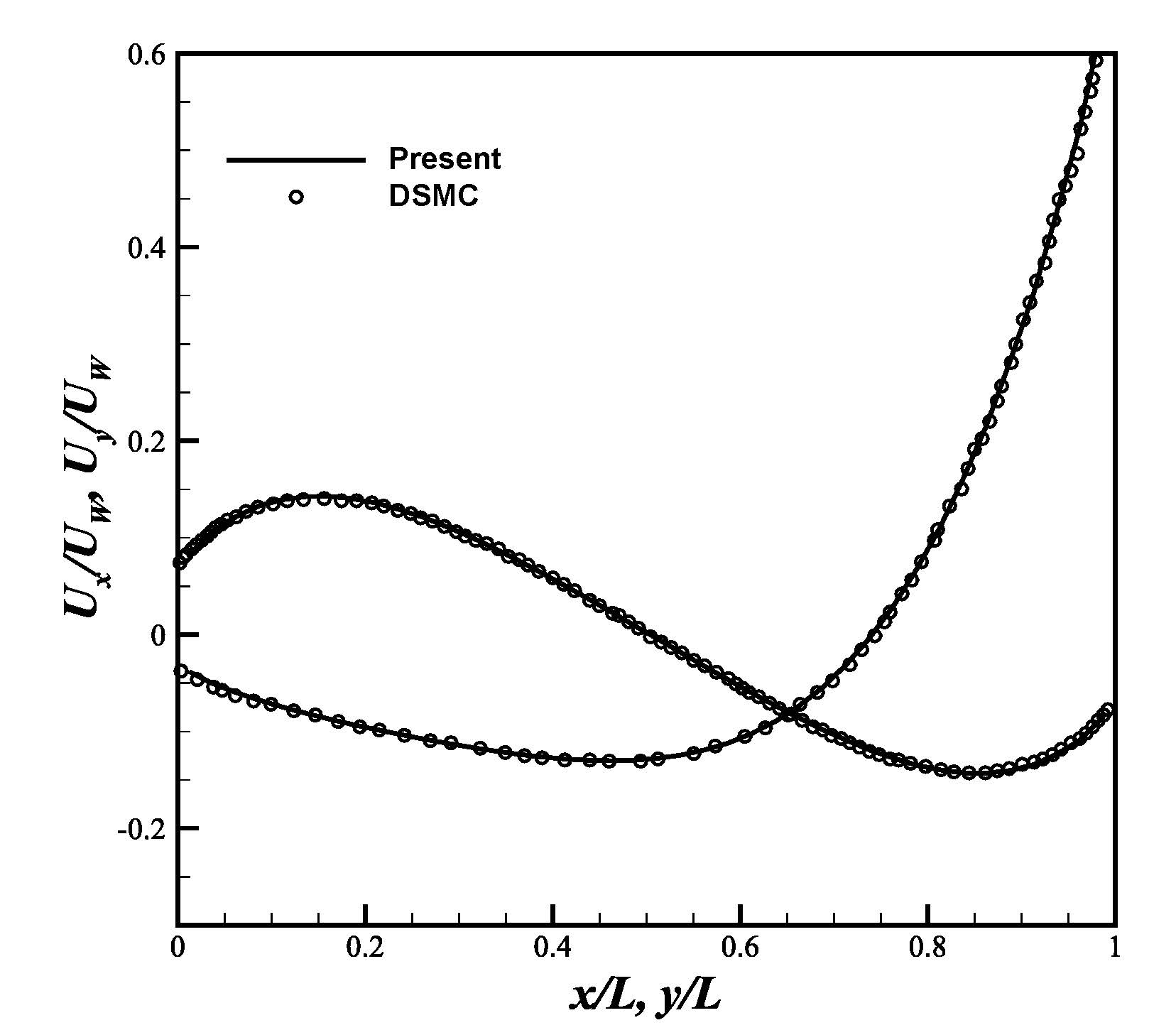}
		\label{Kn0_1}
	}
	\subfigure[]{
		\includegraphics[width=0.4 \textwidth]{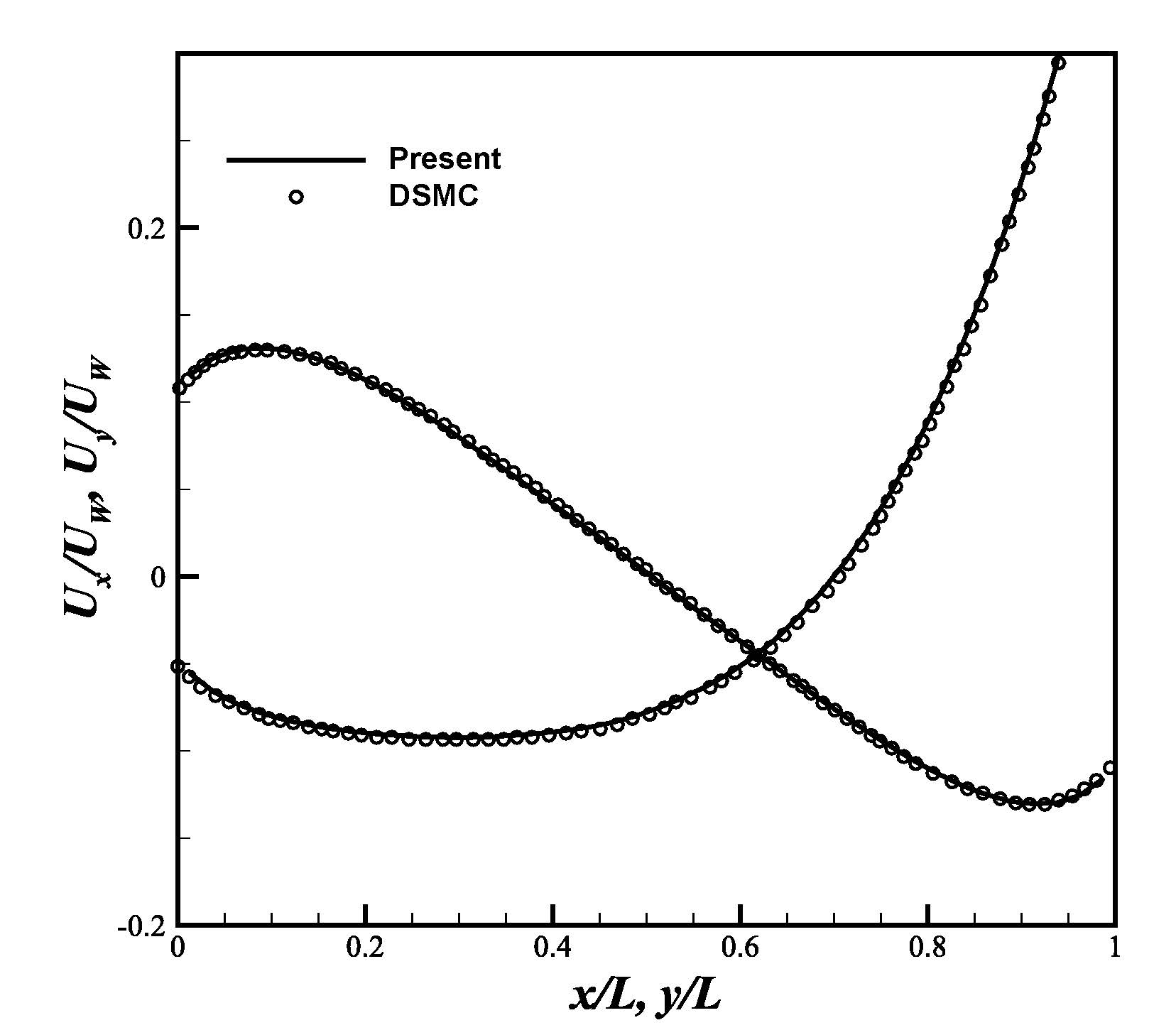}
		\label{Kn1}
	}	
    \subfigure[]{
		\includegraphics[width=0.4 \textwidth]{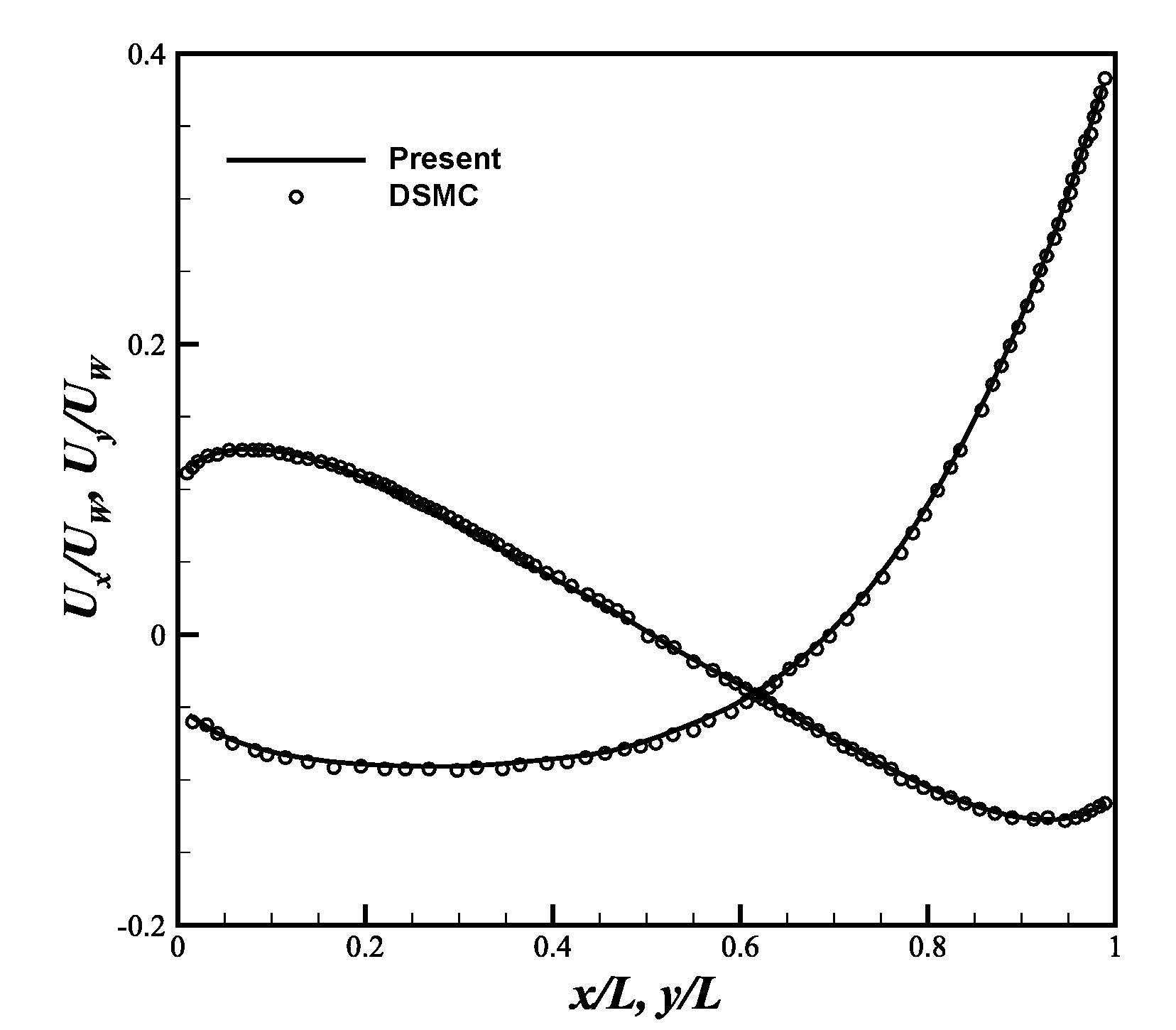}
		\label{Kn2}
	}
	\subfigure[]{
		\includegraphics[width=0.4 \textwidth]{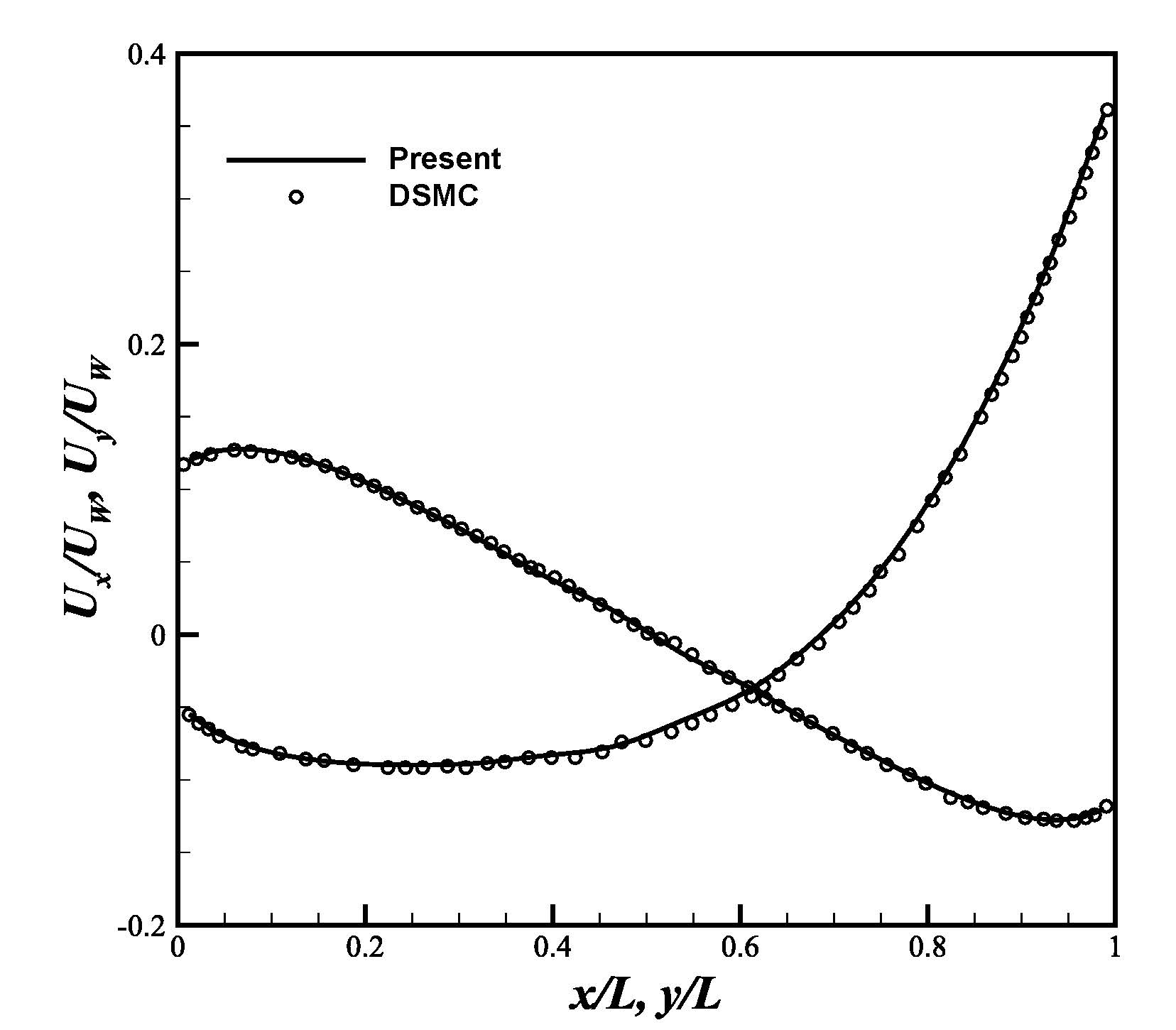}
		\label{Kn8}
	}
    \caption{\label{Microcavity_flow} Comparison results of micro cavity flow case for u-velocity along the central vertical line and v-velocity along the central horizontal line at (a)$Kn = 0.1$, and (b)$Kn = 1$, and (c)$Kn = 2$, and (d)$Kn = 8$.}
\end{figure}

The present SDUGKS is applied to micro cavity flow compared with simulations based on Direct Simulation Monte Carlo (DSMC) Method. In the simulation, the Knudsen numbers are set as $Kn = 0.1, 1, 2, 8$. The velocity of top wall $u_w = 0.1$. The velocity space is discretized into $100 \times 100$ nodes, and distribute uniformly in $[-2, 2] \times [-2, 2]$. The computational domain is divided into $40\times 40$ uniform mesh cells. The length of cavity is $L = 1.0$. On the boundary, the diffuse boundary condition is adopted. \\
The velocity profiles across the cavity center for different Knudsen numbers are shown in Fig.~\ref{Microcavity_flow}, which shows that the present method fits the DSMC results well in all cases. Present method is verified to be an accurate solver for all flow regimes. Its multi-scale property also makes it a reliable tool for all Knudsen number flows study.

\section{Conclusion}\label{Conclusion}
In the present study, we propose a SDUGKS. In the spatial interpolation aspect, an accurate and robust method is expected. And in the previous study, LLSR was proved to be an efficient and accurate method for interpolation in reconstruction step. So, in this work, LLSR is still adopted for spatial interpolation on unstructured mesh. In the proposed scheme, the flux at half time step in the original DUGKS is omitted. The mesoscopic flux of the cell on next time step is predicted by reconstruction of transformed distribution function at interfaces along particle velocity characteristic lines. Macroscopic variables of the cell on next time step are updated by the flux of gas distribution function at interfaces of the cell on current time step according to the conservation law. Conservation is improved and larger time step can be used. The particle nature is retained by reconstruction based on particle velocity characteristic line. The SDUGKS is validated by several test cases. The Couette flow is simulated using unstructured mesh to verify that our new scheme has second order spatial accuracy. The lid-driven cavity flow, laminar flow over the flat plate, flows over a circular cylinder and a NACA 0012 airfoil are carried out to demonstrate the ability of our proposed method to simulate complex flows. Micro cavity flow shows that present method is applicable to rarefied flow simulations. By introducing prediction of gas distribution function of interfaces on next time step in flux evaluation, particle motion is traced by kinetic temporal/spatial reconstruction. Physical evolution process is replaced by numerical marching in a mesh cell within a whole time step. As a result, the stability is preserved with the same accuracy as original DUGKS. A framework rather easy to learn and less complicated to code will also extend its application.

\section*{Acknowledgements}
This work is supported by National Numerical Wind tunnel project, the National Natural Science Foundation of China (No. 11902264), the Natural Science Basic Research Plan in Shaanxi Province of China (Program No. 2019JQ-315), and the 111 Project of China (B17037).

\section*{DATA AVAILABILITY}
The data that support the findings of this study are available from the corresponding author upon reasonable request.

\clearpage
\newpage 




\end{document}